\documentclass[prd, twocolumn, amsmath, floatfix, preprintnumbers, nofootinbib,superscriptaddress,letterpaper]{revtex4-1}
\usepackage[plainpages=false, colorlinks=true, anchorcolor=blue, linkcolor=blue, citecolor=blue, bookmarks=false]{hyperref}
\usepackage[toc,page]{appendix}
\usepackage{graphicx}
\usepackage{amsfonts}
\usepackage{amsmath}
\usepackage{amssymb}
\usepackage{multirow}
\usepackage{enumerate}
\usepackage[normalem]{ulem}
\usepackage[usenames]{xcolor}
\usepackage{float}

\def\cqg{Class. Quantum Grav.}
\def\aap{Astron. Astrophys.}
\def\mnras{Mon. Not. Roy. Astron. Soc.}
\def\apjs{Astrophys. J. Suppl. Ser.}
\def\apjl{Astrophys. J. Lett.}
\def\araa{Ann. Rev. Astron. Astrophys.}

\def\physrep{Phys. Rep.}

\begin{document}
\title{Equation of State Effects on Gravitational Waves from Rotating Core Collapse}
\date{January 10, 2017}
\author{Sherwood Richers}
\email{srichers@tapir.caltech.edu}
\affiliation{TAPIR, Walter Burke Institute for Theoretical Physics,
  California Institute of Technology, Pasadena, CA, USA}
\affiliation{DOE Computational Science Graduate Fellow}
\affiliation{NSF Blue Waters Graduate Fellow}
\affiliation{Los Alamos National Lab, Los Alamos, NM, USA}

\author{Christian D. Ott}
\affiliation{TAPIR, Walter Burke Institute for Theoretical Physics,
  California Institute of Technology, Pasadena, CA, USA}
\affiliation{Yukawa Institute for Theoretical Physics,
  Kyoto University, Kyoto, Japan}

\author{Ernazar Abdikamalov}
\affiliation{Department of Physics, School of Science and Technology, Nazarbayev University, Astana 010000, Kazakhstan}

\author{Evan O'Connor}
\affiliation{Department of Physics, North Carolina State University, Raleigh, NC, USA}
\affiliation{Hubble Fellow}

\author{Chris Sullivan}
\affiliation{National Superconducting Cyclotron Laboratory,
  Michigan State University, East Lansing, MI, USA}
\affiliation{Department of Physics and Astronomy,
  Michigan State University, East Lansing, MI, USA}
\affiliation{Joint Institute for Nuclear Astrophysics: Center for the Evolution of the Elements,
  Michigan State University, East Lansing, MI, USA}

\begin{abstract}
Gravitational waves (GWs) generated by axisymmetric rotating collapse,
bounce, and early postbounce phases of a galactic core-collapse
supernova will be detectable by current-generation gravitational wave
observatories. Since these GWs are emitted from the
quadrupole-deformed nuclear-density core, they may encode information
on the uncertain nuclear equation of state (EOS).  We examine the
effects of the nuclear EOS on GWs from rotating core collapse and
carry out 1824 axisymmetric general-relativistic hydrodynamic
simulations that cover a parameter space of 98 different rotation
profiles and 18 different EOS. We show that the bounce GW signal is
largely independent of the EOS and sensitive primarily to the ratio of
rotational to gravitational energy, $T/|W|$, and at high rotation
rates, to the degree of differential rotation. The GW frequency
($f_\mathrm{peak} \sim 600-1000\,\mathrm{Hz}$) of postbounce core
oscillations shows stronger EOS dependence that can be parameterized
by the core's EOS-dependent dynamical frequency
$\sqrt{G\bar{\rho}_c}$. We find that the ratio of the peak frequency
to the dynamical frequency $f_\mathrm{peak}/\sqrt{G\bar{\rho_c}}$
follows a universal trend that is obeyed by all EOS and rotation
profiles and that indicates that the nature of the core oscillations
changes when the rotation rate exceeds the dynamical frequency. We
find that differences in the treatments of low-density nonuniform
nuclear matter, of the transition from nonuniform to uniform nuclear
matter, and in the description of nuclear matter up to around twice
saturation density can mildly affect the GW signal. More exotic,
higher-density physics is not probed by GWs from rotating core
collapse. We furthermore test the sensitivity of the GW signal to
variations in the treatment of nuclear electron capture during
collapse. We find that approximations and uncertainties in electron
capture rates can lead to variations in the GW signal that are of
comparable magnitude to those due to different nuclear EOS. This
emphasizes the need for reliable experimental and/or theoretical
nuclear electron capture rates and for self-consistent
multi-dimensional neutrino radiation-hydrodynamic simulations of
rotating core collapse.
\end{abstract}

\maketitle

\section{introduction}
Massive stars ($M_\mathrm{ZAMS}\gtrsim10\,M_\odot$) burn their
thermonuclear fuel all the way up to iron-group nuclei at the top of
the nuclear binding energy curve. The resulting iron core is inert and
supported primarily by the pressure of relativistic degenerate
electrons. Once the core exceeds its effective Chandrasekhar mass
(e.g., \cite{bethe:90}), collapse commences.

As the core is collapsing, the density quickly rises, electron
degeneracy increases, and electrons are captured onto protons and
nuclei, causing the electron fraction to decrease. Within a few tenths
of a second after the onset of collapse, the density of the homologous
inner core surpasses nuclear densities. The collapse is abruptly
stopped as the nuclear equation of state (EOS) is rapidly stiffened by
the strong nuclear force, causing the inner core to bounce back and
send a shock wave through the supersonically infalling outer core.

The prompt shock is not strong enough to blow through the entire star;
it rapidly loses energy dissociating accreting iron-group
nuclei and to neutrino cooling. The shock stalls. Determining what
revives the shock and sends it through the rest of the star has been
the bane of core-collapse supernova (CCSN) theory for half a
century. In the \emph{neutrino mechanism} \cite{bethewilson:85}, a
small fraction ($\lesssim 5-10\%$) of the outgoing neutrino luminosity
from the protoneutron star (PNS) is deposited behind the stalled
shock. This drives turbulence and increases thermal pressure. The
combined effects of these may revive the shock \cite{couch:15a} and
the neutrino mechanism can potentially explain the vast majority of
CCSNe (e.g., \cite{bruenn:16}). In the \emph{magnetorotational
  mechanism}
\cite{leblanc:70,bisno:70,burrows:07b,takiwaki:09,moiseenko:06,moesta:14b},
rapid rotation and strong magnetic fields conspire to generate bipolar
jet-like outflows that explode the star and could drive very energetic
CCSN explosions. Such magnetorotational explosions
  could be essential to explaining a class of massive star explosions
  that are about ten times more energetic than regular CCSNe and that
  have been associated with long gamma-ray bursts (GRBs)
  \cite{smith:11,hjorth:11,modjaz:11}. These \emph{hypernovae} make up
  $\gtrsim$$1\%$ of all CCSNe \cite{smith:11}.
  
A key issue for the magnetorotational mechanism is its need for rapid
core spin that results in a PNS with a spin-period of around a
millisecond. Little is known observationally about core rotation in
evolved massive stars, even with recent advances in
asteroseismology~\cite{dupret:09}. On theoretical grounds and on the
basis of pulsar birth spin estimates (e.g.,
\cite{heger:05,ott:06spin,fuller:15spin}), most massive stars are
believed to have slowly spinning cores. Yet, certain astrophysical
conditions and processes, e.g., chemically homogeneous evolution at low
metallicity or binary interactions, might still provide the necessary
core rotation in a fraction of massive stars sufficient to explain
extreme hypernovae and long
GRBs~\cite{woosley:06,yoon:06,demink:13,fryer:05}.

Irrespective of the detailed CCSN explosion mechanism, it is the
repulsive nature of the nuclear force at short distances that causes
core bounce in the first place and that ensures that neutron stars can
be left behind in CCSNe. The nuclear force underlying the nuclear EOS
is an effective quantum many body interaction and a piece of poorly
understood fundamental physics.  While essential for much of
astrophysics involving compact objects, we have only incomplete
knowledge of the nuclear EOS. Uncertainties are particularly large at
densities above a few times nuclear and in the transition regime
between uniform and nonuniform nuclear matter at around nuclear
saturation density \cite{lattimer:12,oertel:17}.

The nuclear EOS can be constrained by experiment (see
\cite{lattimer:12,oertel:17} for recent reviews), through fundamental
theoretical considerations (e.g.,
\cite{hebeler:10,hebeler:13,kolomeitsev:16}), or via astronomical
observations of neutron star masses and radii (e.g.,
\cite{lattimer:12,naetillae:16,oezel:16}). Gravitational wave (GW)
observations \cite{ligo:16detection} with advanced-generation
detectors such as Advanced LIGO~\cite{aligo}, KAGRA~\cite{kagra:12},
and Advanced Virgo~\cite{advirgo:09} open up another observational
window for constraining the nuclear EOS. In the inspiral phase of
neutron star mergers (including double neutron stars and neutron star
-- black hole binaries), tidal forces distort the neutron star
shape. These distortions depend on the nuclear EOS. They measurably
affect the late inspiral GW signal (e.g.,
\cite{bernuzzi:12a,bernuzzi:15,flanagan:08,read:09b}). At merger,
tidal disruption of a neutron star by a black hole leads to a sudden
cut off of the GW signal, which can be used to constrain EOS
properties \cite{vallisneri:00,shibata:08,read:09b}. In the double
neutron star case, a hypermassive metastable or permanently stable
neutron star remnant may be formed. It is triaxial and extremely
efficiently emits GWs with characteristics (amplitudes, frequencies,
time-frequency evolution) that can be linked to the nuclear EOS (e.g,
\cite{radice:17a,bernuzzi:16a,stergioulas:11,bauswein:12b,bauswein:14}).

CCSNe may also provide GW signals that could constrain the nuclear EOS
\cite{dimmelmeier:08,roever:09,marek:09b,kuroda:16b}. In this paper,
we address the question of how the nuclear EOS affects GWs emitted at
core bounce and in the very early post-core-bounce phase ($t -
t_\mathrm{bounce} \lesssim 10\,\mathrm{ms}$) of rotating core
collapse. Stellar core collapse and the subsequent CCSN evolution are
extremely rich in multi-dimensional dynamics that emit GWs with a
variety of characteristics (see \cite{ott:09,kotake:13review} for
reviews). Rotating core collapse, bounce, and early postbounce
evolution are particularly appealing for studying EOS effects because
they are essentially axisymmetric (2D) \cite{ott:07prl,ott:07cqg} and
result in deterministic GW emission that depends on the nuclear EOS,
neutrino radiation-hydrodynamics, and gravity alone. Complicating
processes, such as prompt convection and neutrino-driven convection
set in only later and are damped by rotation (e.g.,
\cite{ott:09,dimmelmeier:08,fh:00}). While rapid rotation will amplify
magnetic field, amplification to dynamically relevant field strengths
is expected only tens of milliseconds after
bounce~\cite{burrows:07b,takiwaki:11,moesta:14b,moesta:15}. Hence,
magnetohydrodynamic effects are unlikely to have a significant impact
on the early rotating core collapse GW signal
\cite{obergaulinger:06a}.

GWs from axisymmetric rotating core collapse, bounce, and the first
ten or so milliseconds of the postbounce phase can, in principle, be
templated to be used in matched-filtering approaches to GW detection
and parameter estimation
\cite{dimmelmeier:08,ott:12a,engels:14,abdikamalov:14}. That is,
without stochastic (e.g., turbulent) processes, the GW signal is
deterministic and predictable for a given progenitor, EOS, and set of
electron capture rates. Furthermore, GWs from rotating core collapse
are expected to be detectable by Advanced-LIGO class observatories
throughout the Milky Way and out to the Magellanic Clouds
\cite{gossan:16}.

Rotating core collapse is the most extensively studied GW emission
process in CCSNe. Detailed GW predictions on the basis of (then 2D)
numerical simulations go back to M\"uller (1982)
\cite{mueller:82}. Early work showed a wide variety of types of
signals
\cite{mueller:82,zwerger:97,moenchmeyer:91,yamadasato:95,kotake:03,ott:04,dimmelmeier:02}. However,
more recent 2D/3D general-relativistic (GR) simulations that included
nuclear-physics based EOS and electron capture
during collapse demonstrated that all GW signals from rapidly rotating
core collapse exhibit a single core bounce followed by PNS
oscillations over a wide range of rotation profiles and progenitor
stars~\cite{ott:07prl,ott:07cqg,dimmelmeier:08,dimmelmeier:07,abdikamalov:14,ott:12a}.
Ott~\emph{et al.} \cite{ott:12a} showed that given the same specific
angular momentum per enclosed mass, cores of different progenitor
stars proceed to give essentially the same rotating core collapse GW
signal. Abdikamalov~\emph{et al.}  \cite{abdikamalov:14} went a step
further and demonstrated that the GW signal is determined primarily by
the mass and ratio of rotational kinetic energy to gravitational
energy ($T/|W|$) of the inner core at bounce.

The EOS dependence of the rotating core collapse GW signal has thus
far received little attention. Dimmelmeier~\emph{et
  al.}~\cite{dimmelmeier:08} carried out 2D GR hydrodynamic rotating
core collapse simulations using two different EOS
(LS180~\cite{lseos:91,lseosweb} and
HShen~\cite{shen:98a,shen:98b,hshen:11,sheneosweb}), four different
progenitors ($11\,M_\odot-40\,M_\odot$), and 16 different rotation
profiles. They found that the rotating core collapse GW signal changes
little between the LS180 and the HShen EOS, but that there may be a
slight ($\sim$$5\%$) trend of the GW spectrum toward higher
frequencies for the softer LS180 EOS. Abdikamalov~\emph{et
  al.}~\cite{abdikamalov:14} carried out simulations with the
LS220~\cite{lseos:91,lseosweb} and the
HShen~\cite{shen:98a,shen:98b,hshen:11,sheneosweb} EOS. However, they
compared only the effects of differential rotation between EOS and did
not carry out an overall analysis of EOS effects.

In this study, we build upon and substantially extend previous work on
rotating core collapse. We perform 2D GR hydrodynamic simulations
using one $12$-$M_\odot$ progenitor star model, 18 different nuclear
EOS, and 98 different initial rotational setups. We carry out a total
of 1824 simulations and analyze in detail the influence of the nuclear
EOS on the rotating core collapse GW signal. The resulting waveform
catalog is an order of magnitude larger than previous GW catalogs for
rotating core collapse and is publicly available at at
\url{https://stellarcollapse.org/Richers\_2017\_RRCCSN\_EOS}.

The results of our study show that the nuclear EOS affects rotating
core collapse GW emission through its effect on the mass of inner core
at bounce and the central density of the postbounce PNS. We
furthermore find that the GW emission is sensitive to the treatment of
the transition of nonuniform to uniform nuclear matter, to the
treatment of nuclei at subnuclear densities, and to the EOS
parameterization at around nuclear saturation density. The interplay
of all of these elements make it challenging for Advanced-LIGO-class
observatories to discern between theoretical models of nuclear matter
in these regimes.  Since rotating core collapse does not probe
densities in excess of around twice nuclear, very little exotic
physics (e.g., hyperons, deconfined quarks) can be probed by its GW
emission.  We also test the sensitivity of our results to variations
in electron capture during collapse. Since the inner core mass at
bounce is highly sensitive to the details of electron capture and
deleptonization during collapse, our results suggest that full GR
neutrino radiation-hydrodynamic simulations with a detailed treatment
of nuclear electron capture (e.g., \cite{sullivan:16,hix:03}) will be
essential for generating truly reliable GW templates for rotating core
collapse.

The remainder of this paper is organized as follows. In
Section~\ref{sec:eos}, we introduce the 18 different nuclear EOS used
in our simulations. We then present our simulation methods in
Section~\ref{sec:methods}. In Section~\ref{sec:results}, we present
the results of our 2D core collapse simulations, investigating the
effects of the EOS and electron capture rates on the rotating core
collapse GW signal. We conclude in Section~\ref{sec:conclusions}. In
Appendix~\ref{app:yeofrho}, we provide fits to electron fraction
profiles obtained from 1D GR radiation-hydrodynamic simulations and,
in Appendix~\ref{app:numerics}, we describe results from supplemental
simulations that test various approximations.

\begin{table*}[t]
  \caption{\textbf{Summary of the employed EOS.} Names of EOS in best agreement
    with the experimental and astrophysical constraints in
    Figure~\ref{fig:constraints} are in bold font. For each EOS, we
    list the underlying model and interaction/parameter set, the
    handling of nuclei in nonuniform nuclear matter, and give the
    principal reference(s). We use CLD for ``compressible liquid drop'', RMF for
    ``relativistic mean field'', and SNA for ``single nucleus
    approximation''. We refer the reader to the individual references
    and to reviews (e.g., \cite{oertel:17,lattimer:12}) for more
    details. Note that we use versions of the EOS provided in tabular
    form that also include contributions from electrons, positrons, and photons at \url{https://stellarcollapse.org/equationofstate}.}

  \begin{tabular}{llll}
  \hline\hline
  Name & Model & Nuclei& Reference\\
  \hline
  LS180                              & CLD, Skyrme                          & SNA, CLD             & \cite{lseos:91}\\
  \textbf{LS220}                     & CLD, Skyrme                          & SNA, CLD             & \cite{lseos:91}\\
  LS375                              & CLD, Skyrme                          & SNA, CLD             & \cite{lseos:91}\\
  HShen                              & RMF, TM1                            & SNA, Thomas-Fermi Approx. & \cite{shen:98a,shen:98b,hshen:11}\\
  HShenH                             & RMF, TM1, hyperons                  & SNA, Thomas-Fermi Approx. & \cite{hshen:11}\\
  GShenNL3                           & RMF, NL3                            & Hartree Approx., Virial Expansion NSE& \cite{gshen:11a}\\
  GShenFSU1.7                        & RMF, FSUGold                        & Hartree Approx., Virial Expansion NSE& \cite{gshen:11b}\\
  \textbf{GShenFSU2.1}               & RMF, FSUGold, stiffened             & Hartree Approx., Virial Expansion NSE& \cite{gshen:11b}\\
  HSTMA                              & RMF, TMA                            & NSE         & \cite{hempel:10,hempel:12}\\
  HSTM1                              & RMF, TM1                            & NSE         & \cite{hempel:10,hempel:12}\\
  HSFSG                              & RMF, FSUGold                        & NSE         & \cite{hempel:10,hempel:12}\\
  HSNL3                              & RMF, NL3                            & NSE         & \cite{hempel:10,hempel:12}\\
  \textbf{HSDD2}                     & RMF, DD2                            & NSE         & \cite{hempel:10,hempel:12}\\
  HSIUF                              & RMF, IUF                            & NSE         & \cite{hempel:10,hempel:12}\\
  \textbf{SFHo}                      & RMF, SFHo                           & NSE         & \cite{steiner:13b}\\
  \textbf{SFHx}                      & RMF, SFHx                           & NSE         & \cite{steiner:13b}\\
  BHB$\Lambda$                       & RMF, DD2-BHB$\Lambda$, hyperons     & NSE         & \cite{banik:14}\\ 
  \textbf{BHB$\mathbf{\Lambda\Phi}$} & RMF, DD2-BHB$\Lambda\Phi$, hyperons & NSE         & \cite{banik:14}\\
  \hline\hline
\end{tabular}
\label{tab:eos}
\end{table*}

\section{Equations of State}
\label{sec:eos}
There is substantial uncertainty in the
behavior of matter at and above nuclear density, and as such, there
are a large number of proposed nuclear EOS that describe the
relationship between matter density, temperature, composition
(i.e.\ electron fraction $Y_e$ in nuclear statistical equilibrium
[NSE]), and energy density and its derivatives. Properties of the EOS
for uniform nuclear matter are often discussed in terms of a
power-series expansion of the binding energy per baryon $E$ at
temperature $T=0$ around the nuclear saturation density $n_s$ of
symmetric matter ($Y_e=0.5$)~(e.g.,
\cite{lattimer:85,hempel:12,oertel:17,lattimer:12}):

\begin{align}
  E(x,\beta) =& -E_0 +\frac{K}{18}x^2 + \frac{K'}{162} x^3 + ... + \mathcal{S}(x,\beta)\,\,,
\label{eq:bulk}
\end{align}
where $x=(n-n_s)/n_s$ for a nucleon number density $n$ and
$\beta=2(0.5-Y_e)$. The saturation density is defined as where
$dE(x,\beta)/dx=0$. The saturation number density
$n_s\approx0.16\,\mathrm{fm}^{-3}$ and the bulk binding energy of
symmetric nuclear matter $E_0\approx16\,\mathrm{MeV}$ are well
constrained from experiments~\cite{lattimer:12,oertel:17} and all EOS
in this work have a reasonable value for both. $K$ is the nuclear
incompressibility, and its density derivative $K'$ is referred to as
the skewness parameter. All nuclear effects of changing $Y_e$ away
from $0.5$ are contained in the symmetry term $\mathcal{S}(x,\beta)$,
which is also expanded around symmetric matter as
\begin{equation}
\mathcal{S}(x,\beta) = \mathcal{S}_2(x)\beta^2+\mathcal{S}_4(x)\beta^4+...\approx \mathcal{S}_2(x)\beta^2\,\,.
\end{equation}
There are only even orders in the expansion due to the charge
invariance of the nuclear interaction. Coulomb effects
  do not come into play at densities above $n_s$, where protons and
  electrons are both uniformly distributed. The $\mathcal{S}_2$ term
is dominant and we do not discuss the higher-order symmetry terms here
(see \cite{lattimer:85,lattimer:12,oertel:17}). $\mathcal{S}_2(x)$ is
itself expanded around saturation density as
\begin{equation}
\mathcal{S}_2(x) = \left(J + \frac{1}{3}Lx + ...\right)\,\,.
\end{equation}
$J$ corresponds to the symmetry term in the Bethe-Weizs\"acker mass
formula~\cite{weizsaecker:35,bethe:36}, so $J$ is what the literature
refers to as ``the symmetry energy`` at saturation density and $L$ is
the density derivative of the symmetry term.

\begin{figure}
\center \includegraphics[width=\linewidth]{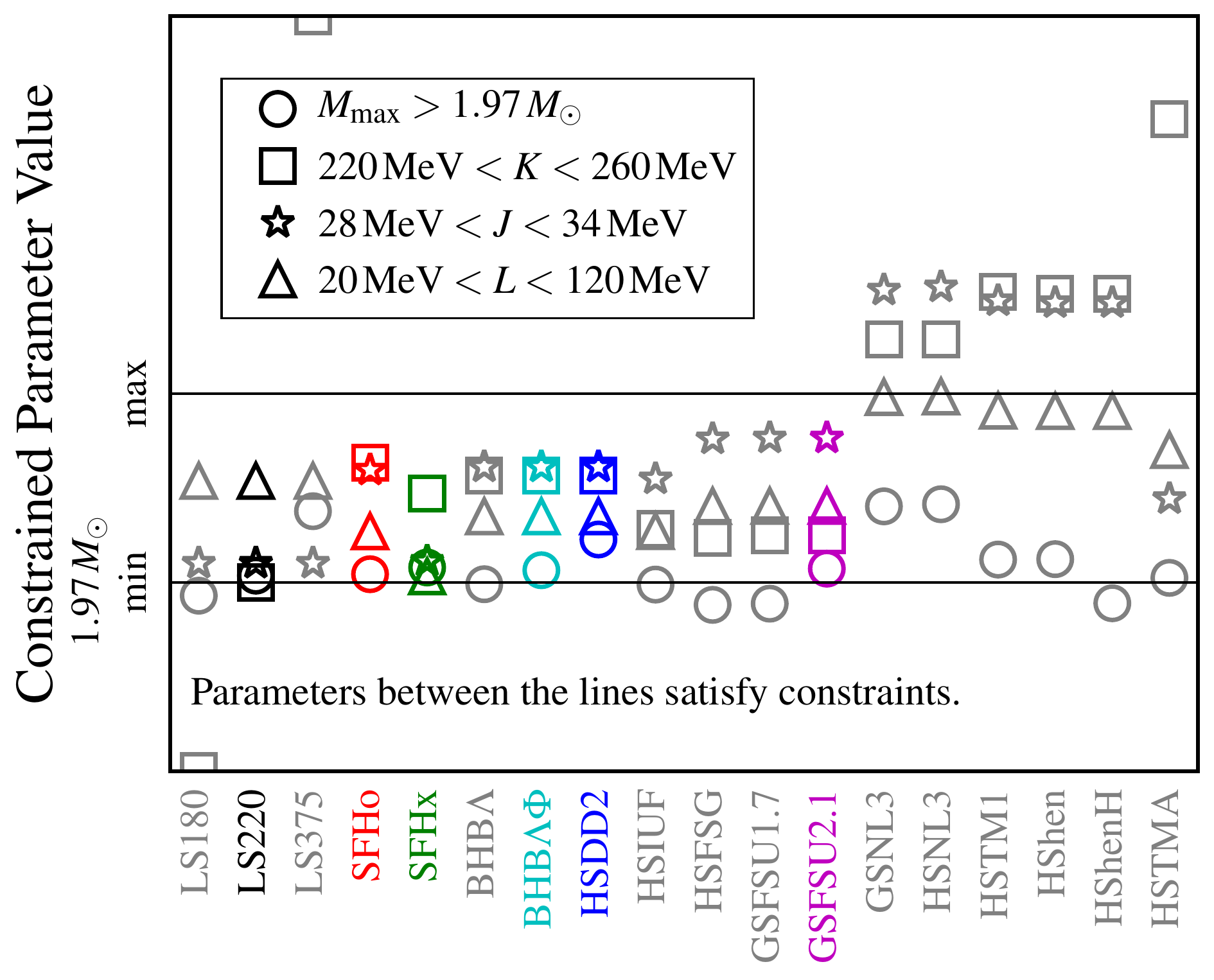}
\caption{\textbf{EOS Constraints from experiment and NS mass
    measurements.} The maximum cold neutron star gravitational mass
  $M_\mathrm{max}$, the incompressibility $K$, symmetry energy $J$,
  and the derivative of the symmetry energy $L$ are plotted. For
  $M_\mathrm{max}$, the bottom of the plot is $0$, the min line is at
  $1.97M_\odot$, and the max line is not used. The other constraints
  are normalized so the listed minima and maxima lie on the min and
  max lines. EOS that are within all of these simple constraints are
  colored, and the color code is consistent throughout the paper. Note
  that there are additional constraints on the NS mass-radius
  relationship, which we show in Figure~\ref{fig:MvsR}, and joint
  constraints on J and L~\cite{kolomeitsev:16} that we do
  not show.}
\label{fig:constraints}
\end{figure}

It is important to note that none of the above parameters can alone
describe the effects an EOS will have on a core collapse
simulation. This can be seen, for example, from the definition of the
pressure,
\begin{equation}
  P(n,Y_e) = n^2 \frac{\partial E(n,Y_e)}{\partial n}\,\,,
\end{equation}
which depends directly on $K$ and the first derivative of
$\mathcal{S}(n)$. Since the matter in core-collapse supernovae and
neutron stars is very asymmetric ($Y_e\neq 0.5$), large values for $J$
and $L$ can imply a very stiff EOS even if $K$ is not particularly
large.

The incompressibility $K$ has been experimentally constrained to $240
\pm 10\,\mathrm{MeV}$~\cite{piekarewicz:10}, though there is some
model dependence in inferring this value, making an error bar of $\pm
20\,\mathrm{MeV}$ more reasonable~\cite{steiner:13b}. A combination of
experiments, theory, and observations of neutron stars suggest that
$28\,\mathrm{MeV} \lesssim J \lesssim 34\,\mathrm{MeV}$ (e.g.,
~\cite{tsang:09}). Several experiments place varying inconsistent
constraints on $L$, but they all lie in the range of $20\,\mathrm{MeV}
\lesssim L \lesssim 120\,\mathrm{MeV}$ (e.g., \cite{carbone:10}). $K'$
and higher order parameters have yet to be constrained by experiment,
though a study of correlations of these higher-order parameters to the
low-order parameters ($K$, $J$, $L$) in theoretical EOS models
provides some estimates~\cite{chen:11}. Additional constraints on the
combination of $J$ and $L$ have been proposed that rule out many of
these EOS (most recently, \cite{kolomeitsev:16}). Finally, the mass of
neutron star PSR~J0348+0432 has been determined to be
$2.01\pm0.04\,M_\odot$ ~\cite{antoniadis:13}, which is the highest
well-constrained neutron star mass observed to date. Any realistic EOS
model must be able to support a cold neutron star of at least this
mass. Indirect measurements of neutron star radii further constrain
the allowable mass-radius region~\cite{naetillae:16}.

In this study, we use the 18 different EOS described in
Table~\ref{tab:eos}. We use tabulated versions that are available from
\url{https://stellarcollapse.org/equationofstate} that also include
contributions from electrons, positrons, and photons. Of the 18 EOS we
use, only SFHo~\cite{steiner:13b,hempeleosweb} appears to reasonably
satisfy all current constraints (including the recent constraint
proposed by \cite{kolomeitsev:16}).

Historically, the EOS of Lattimer \& Swesty~\cite{lseos:91,lseosweb}
(hereafter LS; based on the compressible liquid drop model with a
Skyrme interaction) and of H.~Shen~\emph{et
  al.}~\cite{shen:98a,shen:98b,hshen:11,sheneosweb} (hereafter HShen;
based on a relativistic mean field [RMF] model) have been the most
extensively used in CCSN simulations. The LS EOS is available with
incompressibilities $K$ of 180, 220, and 375 MeV. There is also a
version of the EOS of H.~Shen~\emph{et al.} (HShenH) that includes
effects of $\Lambda$ hyperons, which tend to soften the EOS at high
densities \cite{hshen:11}. Both the LS EOS and the HShen EOS treat
nonuniform nuclear matter in the single-nucleus approximation
(SNA). This means that they include neutrons, protons, alpha particles,
and a single representative heavy nucleus with average mass $\bar{A}$
and charge $\bar{Z}$ number in NSE.

Recently, the number of nuclear EOS available for CCSN simulations has
increased greatly. Hempel~\emph{et
  al.}~\cite{hempel:10,hempel:12,hempeleosweb} developed an EOS that
relies on an RMF model for uniform nuclear matter and nucleons in
nonuniform matter and consistently transitions to NSE with thousands
of nuclei (with experimentally or theoretically determined properties)
at low densities. Six RMF EOS by Hempel~\emph{et
  al.}~\cite{hempel:10,hempel:12,hempeleosweb} (hereafter HS) are
available with different RMF parameter sets (TMA, TM1, FSU Gold, NL3,
DD2, and IUF). Based on the Hempel model, the EOS by Steiner~\emph{et
  al.}~\cite{steiner:13b,hempeleosweb} require that experimental and
observational constraints are satisfied. They fit the free parameters
to the maximum likelihood neutron star mass-radius curve (SFHo) or
minimize the radius of low-mass neutron stars while still satisfying
all constraints known at the time (SFHx).  SFH\{o,x\} differ from the
other Hempel EOS only in the choice of RMF parameters.

The EOS by Banik~\emph{et al.}~\cite{banik:14,hempeleosweb} are based
on the Hempel model and the RMF DD2 parameterization, but also include
$\Lambda$ hyperons with (BHB$\Lambda \phi$) and without (BHB$\Lambda$)
repulsive hyperon-hyperon interactions.

The EOS by G.~Shen~\emph{et
  al.}~\cite{gshen:11a,gshen:11b,gsheneosweb} are also based on RMF
theory with the NL3 and FSU Gold parameterizations. The GShenFSU2.1
EOS is stiffened at currently unconstrained super-nuclear densities
to allow a maximum neutron star mass that agrees with observations.
G.~Shen~\emph{et al.}\ paid particular attention to the transition
region between uniform and nonuniform nuclear matter where they
carried out detailed Hartree calculations~\cite{gshen:10a}. At lower
densities they employed an EOS based on a virial expansion that
self-consistently treats nuclear force contributions to the
thermodynamics and composition and includes nucleons and nuclei
\cite{gshen:10b}. It reduces to NSE at densities where the strong
nuclear force has no influence on the EOS.

\begin{figure}
\center \includegraphics[width=\linewidth]{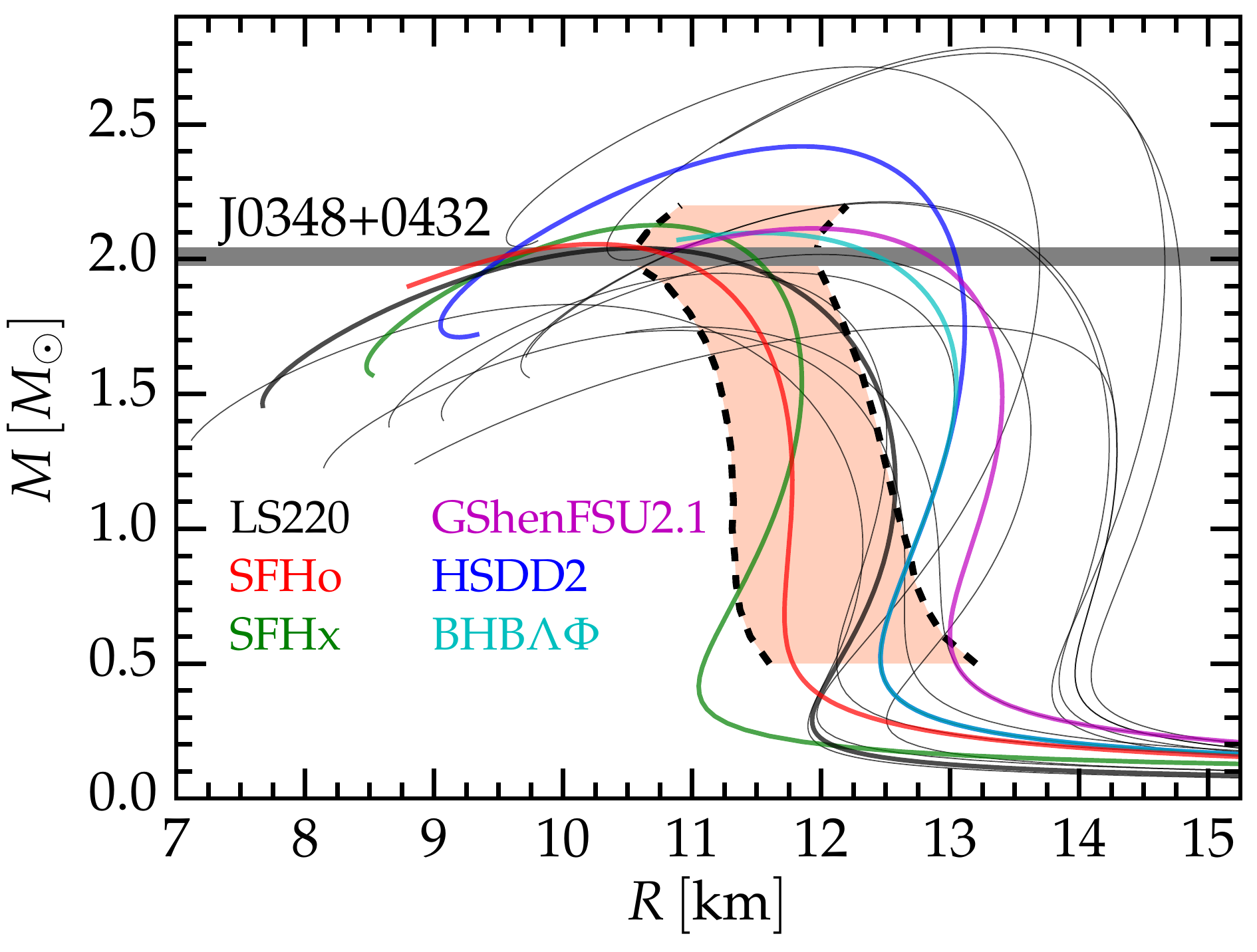}
\caption{\textbf{Neutron star mass-radius relations}. The relationship
  between the gravitational mass and radius of a cold neutron star is
  plotted for each EOS. The EOS employed in this study cover a wide
  swath of parameter space. EOS that lie within the constraints
  depicted in Figure~\ref{fig:constraints} are colored, and the color
  code is consistent throughout the paper. We show the $2\sigma$
  mass-radius constraints from ``model A'' of \cite{naetillae:16} as a
  shaded region between two dashed lines. These constraints were
  obtained from a Bayesian analysis of observations of type-I X-ray
  bursts in combination with theoretical constraints on nuclear
  matter. The EOS that agree best with these constraints are SFHo,
  SFHx, and LS220.}
\label{fig:MvsR}
\end{figure}

Few of these EOS obey all available experimental and observational
constraints. In Figure~\ref{fig:constraints} we show where each EOS
lies within the uncertainties for experimental constraints on nuclear
EOS parameters and the observational constraint on the maximum neutron
star mass. We color the EOS that satisfy the constraints, and use the
same colors consistently throughout the paper. 

The mass-radius curves of zero-temperature neutron stars in
neutrino-less $\beta$-equilibrium predicted by each EOS are shown in
Figure~\ref{fig:MvsR}. We mark the mass range for PSR~J0348+0432 with
a horizontal bar. We also include the $2\sigma$ semi-empirical
mass-radius constraints of ``model A'' of N\"atill\"a~\emph{et
  al.}~\cite{naetillae:16}. They were obtained via a Bayesian analysis
of type-I X-ray burst observations. This analysis assumed a particular
three-body quantum Monte Carlo EOS model near saturation density by
\cite{gandolfi:12} and a parameterization of the super-nuclear EOS
with a three-piece piecewise polytrope
\cite{steiner:13a,steiner:15}. Similar constraints are available from
other groups (see, e.g.,
\cite{oezel:16,oezel:16b,miller:16,guillot:14}).

Throughout this paper, we use the SFHo EOS as a fiducial standard for
comparison, since it represents the most likely fit to known
experimental and observational constraints. While many of the
considered EOS do not satisfy multiple constraints, we still include
them in this study for two reasons: (1) a larger range of EOS will
allow us to better understand and possibly isolate causes of trends
in the GW signal with EOS properties and, (2), many
constraint-violating EOS likely give perfectly reasonable
thermodynamics for matter under collapse and PNS conditions even if
they may be unrealistic at higher densities or lower temperatures.

\section{Methods}
\label{sec:methods}
As the core of a massive star is collapsing, electron capture and the
release of neutrinos drives the matter to be increasingly
neutron-rich. The electron fraction $Y_e$ of the inner core in the
final stage of core collapse has an important role in setting the mass
of the inner core, which, in turn, influences characteristics of the
emitted GWs.  Multidimensional neutrino radiation hydrodynamics to
account for these neutrino losses during collapse is still too
computationally expensive to allow a large parameter study of
axisymmetric (2D) simulations. Instead, we follow the proposal by
Liebend\"orfer~\cite{liebendoerfer:05fakenu} and approximate this
prebounce deleptonization of the matter by parameterizing the electron
fraction $Y_e$ as a function of only density (see
Appendix~\ref{app:GR1Dtests} for tests of this approximation). Since
the collapse-phase deleptonization is EOS dependent, we extract the
$Y_e(\rho)$ parameterizations from detailed spherically symmetric (1D)
nonrotating GR radiation-hydrodynamic simulations and apply them to
rotating 2D GR hydrodynamic simulations. We motivate using the
$Y_e(\rho)$ approximation also for the rotating case by the fact that
electron capture and neutrino-matter interactions are local and
primarily dependent on density in the collapse phase
\cite{liebendoerfer:05fakenu}. Hence, geometry effects due to the
rotational flattening of the collapsing core can be assumed to be
relatively small. This, however, has yet to be demonstrated with full
multi-dimensional radiation-hydrodynamic simulations. Furthermore, the
$Y_e(\rho)$ approach has been used in many previous studies of
rotating core collapse (e.g.,
\cite{dimmelmeier:07,dimmelmeier:08,abdikamalov:10,abdikamalov:14})
and using it lets us compare with these past results. We ignore the
magnetic field throughout this work, since are expected to grow to
dynamical strengths on timescales longer than the first
$\sim10\,\mathrm{ms}$ after core bounce that we
investigate~\cite{burrows:07b,takiwaki:11,moesta:14b,moesta:15}.

\subsection{1D Simulations of Collapse-Phase Deleptonization with {\tt GR1D}}
\label{sec:yeofrho}

We run spherically symmetric GR radiation hydrodynamic core collapse
simulations of a nonrotating $12M_\odot$ progenitor (Woosley \emph{et
  al.}~\cite{woosley:07}, model s12WH07) in our open-source code {\tt
  GR1D}~\cite{oconnor:15a}, once for each of our 18 EOS. The fiducial
radial grid consists of 1000 zones extending out to
$2.64\times10^4\,\mathrm{km}$, with a uniform grid spacing of
$200\,\mathrm{m}$ out to $20\,\mathrm{km}$ and logarithmic spacing
beyond that. We test the resolution in Appendix~\ref{app:GR1Dtests}.

The neutrino transport is handled with a two-moment scheme with 24
logarithmically-spaced energy groups from $0$ to
$287\,\mathrm{MeV}$. This allows us to treat the effects of neutrino
absorption and emission explicitly and self-consistently. The neutrino
interaction rates are calculated by {\tt NuLib}~\cite{oconnor:15a} and
include absorption onto and emission from nucleons and nuclei
including neutrino blocking factors, elastic scattering off nucleons
and nuclei, and inelastic scattering off electrons. We neglect
bremsstrahlung and neutrino pair creation and annihilation, since they
are unimportant during collapse and shortly after core bounce
(e.g.,~\cite{lentz:12b}). To ensure a consistent treatment of electron
capture for all EOS, the rates for absorption, emission, and
scattering from nuclei are calculated using the SNA. To test this
approximation, in Section~\ref{sec:ecapture}, we run additional
simulations with experimental and theoretical nuclear electron capture
rates instead included individually for each of the heavy nuclei in an
NSE distribution. In Appendix~\ref{app:GR1Dtests}, we test the neutrino
energy resolution and the resolution of the interaction rate table.

\begin{figure}
\center
\includegraphics[width=\linewidth]{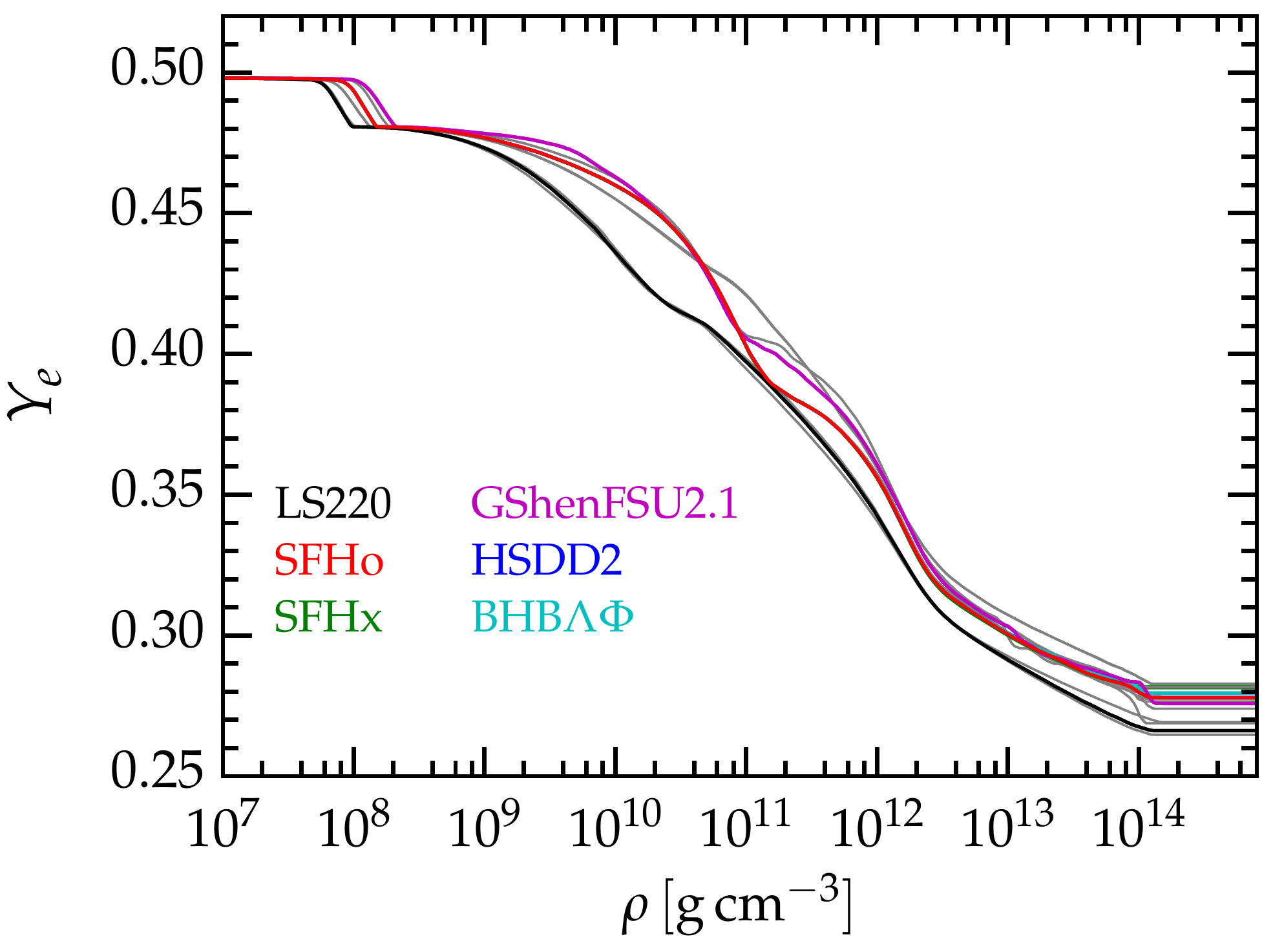}
\caption{\textbf{$\mathbf{Y_e(\rho)}$ Deleptonization Profiles.} For
  each EOS, radial profiles of the electron fraction $Y_e$ as a
  function of density $\rho$ are taken from spherically-symmetric \texttt{GR1D}
  radiation hydrodynamics simulations using two-moment neutrino
  transport at the point in time when the central $Y_e$ is smallest
  (roughly at core bounce) and are plotted here. We manually extend
  the curves out to high densities with a constant $Y_e$ to ensure
  that simulations never encounter a density outside the range provided
  in these curves. In the 2D simulations, $Y_e$ is determined by the
  density and one of these curves until core bounce.}
\label{fig:yeofrho}
\end{figure}
To generate the $Y_e(\rho)$ parameterizations, we take a fluid
snapshot at the time when the central $Y_e$ is at a minimum
($\sim0.5\,\mathrm{ms}$ prior to core bounce) and create a list of the
$Y_e$ and $\rho$ at each radius. We then manually enforce that $Y_e$
decreases monotonically with increasing $\rho$. The resulting profiles
are shown in Figure~\ref{fig:yeofrho}.

\subsection{2D Core Collapse Simulations with {\tt CoCoNuT}}
\label{sec:coconut}
\begin{table}
\caption{\textbf{Rotation Profiles.} A list of the differential
  rotation $A$ and maximum rotation rate $\Omega_0$ parameters used in
  generating rotation profiles. The $\Omega_0$ ranges imply a rotation
  profile at each $0.5\,\mathrm{rad\,s}^{-1}$ interval. In total, we
  use 98 rotation profiles.}
\begin{tabular}{crcc}
  \hline\hline
Name & $A\,[\mathrm{km}]$ & $\Omega_0\,[\mathrm{rad\,s}^{-1}]$ & \# of Profiles\\
\hline
A1 & 300 & 0.5 - 15.5 & 31 \\
A2 & 467 & 0.5 - 11.5 & 23 \\
A3 & 634 & 0.0 - \phantom{0}9.5 & 20 \\
A4 & 1268 & 0.5 - \phantom{0}6.5 & 13 \\
A5 & 10000 & 0.5 - \phantom{0}5.5 & 11 \\
\hline\hline
\end{tabular}
\label{tab:rotation}
\end{table}

\begin{table}
\caption{\textbf{No Collapse List}. We list the simulations that do
  not undergo core collapse within $1\,\mathrm{s}$ of simulation time
  due to sufficiently large centrifugal support already at the onset
  of collapse. These simulations are excluded from further analysis.}
\begin{tabular}{rcl}
  \hline\hline
$A\,[\mathrm{km}]$ & $\Omega_0\,[\mathrm{rad\,s^{-1}}]$ & EOS\\
\hline
300 & 15.5 & GShenNL3 \\
467 & 10.0 & GShenNL3 \\
    & 10.5 & GShenNL3 \\
    & 11.0 & GShen\{NL3,FSU2.1,FSU1.7\} \\
    & 11.5 & GShen\{NL3,FSU2.1,FSU1.7\} \\
634 & 8.0 & GShenNL3 \\
    & 8.5 & GShen\{NL3,FSU2.1,FSU1.7\} \\
    & 9.0 & GShen\{NL3,FSU2.1,FSU1.7\} \\
    & 9.5 & GShen\{NL3,FSU2.1,FSU1.7\} \\
    &     & LS\{180,220,375\} \\
1268 & 5.5 & GShenNL3 \\
     & 6.0 & GShen\{NL3,FSU2.1,FSU1.7\}\\
     & 6.5 & GShen\{NL3,FSU2.1,FSU1.7\}\\
     &     & LS\{180,220,375\}\\
10000 & 4.0 & GShen\{NL3,FSU2.1,FSU1.7\}\\
      & 4.5 & GShen\{NL3,FSU2.1,FSU1.7\}\\
      & 5.0 & GShen\{NL3,FSU2.1,FSU1.7\}\\
      &     & LS\{180,220,375\}\\
& 5.5 & all but HShen,HShenH\\
\hline\hline
\end{tabular}
\label{tab:nobounce}
\end{table}

We perform axisymmetric (2D) core collapse simulations using the {\tt
  CoCoNuT} code~\cite{dimmelmeier:02,dimmelmeier:05} with conformally
flat GR. We use a setup identical to that in Abdikamalov \emph{et
  al.}~\cite{abdikamalov:14}, but we review the key details here for
completeness. We generate rotating initial conditions for the 2D
simulations from the same $12\,M_\odot$ progenitor by imposing a
rotation profile on the precollapse star according to
(e.g.,~\cite{zwerger:97})
\begin{equation}
  \label{eq:rotation}
  \Omega(\varpi) = \Omega_0
  \left[1+\left(\frac{\varpi}{A}\right)^2\right]^{-1}\,\,,
\end{equation}
where $A$ is a measure of the degree of differential rotation,
$\Omega_0$ is the maximum initial rotation rate, and $\varpi$ is the
distance from the axis of rotation. Following Abdikamalov \emph{et
  al.}~\cite{abdikamalov:14}, we generate a total of 98 rotation
profiles using the parameter set listed in Table~\ref{tab:rotation},
chosen to span the full range of rotation rates slow enough to allow
the star to collapse. All 98 rotation profiles are simulated using
each of the 18 EOS for a total of 1764 2D core collapse
simulations. However, the 60 simulations listed in
Table~\ref{tab:nobounce} do not result in core collapse within
$1\,\mathrm{s}$ of simulation time due to centrifugal support and are
excluded from the analysis.

{\tt CoCoNuT} solves the equations of GR hydrodynamics on a
spherical-polar mesh in the Valencia formulation~\cite{font:08}, using
a finite volume method with piecewise parabolic
reconstruction~\cite{colella:84} and an approximate HLLE Riemann
solver~\cite{einfeldt:88}. Our fiducial fluid mesh has 250
logarithmically spaced radial zones out to $R=3000\,\mathrm{km}$ with
a central resolution of $250\,\mathrm{m}$, and 40 equally spaced
meridional angular zones between the equator and the pole. We assume
reflection symmetry at the equator. The GR CFC equations are solved
spectrally using 20 radial patches, each containing 33 radial
collocation points and 5 angular collocation points (see Dimmelmeier
\emph{et al.}~\cite{dimmelmeier:05}). We perform resolution tests in
Appendix~\ref{app:coconut_tests}.

The effects of neutrinos during the collapse phase are treated with a
$Y_e(\rho)$ parameterization as described above and
in~\cite{liebendoerfer:05fakenu,dimmelmeier:08}. After core bounce, we
employ the neutrino leakage scheme described in~\cite{ott:12a} to
approximately account for neutrino heating, cooling, and
deleptonization, though Ott \emph{et al.}~\cite{ott:12a} have shown
that neutrino leakage has a very small effect on the bounce and early
postbounce GW signal.

We allow the simulations to run for $50\,\mathrm{ms}$ after core
bounce, though in order to isolate the bounce and post-bounce
oscillations from prompt convection, we use only about
$10\,\mathrm{ms}$ after core bounce. Gravitational waveforms are
calculated using the quadrupole formula as given in Equation~A4
of~\cite{dimmelmeier:02}. All of the waveforms and reduced data used
in this study along with the analysis scripts are available at
\url{https://stellarcollapse.org/Richers\_2017\_RRCCSN\_EOS}.

\section{Results}
\label{sec:results}
\begin{figure*}
  \center
  \includegraphics[height=2.527in]{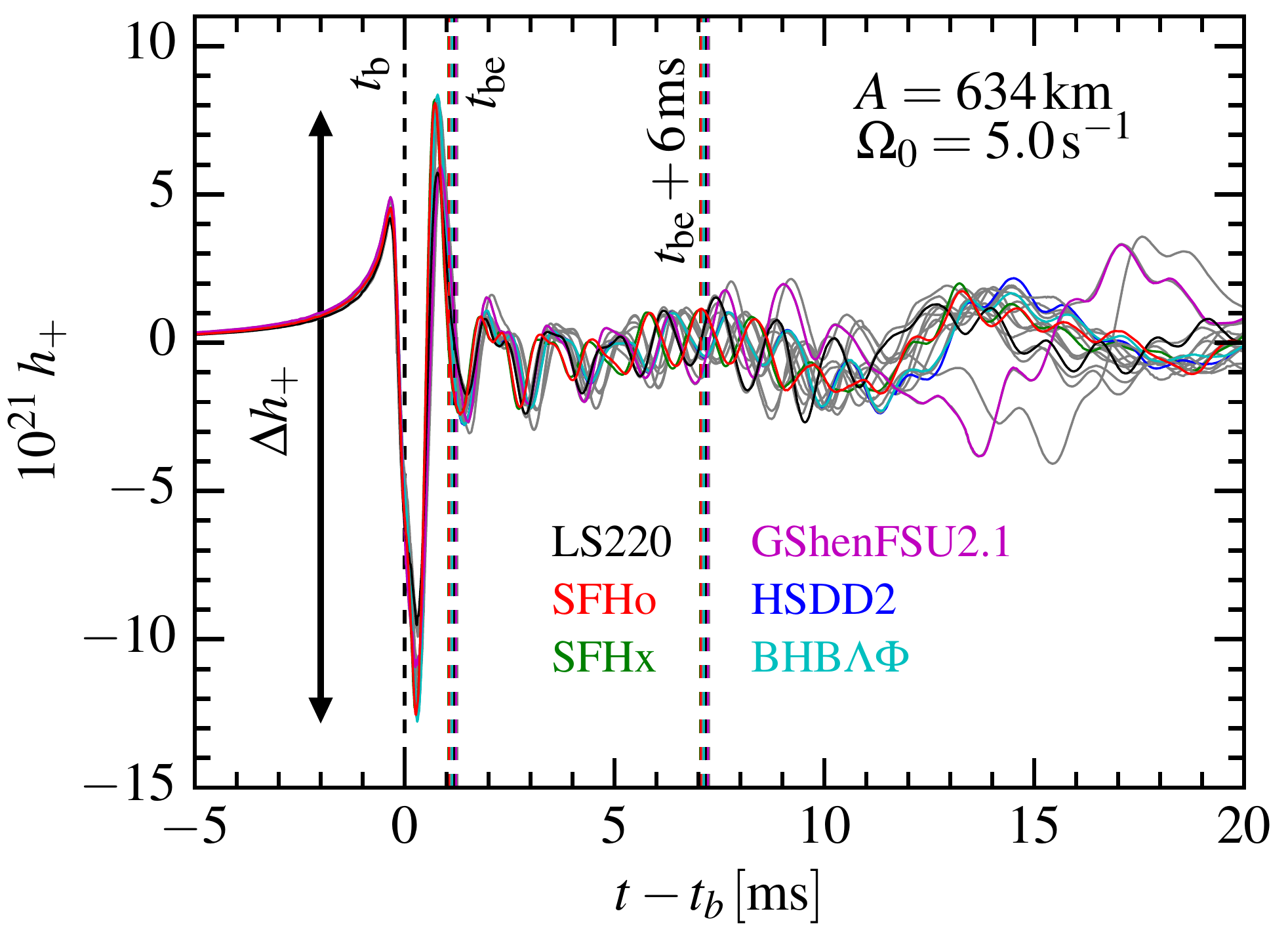}
  \includegraphics[height=2.527in]{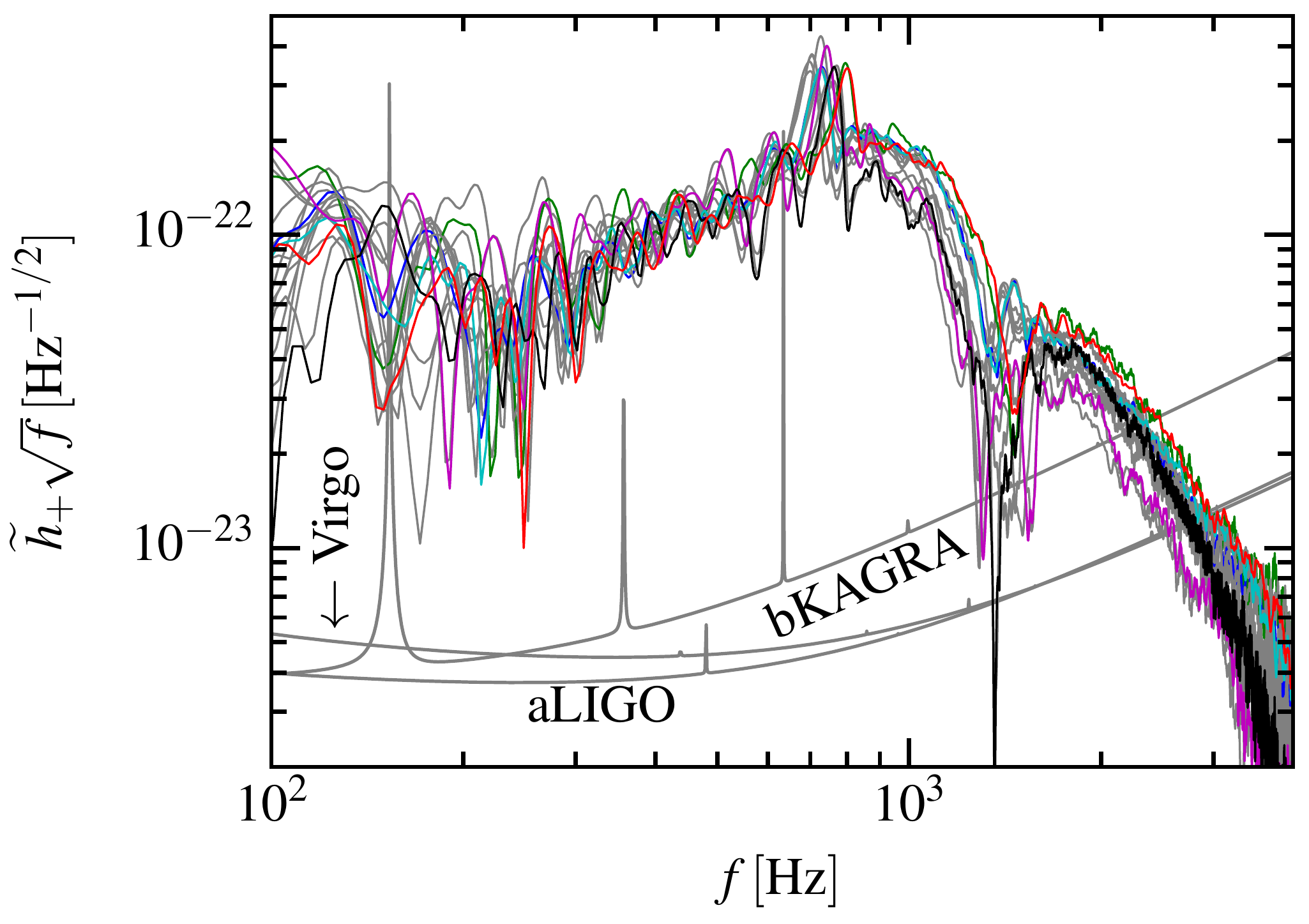}
  \caption{\textbf{EOS Variability in Waveforms.} The time-domain
    waveforms (left panel) and Fourier transforms scaled by $\sqrt{f}$
    (right panel) of signals from all 18 EOS for the
    $\mathrm{A}=634\,\mathrm{km}$, $\Omega=5.0\,\mathrm{rad\,s}^{-1}$
    rotation profile (moderately rapidly rotating, $T/|W|=0.069-0.074$
    at core bounce, depending on the EOS) are plotted assuming a
    distance of $10\,\mathrm{kpc}$ and optimal orientation, along with
    the Advanced LIGO~\cite{aligo,LIGO-sens-2010},
    VIRGO~\cite{advirgo:09}, and KAGRA in the zero detuning VRSE
    configuration~\cite{kagra:12,kagranoise:09} design sensitivity
    curves. $t_b$ is the time of core bounce, $t_\mathrm{be}$ is the
    end of the bounce signal and the beginning of the post-bounce
    signal. We use data only until $t_\mathrm{be}+6\,\mathrm{ms}$ to
    exclude the GW signal from prompt convection from our
    analysis. The differences in post-bounce oscillation rates can be
    seen both in phase decoherence of the waveform and the peak
    location of the Fourier transform. The colored curves correspond
    to EOS that satisfy the constraints depicted in
    Figure~\ref{fig:constraints}.}
  \label{fig:fourier}
\end{figure*}
\begin{figure}
\center \includegraphics[width=\linewidth]{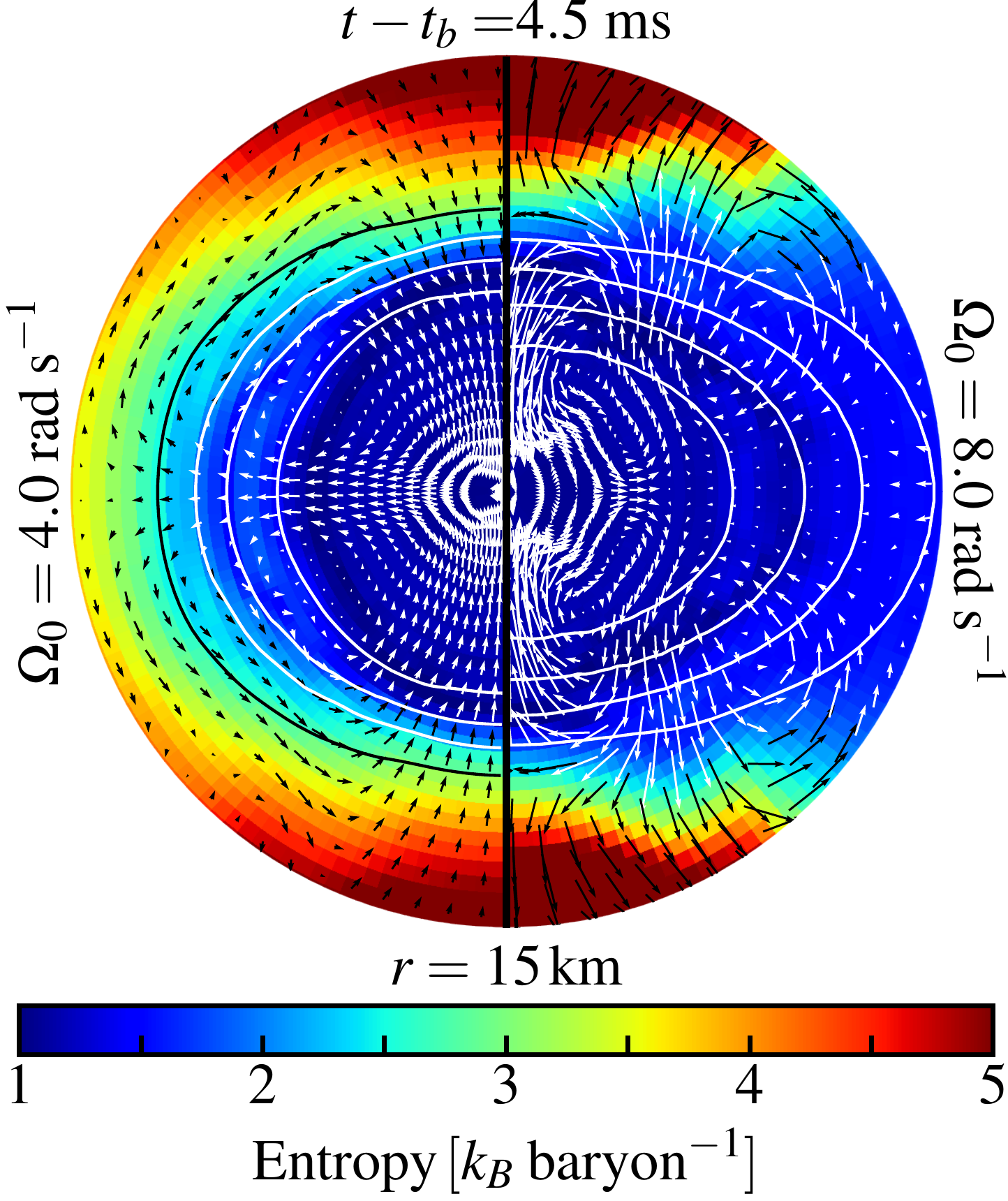}
\caption{\textbf{Velocity Field.} We plot the entropy, density, and
  velocity for the $\Omega_0=4.0\,\mathrm{rad\,s}^{-1}$ (left) and
  $\Omega_0=8.0\,\mathrm{rad\,s}^{-1}$ (right) simulations with
  $A=634\,\mathrm{km}$ at $4.5\,\mathrm{ms}$ after core bounce. The
  color map shows entropy. Blue regions belong to the unshocked inner
  core. The density contours show densities of
  $10^{\{13.5,13.75,14.0,14.25\}}\,\mathrm{g\,cm}^{-3}$ from outer to
  inner. The vectors represent only the poloidal velocity (i.e. the
  rotational velocity is ignored) and are colored for visibility. At
  low rotation rates (left) the flow in the inner core is largely
  quadrupolar. At high rotation rates (right), rotation significantly
  deforms the inner core and couples $\ell=2, m=0$ quadrupole
  oscillations to other modes.}
\label{fig:colormap}
\end{figure}

We begin by briefly reviewing the general properties of the GW signal
from rapidly rotating axisymmetric core collapse, bounce, and the
early postbounce phase. The GW strain can be approximately computed as
(e.g.,~\cite{finnevans:90,blanchet:90})
\begin{equation}
  \label{eq:quadrupole}
  h_+ \approx \frac{2G}{c^4 D} \ddot{I}\,\,,
\end{equation}
where $G$ is the gravitational constant, $c$ is the speed of light,
$D$ is the distance to the source, and $I$ is the mass quadrupole
moment.  In the left panel of Figure~\ref{fig:fourier} we show a
superposition of 18 gravitational waveforms for the
$A3=634\,\mathrm{km}$, $\Omega_0=5.0\,\mathrm{rad\,s}^{-1}$ rotation
profile using each of the 18 EOS and assuming a distance of
$10\,\mathrm{kpc}$ and optimal source-detector orientation.

As the inner core enters the final phase of collapse, its collapse
velocity greatly accelerates, reaching values of
$\sim$$0.3\,\mathrm{c}$. At bounce, the inner core suddenly (within
$\sim$$1\,\mathrm{ms}$) decelerates to near zero velocity and then
rebounds into the outer core. This causes the large spike in $h_+$
seen around the time of core bounce $t_b$. We determine $t_b$ as the
time when the entropy along the equator exceeds
$3\,k_b\,\mathrm{baryon}^{-1}$, indicating the formation of the bounce
shock. The rotation causes the shock to form in the equatorial
direction a few tenths of a millisecond after the shock forms in the
polar direction.

The bounce of the rotationally-deformed core excites postbounce
``ring-down'' oscillations of the PNS that are a complicated mixture
of multiple modes. They last for a few cycles after bounce, are damped
hydrodynamically \cite{fuller:15a}, and cause the postbounce
oscillations in the GW signal that are apparent in the left panel of
Figure~\ref{fig:fourier}. The dominant oscillation has been identified
as the $\ell=2,m=0$ (i.e., quadrupole) fundamental mode~(i.e., no radial
nodes)~\cite{ott:12a,fuller:15a}. The quadrupole oscillations can be
seen in the postbounce velocity field that we plot in the left panel
of Figure~\ref{fig:colormap}.  With increasing rotation rate, changes
in the mode structure and nonlinear coupling with other modes result
in the complex flow geometries shown in the right panel of
Figure~\ref{fig:colormap}. The density contours in
Figure~\ref{fig:colormap} also visualize how the PNS becomes more
oblate and less dense with increasing rotation rate.

After the PNS has rung down, other fluid dynamics, notably prompt
convection, begin to dominate the GW signal, generating a stochastic
GW strain whose time domain evolution is sensitive to the
perturbations from which prompt convection grows
(e.g.,~\cite{ott:09,marek:09b,kotake:09,abdikamalov:14}). We exclude
the convective part of the signal from our analysis. For our analysis,
we delineate the end of the bounce signal and the start of the
postbounce signal at $t_{be}$, defined as the time of the third zero
crossing of the GW strain. We also isolate the postbounce PNS
oscillation signal from the convective signal by considering only the
first $6\,\mathrm{ms}$ after $t_{be}$.

In the right panel of Figure~\ref{fig:fourier}, we show the Fourier
transforms of each of the time-domain waveforms shown in the left
panel, multiplied by $\sqrt{f}$ for comparison with GW detector
sensitivity curves. The bounce signal is visible in the broad bulge in
the range of $200-1500\,\mathrm{Hz}$. The postbounce oscillations
produce a peak in the spectrum of around $700-800\,\mathrm{Hz}$, the
center of which we call the peak frequency $f_\mathrm{peak}$. Both the
peak frequency and the amplitude of the bounce signal in general
depend on both the rotation profile and the EOS.

\subsection{The Bounce Signal}
\label{sec:bounce}
\begin{figure}
\center \includegraphics[width=\linewidth]{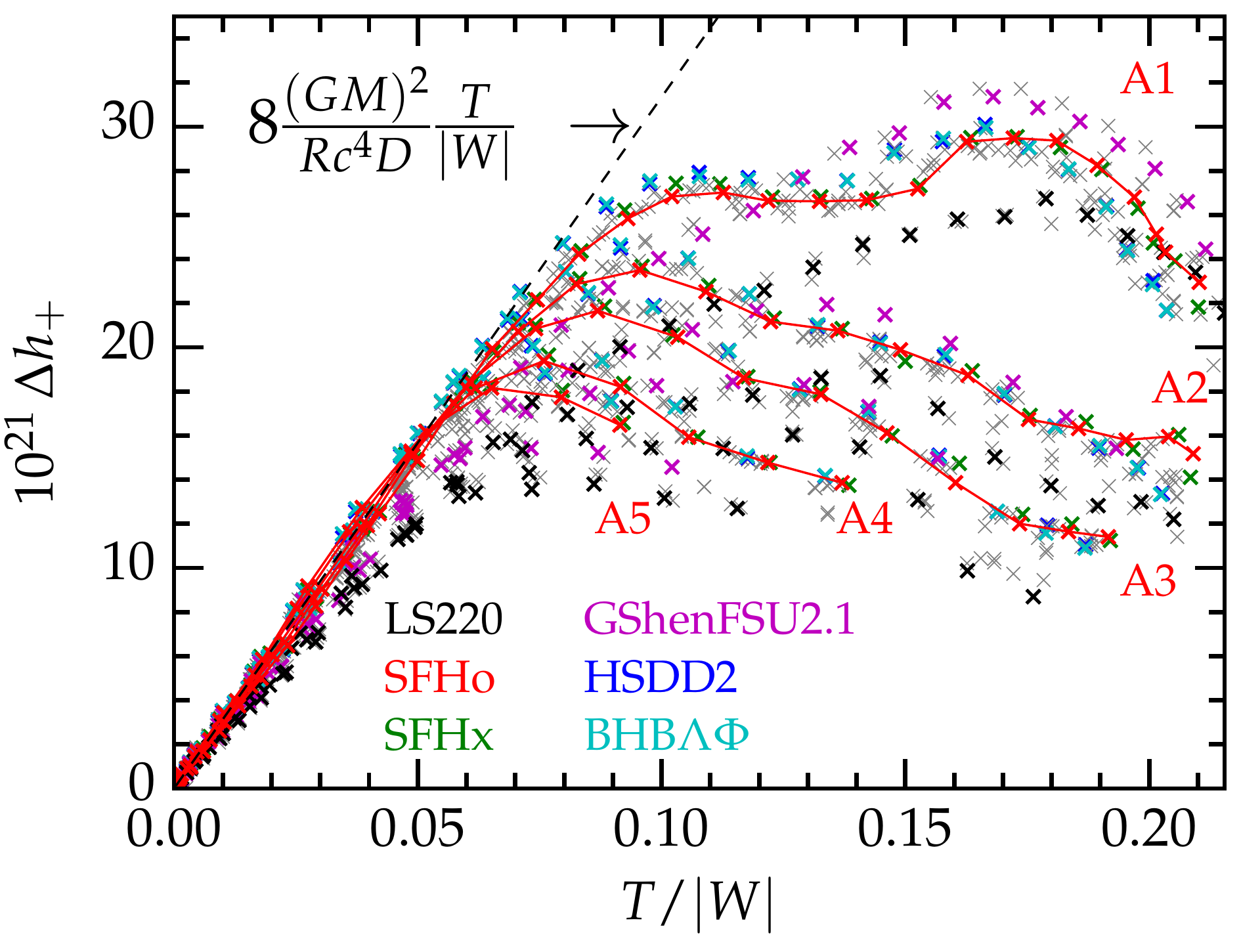}
\caption{\textbf{Bounce Signal Amplitude.} We plot the difference
  between the maximum and minimum strain $\Delta h_+$ before
  $t_\mathrm{be}$ assuming $D=10\,\mathrm{kpc}$ and optimal
  source-detector orientation as a function of the ratio of rotational
  to gravitational energy $T/|W|$ of the inner core at bounce. Each 2D
  simulation is a single point and the SFHo simulations with the same
  differential rotation parameter $A$ are connected to guide the
  eye. $A1-A5$ corresponds to $A=300,467,634,1268,10000\,\mathrm{km}$,
  respectively. Simulations with all EOS and values of $A$ behave
  similarly for $T/|W| \lesssim 0.06$, but branch out when rotation
  becomes dynamically important. We plot a dashed line representing
  the expected perturbative behavior with $T/|W|$, using
  representative values of $M=0.6M_\odot$ and $R=65\,\mathrm{km}$. All
  1704 collapsing simulations are included in this figure.}
\label{fig:amplitude}
\end{figure}

The bounce spike is the loudest component of the GW signal. In
Figure~\ref{fig:amplitude}, we plot $\Delta h_+$, the difference
between the highest and lowest points in the bounce signal strain, as
a function of the ratio of rotational kinetic energy to gravitational
potential energy $T/|W|$ of the inner core at core bounce (see the
beginning Section~\ref{sec:results} for details of our definition of
core bounce). We assume a distance of $10\,\mathrm{kpc}$ and optimal
detector orientation. Just as in Abdikamalov \emph{et
  al.}~\cite{abdikamalov:14}, we see that at low rotation rates, the
amplitude increases linearly with rotation rate, with a similar slope
for all EOS. At higher rotation rates, the curves diverge from this
linear relationship due to centrifugal support as the angular velocity
$\Omega$ at bounce approaches the Keplerian angular velocity. Rotation
slows the collapse, softening the violent EOS-driven bounce and
resulting in a smaller acceleration of the mass quadrupole
moment. However, the value of $T/|W|=0.06-0.09$ at which simulations
diverge from the linear relationship depends on the value of the
differential rotation parameter $A$. Stronger differential rotation
affords less centrifugal support at higher rotation energies, allowing
the linear behavior to survive to higher rotation rates.

The linear relationship between the bounce amplitude and $T/|W|$ of
the inner core at bounce can be derived in a perturbative,
order-of-magnitude sense. The GW amplitude depends on the second time
derivative of the mass quadrupole moment $I\sim M (x^2 - z^2)$, where
$M$ is the mass of the oscillating inner core and $x$ and $z$ are the
equatorial and polar equilibrium radii, respectively. If we treat the
inner core as an oblate sphere, we can call the radius of the inner
core in the polar direction $z=R$ and the larger radius of the inner
core in the equatorial direction (due to centrifugal support)
$x=R+\delta R$.  To first order in $\delta R$, the mass quadrupole
moment becomes
\begin{equation}
  I\sim M ((R+\delta R)^2 - R^2) \sim M R (\delta R)\,\,.
\end{equation}
The difference between polar and equatorial radii in our simplified
scenario can be determined by noting that the surface of a rotating
sphere in equilibrium is an isopotential surface with a potential of
$-\varpi^2\Omega^2/2 - GM/r$, where $\varpi$ is the distance to the
rotation axis, $r$ is the distance to the origin, $\Omega$ is the
angular rotation rate, and $G$ is the gravitational constant. Setting
the potential at the equator and poles equal to each other yields
\begin{equation}
  (R+\delta R)^2 \Omega^2 + \frac{GM}{(R+\delta R)} = \frac{GM}{R}\,\,.
\end{equation}
Assuming differences between equatorial and polar radii are small, we
can take only the $O(\delta R/R)$ terms to get $\delta(\varpi^2\Omega^2)
\sim R^2\Omega^2 \sim GM(\delta R) / R^2$. Solving for $\delta R$,
\begin{equation}
\delta R \sim \Omega^2 R^4/ GM\,\,.
\end{equation} 
The timescale of core bounce is the dynamical time
$t_\mathrm{dyn}^{-2} \sim G \rho \sim GM/R^3$. In this
order-of-magnitude estimate we can replace time derivatives in
Equation~\ref{eq:quadrupole} with division by the dynamical time. We
can also approximate $T/|W|\sim R^3 \Omega^2 / GM$. This results in
\begin{equation}
  \label{eq:amplitude}
  h_+ \sim \frac{GM \Omega^2 R^2}{c^4 D} \sim \frac{T}{|W|} \frac{(GM)^2}{R c^4 D}\,\,.
\end{equation}
Though the mass and polar radius of the PNS depend on rotation as
well, the dependence is much weaker (in the slow rotating limit)
\cite{abdikamalov:14}, and $T/|W|$ contains all of the first-order
rotation effects used in the derivation. Hence, \emph{in the linear
  regime, the bounce signal amplitude should depend approximately
  linearly on $T/|W|$}, which is reflected by
Figure~\ref{fig:amplitude}.

\begin{table}
  \caption{\textbf{Bounce Amplitude Linear Fits.} We calculate a
    linear least squares fit for the bounce amplitudes in
    Figure~\ref{fig:amplitude} to the function $\Delta h_+ = m (T/|W|)
    + b$. We only include data with $T/|W|\leq 0.04$. All fitted lines
    have a y-intercept $b$ of approximately $0$ and slopes $m$ in the
    range of $237-315\times10^{-21}$. The three LS EOS have the
    shallowest slopes and the ten Hempel-based EOS (HS, SFH, and BHB)
    have the steepest. The $m_\mathrm{predicted}$ column shows the
    predicted slope of $m_\mathrm{predicted} =
    T/|W|\times8(GM)^2/Rc^4D$ using the mass and radius of the
    nonrotating inner core at bounce. We choose the arbitrary factor
    of $8$ to make the predicted and actual SFHo slopes match. We list
    the mass of the nonrotating inner core at bounce
    ($M_\mathrm{IC,b,0}$) for each EOS in the last
    column. The SFHo\_ecap\{0.1,1.0,10.0\} rows use
      detailed electron capture rates in the {\tt GR1D} simulations
      for the $Y_e(\rho)$ profile (see Section~\ref{sec:ecapture}).}
    \begin{tabular}{lcccc}
  \hline\hline
  EOS & $m$ & $b$ & $m_\mathrm{predicted}$ & $M_\mathrm{IC,b,0}$\\
      & $[10^{-21}]$ & $[10^{-21}]$ & $[10^{-21}]$ & $[M_\odot]$\\
\hline
BHBL & 318 & -0.03 & 321 & 0.598\\
BHBLP & 317 & \phantom-0.02 & 322 & 0.599\\
HSDD2 & 316 & \phantom-0.00 & 322 & 0.599\\
SFHo & 306 & \phantom-0.03 & 304 & 0.582\\
HSFSG & 306 & -0.00 & 325 & 0.602\\
SFHx & 305 & \phantom-0.09 & 303 & 0.581\\
HSIUF & 304 & \phantom-0.06 & 316 & 0.593\\
HSNL3 & 298 & \phantom-0.07 & 324 & 0.600\\
HSTMA & 295 & \phantom-0.15 & 315 & 0.593\\
HSTM1 & 295 & \phantom-0.18 & 314 & 0.591\\
HShenH & 281 & \phantom-0.28 & 311 & 0.604\\
HShen & 280 & \phantom-0.29 & 310 & 0.604\\
SFHo\_ecap0.1 & 274 & \phantom-0.22 & 262 & 0.562\\
GShenNL3 & 267 & \phantom-0.32 & 298 & 0.592\\
GShenFSU1.7 & 264 & \phantom-0.24 & 294 & 0.587\\
GShenFSU2.1 & 263 & \phantom-0.24 & 293 & 0.587\\
LS180 & 242 & \phantom-0.16 & 245 & 0.536\\
LS375 & 237 & \phantom-0.15 & 284 & 0.562\\
LS220 & 237 & \phantom-0.20 & 258 & 0.543\\
SFHo\_ecap1.0 & 210 & \phantom-0.08 & 207 & 0.506\\
SFHo\_ecap10.0 & 174 & \phantom-0.03 & 198 & 0.482\\
\hline\hline

  \end{tabular}
  \label{tab:ampfits}
\end{table}

Differences between EOS in the bounce signal $\Delta h_+$ enter
through the mass and radius of the inner core at bounce
(cf.\ Equation~\ref{eq:amplitude}). Neither $M$ nor $R$ of the inner
core are particularly well defined quantities since they vary rapidly
around bounce -- all quantitative results we state depend on our
definition of the bounce time and Equation~\ref{eq:amplitude} is
expected to be accurate only to an order of magnitude.  With that in
mind, in order to test how well Equation~\ref{eq:amplitude} matches
our numerical results, we generate fits to functionals of the form
$h_+ = m(T/|W|) + b$. $b$ is simply the y-intercept of the line, which
should be approximately 0 based on Equation~\ref{eq:amplitude}. $m$ is
the slope of the line, which we expect to be $m_\mathrm{predicted}=8(G
M_\mathrm{IC,b,0})^2/R_\mathrm{IC,b,0}c^4D$ based on
Equation~\ref{eq:amplitude}, using the mass and radius of the
nonrotating PNS at bounce. We include the arbitrary factor of 8 to
make the order-of-magnitude predicted slopes similar to the fitted
slopes. In Table~\ref{tab:ampfits} we show the results of the linear
least-squares fits to results of slowly rotating collapse below
$T/|W|\leq0.04$ for each EOS. Though $m_\mathrm{predicted}$ is of the
same order of magnitude as $m$, significant differences exist. This is
not unexpected, considering that our model does not account for
nonuniform density distribution and the increase of the inner core
mass with rotation, which can significantly affect the quadrupole
moment.

At a given inner core mass, the structure (i.e.\ radius) of the inner
core is determined by the EOS. Furthermore, the mass of the inner core
is highly sensitive to the electron fraction $Y_e$ in the final stages
of collapse. In the simplest approximation, it scales with
$M_\mathrm{IC}\sim Y_e^2$~\cite{yahil:83}, which is due to the
electron EOS that dominates until densities near nuclear density are
reached. The inner-core $Y_e$ in the final phase of collapse is set by
the deleptonization history, which varies between EOS
(Figure~\ref{fig:yeofrho}). In addition, contributions of the
nonuniform nuclear matter EOS play an additional $Y_e$-independent
role in setting $M_\mathrm{IC}$. For example, we see from
Figure~\ref{fig:yeofrho} that the LS220 EOS yields a bounce $Y_e$ of
$\sim$0.278, while the GShenFSU2.1 EOS results in
$\sim$0.267. Naively, relying just on the $Y_e$ dependence of
$M_\mathrm{IC}$, we would expect LS220 to yield a larger inner core
mass.  Yet, the opposite is the case: our simulations show that the
nonrotating inner core mass at bounce for the GShenFSU2.1 EOS is
$\sim$$0.59\,M_\odot$ while that for the LS220 EOS is
$\sim$$0.54\,M_\odot$.

\begin{table}
  \caption{\textbf{Example Quantitative Results for the Bounce
      Signal.} We present results for the bounce signals of models
    with differential rotation parameter $A3 = 634\,\mathrm{km}$, a
    representative set of initial rotation rates ($2.5$, $5.0$, and
    $7.5\,\mathrm{rad\, s}^{-1}$), and the six EOS in best agreement
    with current constraints (cf.~Section~\ref{sec:eos}). The models
    are grouped by rotation rate. $\rho_\mathrm{c,b}$ is the central
    density at bounce (time averaged from $t_b$ to
    $t_b+0.2\,\mathrm{ms}$), $T/|W|$ is the
    ratio of rotational kinetic energy to gravitational energy of the
    inner core at bounce, and $M_\mathrm{IC,b}$ is its gravitational
    mass at bounce. $\Delta h_+$ is the difference between the highest
    and lowest points in the bounce spike at a distance of
    $10\,\mathrm{kpc}$. Note that $\rho_\mathrm{c}$,
    $T/|W|$, and $M_\mathrm{IC}$ all vary rapidly around
    core bounce and their exact values are rather sensitive to the
    definition of the time of bounce. The quantities summarized here
    for this set of models are available for all models at
    \url{https://stellarcollapse.org/Richers\_2017\_RRCCSN\_EOS}.}
\begin{tabular}{lcccc}
  \hline
  \hline
  Model & $\rho_\mathrm{c,b}$ & $T/|W|$ & $M_\mathrm{IC,b}$ &$\Delta h_+$\\
  & [$10^{14}\,\mathrm{g\,cm}^{-3}$] & & [$M_\odot$] & [$10^{-21}$]\\
  \hline
  A3$\Omega$2.5-LS220 & 3.976 & 0.020 & 0.589 & \phantom04.7 \\
  A3$\Omega$2.5-SFHo & 4.262 & 0.020 & 0.624 & \phantom06.1 \\
  A3$\Omega$2.5-SFHx & 4.252 & 0.020 & 0.610 & \phantom06.1 \\
  A3$\Omega$2.5-GShenFSU2.1 & 3.612 & 0.020 & 0.634 & \phantom05.2 \\
  A3$\Omega$2.5-HSDD2 & 3.582 & 0.019 & 0.629 & \phantom05.9 \\
  A3$\Omega$2.5-BHB$\Lambda\Phi$ & 3.583 & 0.019 & 0.629 & \phantom06.0 \\
  \hline
  A3$\Omega$5.0-LS220 & 3.581 & 0.071 & 0.673 & 15.3 \\
  A3$\Omega$5.0-SFHo & 3.868 & 0.074 & 0.708 & 20.8 \\
  A3$\Omega$5.0-SFHx & 3.857 & 0.074 & 0.705 & 21.0 \\
  A3$\Omega$5.0-GShenFSU2.1 & 3.376 & 0.072 & 0.729 & 17.1 \\
  A3$\Omega$5.0-HSDD2 & 3.314 & 0.071 & 0.712 & 21.3 \\
  A3$\Omega$5.0-BHB$\Lambda\Phi$ & 3.321 & 0.071 & 0.709 & 21.3 \\
  \hline
  A3$\Omega$7.5-LS220 & 2.940 & 0.141 & 0.784 & 15.5 \\
  A3$\Omega$7.5-SFHo & 3.183 & 0.146 & 0.829 & 16.1 \\
  A3$\Omega$7.5-SFHx & 3.237 & 0.147 & 0.831 & 16.0 \\
  A3$\Omega$7.5-GShenFSU2.1 & 2.878 & 0.143 & 0.838 & 17.3 \\
  A3$\Omega$7.5-HSDD2 & 2.763 & 0.142 & 0.835 & 17.1 \\
  A3$\Omega$7.5-BHB$\Lambda\Phi$ & 2.763 & 0.142 & 0.835 & 17.1 \\
\hline
\hline
\end{tabular}
\label{tab:bounce}
\end{table}

We further investigate the EOS-dependence of the bounce GW signal by
considering a representative quantitative example of models with
precollapse differential rotation parameter $A3 = 634\,\mathrm{km}$,
computed with the six EOS identified in Section~\ref{sec:eos} as most
compliant with constraints. In Table~\ref{tab:bounce}, we summarize
the results for these models for three precollapse rotation rates,
$\Omega_0 = \{2.5, 5.0, 7.5\}\,\mathrm{rad\,s}^{-1}$, probing different
regions in Figure~\ref{fig:amplitude}.

At $\Omega_0 = 2.5\,\mathrm{rad\,s}^{-1}$, all models reach $T/|W|$ of
$\sim$$0.02$, hence are in the linear regime where
Equation~\ref{eq:amplitude} holds. The LS220 EOS model has the
smallest inner core mass and results in the smallest bounce GW
amplitude of all EOS (cf.\ also Figure~\ref{fig:amplitude}). The SFHx
and the GShenFSU2.1 EOS models have roughly the same inner core masses
($\sim$$0.64-0.65\,M_\odot$), but the SFHx EOS is considerably softer,
resulting in higher bounce density and correspondingly smaller radius,
and thus larger $\Delta h_+$, $6.7 \times 10^{-21}$ (at
$10\,\mathrm{kpc}$) vs.\ $5.4 \times 10^{-21}$ for the GShenFSU2.1
EOS. We also note that the HSDD2 and the BHB$\Lambda\Phi$ EOS models
give nearly identical results. They employ the same low-density EOS
and the same RMF DD2 parameterization and their only difference is
that $BHB\Lambda\Phi$ includes softening hyperon contributions that
appear above nuclear density. However, at the densities reached in our
core collapse simulations with these EOS
($\sim$$3.6\times10^{14}\,\mathrm{g\,cm}^{-3}$), the hyperon fraction
barely exceeds $\sim$1\% \cite{banik:14} and thus has a negligible
effect on dynamics and GW signal.

The models at $\Omega_0 = 5.0\,\mathrm{rad\,s}^{-1}$ listed in
Table~\ref{tab:bounce} reach $T/|W| \sim 0.071-0.076$ and begin to
deviate from the linear relationship of Equation~\ref{eq:amplitude}.
However, their bounce amplitudes $\Delta h_+$ still follow the same
trends with EOS (and resulting inner core mass and bounce density) as
their more slowly spinning counterparts.

Finally, the rapidly spinning models with $\Omega_0 =
7.5\,\mathrm{rad\,s}^{-1}$ listed in Table~\ref{tab:bounce} result in
$T/|W| \sim 0.141-0.152$ and are far outside the linear
regime. Centrifugal effects play an important role in their bounce
dynamics, substantially decreasing their bounce densities and
increasing their inner core masses. Increasing rotation, however,
tends to decrease the EOS-dependent relative differences in $\Delta
h_+$.  At $\Omega_0 = 5\,\mathrm{rad\,s}^{-1}$, the standard deviation
of $\Delta h_+$ is $\sim$12.5\% of the mean value, while at $\Omega_0
= 7.5\,\mathrm{rad\,s}^{-1}$, it is only $\sim$3\%. This is also
visualized by Figure~\ref{fig:amplitude} in which the rapidly rotating
models cluster rather tightly around the A3 branch (third from the
bottom). In general, for any value of $A$, the EOS-dependent spread on
a given differential rotation branch is smaller than the spread
between branches.

\emph{Conclusions:} In the Slow Rotation regime ($T/|W|\lesssim 0.06$)
the bounce GW amplitude varies linearly with $T/|W|$
(Equation~\ref{eq:amplitude}), in agreement with previous works. Small
differences in this linear slope are due primarily to differences in
the inner core mass at bounce induced by different EOS. In the Rapid
Rotation regime ($0.06\lesssim T/|W|\lesssim 0.17$) the core is
centrifugally supported at bounce and the bounce GW signal depends
much more strongly on the amount of precollapse differential rotation
than on the EOS. In the Extreme Rotation regime ($T/|W|\gtrsim 0.17$)
the core undergoes a centrifugally-supported bounce and the GW bounce
signal weakens.

\subsection{The Postbounce Signal from PNS Oscillations}
\label{sec:postbounce}
\begin{figure}
  \center \includegraphics[width=\linewidth]{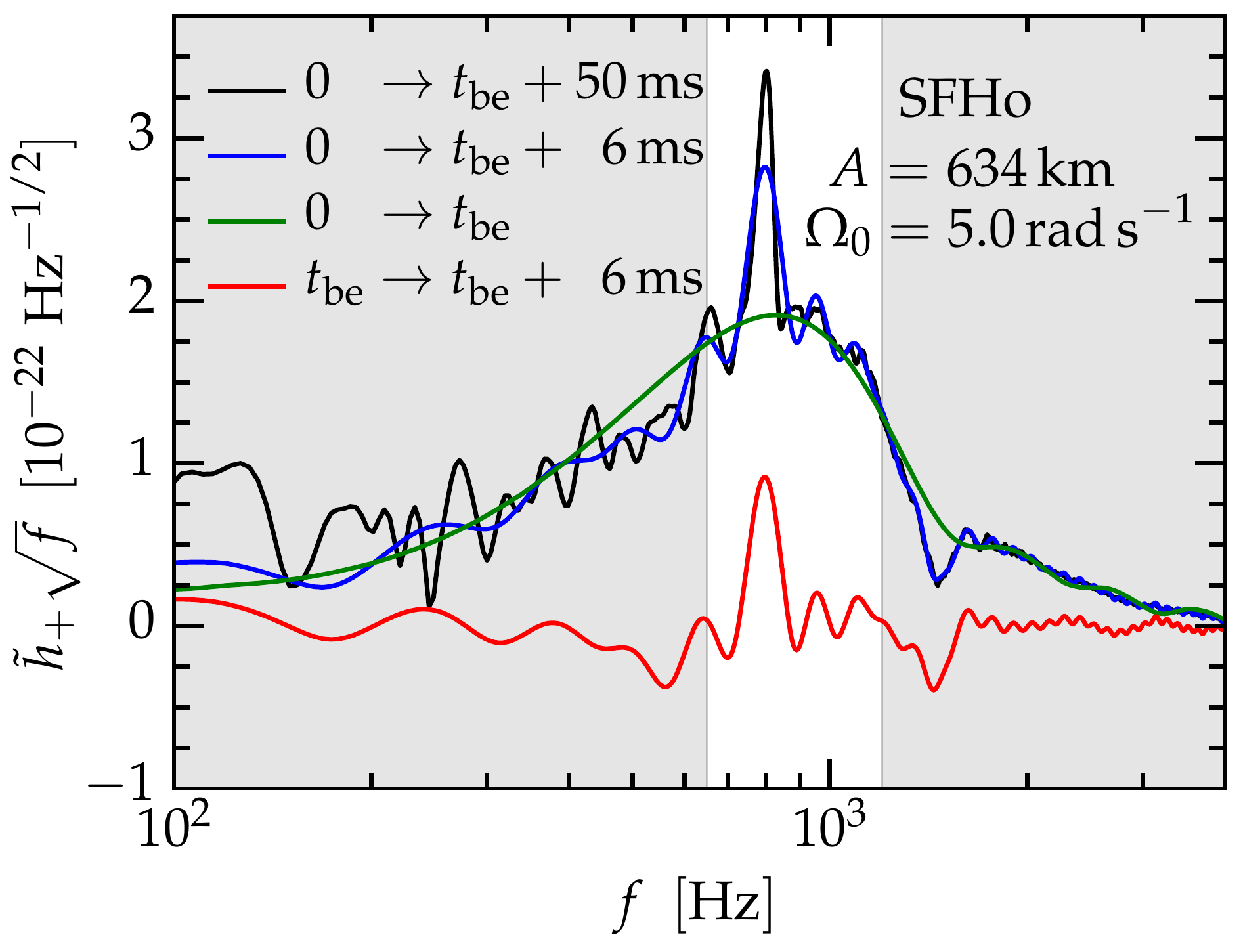}
\caption{\textbf{Peak Frequency Determination.} The full GW spectrum
  for the $A3=634\,\mathrm{km}$, $\Omega=5.0\,\mathrm{rad\,s}^{-1}$
  SFHo simulation is plotted in black. To prevent convection
  contributions from entering into the analysis, we cut the GW signal
  at $6\,\mathrm{ms}$ after the end of core bounce
  ($t_\mathrm{be}+6\,\mathrm{ms}$, blue line). The green line is the
  spectrum for the time series through the end of core bounce. To
  remove the bounce signal from the spectrum, we subtract the green
  line from the blue line to get the red line. The maximum of the red
  line within the depicted window of $600-1075\,\mathrm{Hz}$
  determines the peak frequency $f_\mathrm{peak}$.}
\label{fig:fpeak_reduction}
\end{figure}

The observable of greatest interest in the postbounce GW signal is the
oscillation frequency of the PNS, which may encode EOS information. To
isolate the PNS oscillation signal from the earlier bounce and the
later convective contributions, we separately Fourier transform the GW
signal calculated from GWs up to $t_\mathrm{be}$ (the end of the
bounce signal, defined as the third zero-crossing after core bounce as
in Figure~\ref{fig:fourier}) and from GWs up to
$t_\mathrm{be}+6\,\mathrm{ms}$ (empirically chosen to produce reliable
PNS oscillation frequencies).  We begin with a simulation with
intermediate rotation and subtract the former bounce spectrum from the
latter full spectrum and we take the largest spectral feature in the
window of $600$ to $1075\,\mathrm{Hz}$ to be the $\ell=2$ f-mode peak
frequency $f_\mathrm{peak}$~\cite{ott:12a,fuller:15a}. The spectral
windows for simulations with the same value of $A$ and adjacent values
of $\Omega_0$ are centered at this frequency and have a width of
$75\,\mathrm{Hz}$. This process is repeated outward from the
intermediate simulation and allows us to more accurately isolate the
correct oscillation frequency in slowly and rapidly rotating regimes
where picking out the correct spectral feature is difficult. This
procedure is visualized in Figure~\ref{fig:fpeak_reduction}. Note that
there are only around five to ten postbounce oscillation cycles before
the oscillations damp, so the peak has a finite width of about
$100\,\mathrm{Hz}$. However, our analysis in this section shows that
the peak frequency is known far better than that.

\begin{figure}
\center
\includegraphics[width=\linewidth]{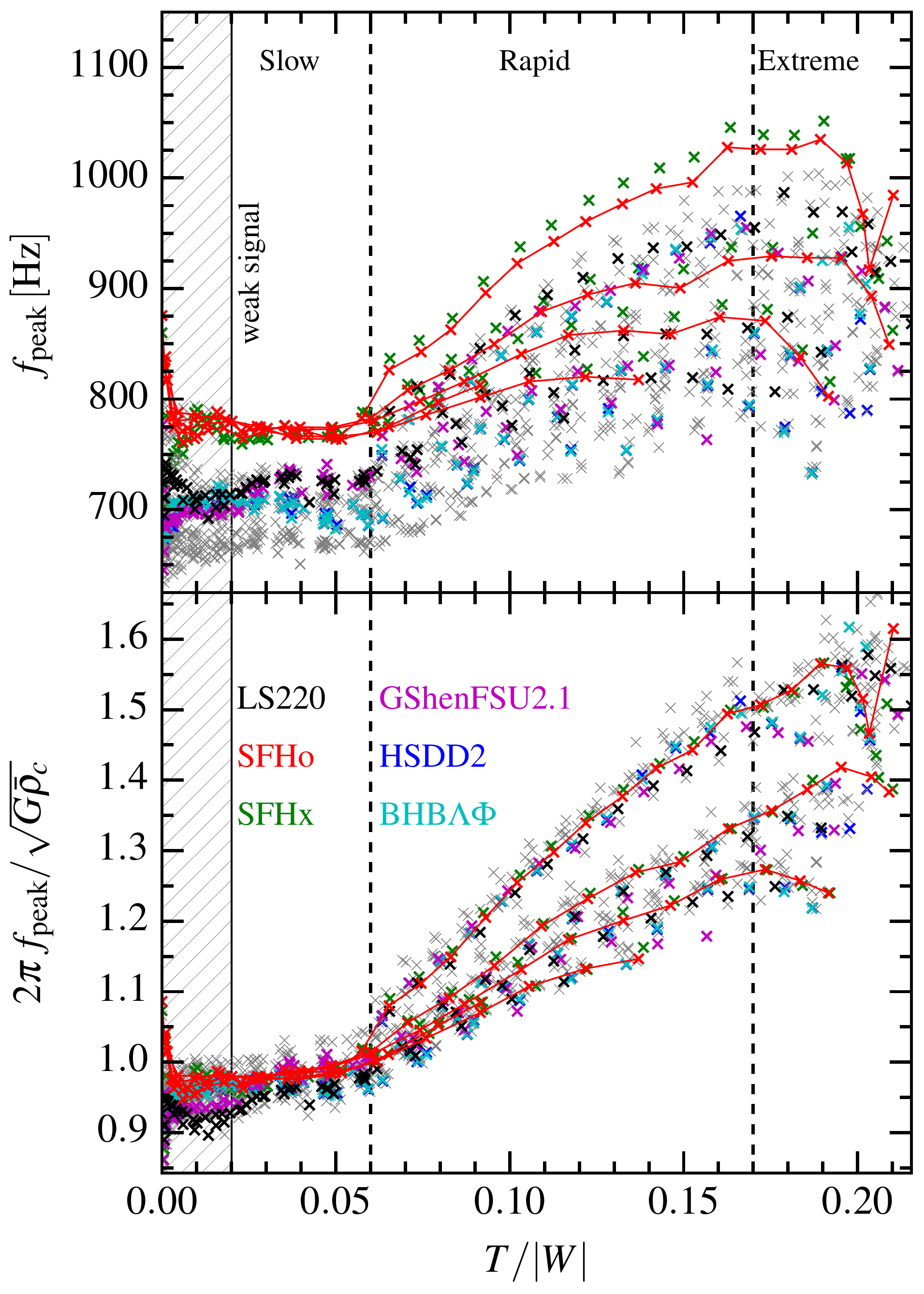}
\caption{\textbf{Peak Frequencies.}  \textit{Top:} The peak
  frequencies of GWs emitted by postbounce PNS oscillations in all
  1704 collapsing simulations are plotted as a function of the ratio
  of rotational to gravitational energy $T/|W|$ of the inner core at
  bounce. Red lines connect SFHo simulations with the same
  differential rotation parameter$A$. There is a large spread in the
  peak frequencies due to both the EOS and due to differential
  rotation. \textit{Bottom:} We can remove most of the effects of the
  different EOS by normalizing the peak frequency by the dynamical
  frequency $\sqrt{G\bar{\rho}_c}$ and multiply by $2\pi$ to make it
  an angular frequency. However, significant differences due to
  differing amounts of differential rotation remain for rapidly
  spinning models. The transition from slow to rapid rotation regimes
  occurs at $T/|W|\approx 0.06$ and it becomes difficult for our
  analysis scripts to find the $\ell=2$ f-mode peak at
  $T/|W|\gtrsim0.17$. Each panel contains 1704 data points, and there
  are 1487 good points with $T/|W|<0.17$.}
\label{fig:fpeak}
\end{figure}

In the top panel of Figure~\ref{fig:fpeak}, we plot the GW peak
frequency $f_\mathrm{peak}$ as a function of $T/|W|$ (of the inner
core at bounce) for each of our 1704 collapsing cores. We identify
three regimes of rotation and $f_\mathrm{peak}$ systematics in this
figure:

(\textbf{Slow Rotation} Regime) In slowly rotating cores, $T/|W|
\lesssim 0.06$, $f_\mathrm{peak}$ shows little variation with
increasing rotation rate or degree of differential rotation.  Note
that our analysis is unreliable in the very slow rotation limit
($T/|W| \lesssim 0.02$). There, the PNS oscillations are only weakly
excited and the corresponding GW signal is very weak. This is a
consequence of the fact that our nonlinear hydrodynamics approach is
noisy and not made for the perturbative regime.
  
(\textbf{Rapid Rotation} Regime) In the rapidly rotating regime, $0.06
\lesssim T/|W| \lesssim 0.17$, $f_\mathrm{peak}$ increases with
increasing rotation rate and initially more differentially rotating
cores have systematically higher $f_\mathrm{peak}$.

(\textbf{Extreme Rotation} Regime) At $T/|W| \gtrsim 0.17$, bounce and
the postbounce dynamics become centrifugally dominated, leading to
very complex PNS oscillations involving multiple nonlinear modes with
comparable amplitudes. This makes it difficult to unambiguously define
$f_\mathrm{peak}$ in this regime and our analysis becomes
unreliable. Excluding all models with $T/|W|\gtrsim0.17$ leaves us
with 1487 simulations with a reliable determination of
$f_\mathrm{peak}$.

Figure~\ref{fig:fpeak} shows that the different EOS lead to a
${\sim150\,\mathrm{Hz}}$ variation in $f_\mathrm{peak}$. The peak
frequency is expected to scale with the PNS dynamical frequency
(e.g.,~\cite{fuller:15a}). That is,
\begin{equation}
  f_\mathrm{peak} \sim \Omega_\mathrm{dyn} = \sqrt{G\rho_c}\,\,,
\label{eq:dyn}
\end{equation}
where $G$ is the gravitational constant and $\rho_c$ is the central
density. In the bottom panel of Figure~\ref{fig:fpeak}, we normalize
the observed peak frequency by the dynamical frequency, using the
central density averaged over $6\,\mathrm{ms}$ after the end of the
bounce signal (the same time interval from which we extract
$f_\mathrm{peak}$). The scatter between different EOS is drastically
reduced, and thus the effect of the EOS on the peak frequency is
essentially parameterized by the PNS central density, which is a
reflection of the compactness of the PNS core.

\begin{table}
  \caption{\textbf{GW Peak Frequencies of PNS Oscillations in the Slow
      Rotation Regime}. $\langle f_\mathrm{peak} \rangle$ is the peak
    frequency for each EOS averaged over all simulations with
    $0.02\leq T/|W|\leq0.06$. $\sigma_{f_\mathrm{peak}}$ is its
    standard deviation and provides a handle on how much
    $f_\mathrm{peak}$ varies in the Slow Rotation regime.  We also
    provide the average dynamical frequency $\langle f_\mathrm{dyn}
    \rangle = \langle \sqrt{G\bar{\rho}_c}/2\pi \rangle$, the averaged
    central density $\langle \bar{\rho}_c \rangle$, and the averaged
    gravitational mass of the inner core at bounce $\langle
    M_\mathrm{IC,b}\rangle$, all averaged over simulations with
    $0.02\leq T/|W|\leq0.06$. The
      SFHo\_ecap\{0.1,1.0,10.0\} rows use detailed electron capture
      rates in the {\tt GR1D} simulations for the $Y_e(\rho)$ profile
      (see Section~\ref{sec:ecapture}). Note that despite some
    outliers there is an overall EOS-dependent trend that softer EOS
    (producing higher $\bar{\rho}_c$) have higher
    $f_\mathrm{peak}$. Also note that $f_\mathrm{peak}$ is for all EOS
    quite close to the dynamical frequency of the PNS,
    $f_\mathrm{dyn}$.}
    \begin{tabular}{lccccc}
  \hline\hline
  EOS & $\langle f_\mathrm{peak}\rangle$ & $\sigma_{f_\mathrm{peak}}$ & $\langle f_\mathrm{dyn}\rangle$ & $\langle \bar{\rho}_c \rangle$ & $\langle M_\mathrm{IC,b} \rangle$\\
      & $[\mathrm{Hz}]$                  & $[\mathrm{Hz}]$            & $[\mathrm{Hz}]$                 & $[10^{14}\,\mathrm{g\,cm}^{-3}]$ & $[M_\odot]$\\
  \hline
  SFHo\_ecap0.1 & 871 & \phantom07.9 & 795 & 3.74 & 0.656 \\
  SFHo\_ecap1.0 & 846 & \phantom09.4 & 778 & 3.58 & 0.573 \\
  SFHo\_ecap10.0 & 790 & 10.5 & 760 & 3.42 & 0.532 \\
  SFHo & 772 & \phantom05.6 & 784 & 3.64 & 0.650 \\
  SFHx & 769 & \phantom07.3 & 785 & 3.64 & 0.648 \\
  LS180 & 727 & \phantom07.4 & 767 & 3.48 & 0.611 \\
  HSIUF & 725 & \phantom08.6 & 747 & 3.30 & 0.656 \\
  LS220 & 724 & \phantom06.2 & 756 & 3.38 & 0.616 \\
  GShenFSU2.1 & 723 & 10.9 & 734 & 3.19 & 0.664 \\
  GShenFSU1.7 & 722 & 10.6 & 735 & 3.20 & 0.665 \\
  LS375 & 709 & \phantom08.0 & 729 & 3.15 & 0.626 \\
  HSTMA & 704 & \phantom05.6 & 702 & 2.91 & 0.661 \\
  HSFSG & 702 & \phantom07.6 & 731 & 3.16 & 0.662 \\
  HSDD2 & 701 & \phantom08.2 & 723 & 3.09 & 0.660 \\
  BHB$\Lambda$ & 700 & \phantom08.3 & 723 & 3.09 & 0.660 \\
  BHB$\Lambda\Phi$ & 700 & \phantom08.4 & 722 & 3.09 & 0.659 \\
  GShenNL3 & 699 & 11.9 & 691 & 2.83 & 0.671 \\
  HSTM1 & 675 & \phantom05.1 & 688 & 2.80 & 0.659 \\
  HShenH & 670 & \phantom06.8 & 694 & 2.85 & 0.678 \\
  HShen & 670 & \phantom06.4 & 694 & 2.85 & 0.678 \\
  HSNL3 & 669 & \phantom03.8 & 681 & 2.75 & 0.660 \\
  \hline\hline

  \end{tabular}
  \label{tab:avgf}
\end{table}

In the Slow Rotation regime, the parameterization of $f_\mathrm{peak}$
with $\sqrt{G\bar{\rho}_c}$ works particularly well, because
centrifugal effects are mild and there is no dependence on the
precollapse degree of differential rotation. In Table~\ref{tab:avgf},
we list $f_\mathrm{peak}$ and $\bar{\rho}_c$ averaged over simulations
with $0.02\leq T/|W|\leq0.06$ and broken down by EOS. We also provide
the standard deviation for $f_\mathrm{peak}$, average dynamical
frequency, average time-averaged central density, and the average
inner core mass at bounce for each EOS.  These quantitative results
further corroborate that $f_\mathrm{peak}$ and $\bar{\rho}_c$ are
closely linked. As expected from our analysis of the bounce signal in
Section~\ref{sec:bounce}, hyperons have no effect: HShen and HShenH
yield the same $f_\mathrm{peak}$ and $\bar{\rho}_c$ and so do HSDD2,
BHB$\Lambda$, and BHB$\Lambda\Phi$.

The results summarized by Table~\ref{tab:avgf} also suggest that the
subnuclear, nonuniform nuclear matter part of the EOS may play an
important role in determining $f_\mathrm{peak}$ and PNS
structure. This can be seen by comparing the results for EOS with the
same high-density uniform matter EOS but different treatment of
nonuniform nuclear matter.  For example, GShenNL3 and HSNL3 both
employ the RMF NL3 model for uniform matter, but differ in their
descriptions of nonuniform matter (cf.~Section~\ref{sec:eos}). They
yield $f_\mathrm{peak}$ that differ by
$\sim$$30\,\mathrm{Hz}$. Similarly, GShenFSU1.7 (and GShenFSU2.1)
produce $\sim$$15\,\mathrm{Hz}$ higher peak frequencies than HSFSG.
Interestingly, the difference between HShen and HSTM1 (both using RMF
TM1) in $f_\mathrm{peak}$ is much smaller even though they have
substantially different averaged $\bar{\rho}_c$ and $M_\mathrm{IC,b}$.

\begin{figure}
\center
\includegraphics[width=\linewidth]{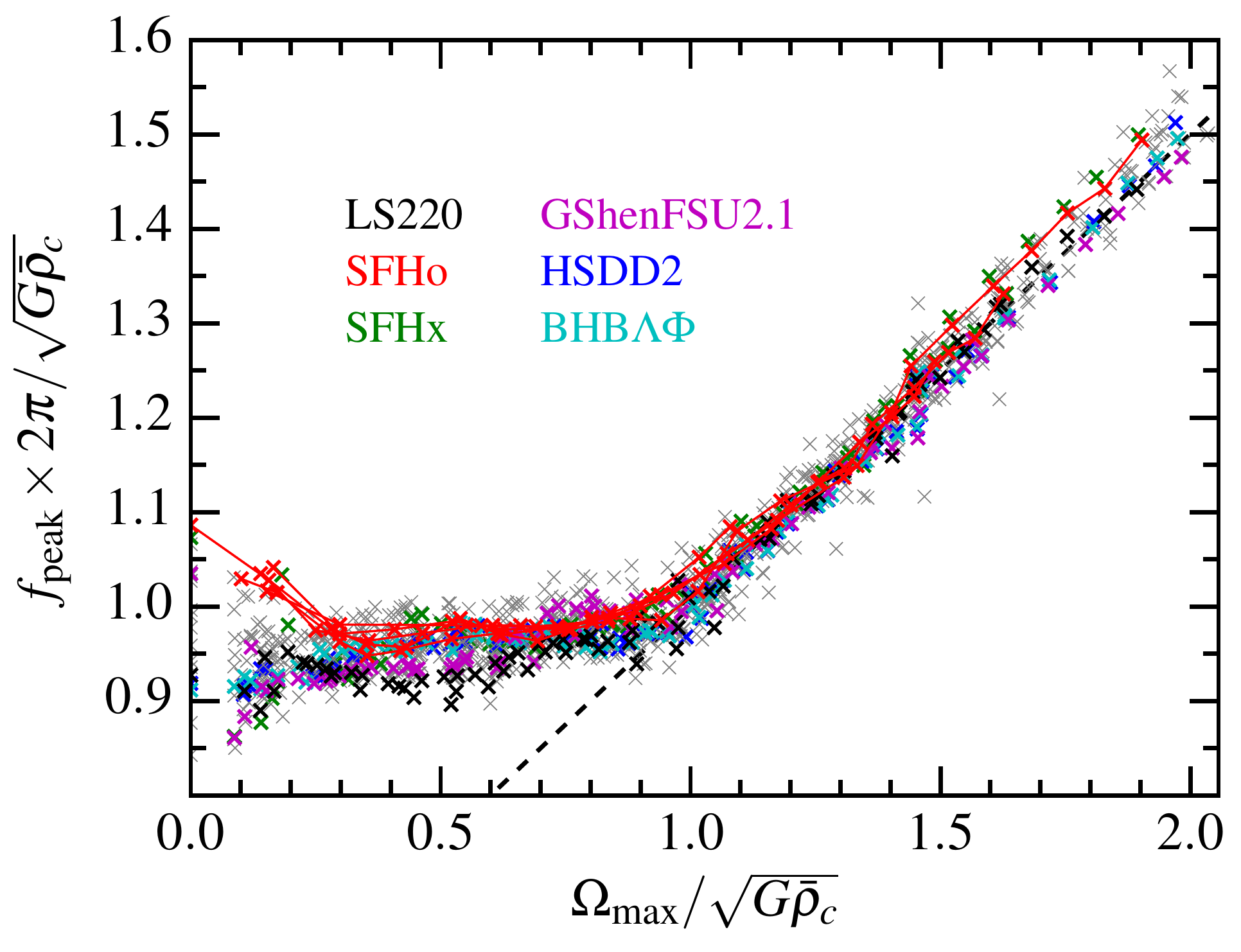}
\caption{\textbf{Universal Relation:} All differential rotation
  parameters and EOS result in simulations that obey the same
  relationship between the normalized peak frequency and the
  normalized maximum rotation rate $\Omega_\mathrm{max}$. The kink in
  the plot where $\Omega_\mathrm{max}=\sqrt{G\bar{\rho}_c}$
  corresponds to $T/|W|\approx 0.06$. The dashed line is described by
  $2\pi f_\mathrm{peak}/\sqrt{G\bar{\rho}_c} =
  0.5(1+\Omega_\mathrm{max}/\sqrt{G\bar{\rho}_c})$. This figure
  includes all 1487 simulations with $T/|W|<0.17$.}
\label{fig:fpeakmod_vs_wmaxmod}
\end{figure}

Figure~\ref{fig:fpeak} shows that $f_\mathrm{peak}$ is roughly
constant in the Slow Rotation regime, but increases with faster
rotation in the Rapid Rotation regime.  Centrifugal support, leads to
a monotonic decrease of the PNS density with increasing rotation
(cf.~Figure~\ref{fig:colormap}).  Thus, naively and based on
Equation~(\ref{eq:dyn}), we would expect a decrease $f_\mathrm{peak}$
with increasing rotation rate. We observe the opposite and this
warrants further investigation.

Figure~\ref{fig:fpeak} also shows that in the Rapid Rotation regime
the precollapse degree of differential rotation determines how quickly
the peak frequency increases with $T/|W|$, suggesting that $T/|W|$ may
not be the best measure of rotation for the purposes of understanding
the behavior of $f_\mathrm{peak}$. Instead, in
Figure~\ref{fig:fpeakmod_vs_wmaxmod}, we plot the normalized peak
frequency as a function of a different measure of rotation,
$\Omega_\mathrm{max}$ (normalized by $\sqrt{G\bar{\rho}_c}$), the
highest equatorial angular rotation rate achieved at any time outside
of a radius of $5\,\mathrm{km}$. We impose this limit to prevent
errors from dividing by small radii in $\Omega = v_\phi/r$.  This is a
convenient way to measure the rotation rate of the configuration
without needing to refer to a specific location or time. This produces
an interesting result (Figure~\ref{fig:fpeakmod_vs_wmaxmod}):
\emph{all our simulations for which we are able to reliably calculate
  the peak frequency follow the same relationship} in which the
normalized peak frequency is essentially independent of rotation at
lower rotation rates (Slow Rotation), followed by a linear increase
with rotation rate at higher rotation rates (Rapid Rotation). Note
that the transition between these regimes and the two parts of
Figure~\ref{fig:fpeakmod_vs_wmaxmod} occurs just when
$\Omega_\mathrm{max} \approx \sqrt{G\bar{\rho}_c}$.

\begin{figure}
\center \includegraphics[width=\linewidth]{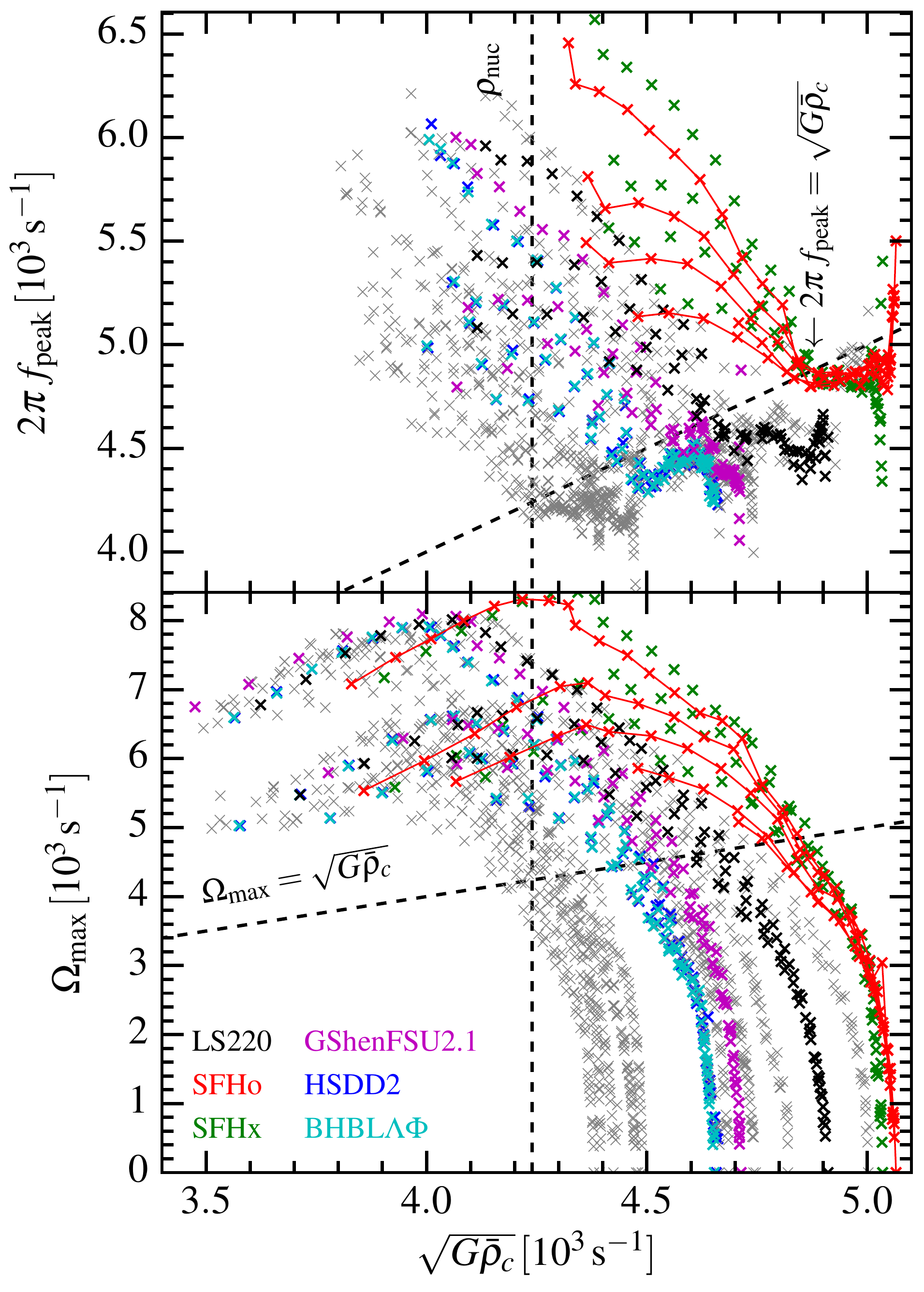}
\caption{\textbf{Demystifying the Universal Relation.} To better
  understand the relation in Figure~\ref{fig:fpeakmod_vs_wmaxmod}, we
  plot the peak frequency $f_\mathrm{peak}$ and the maximum rotation
  rate $\Omega_\mathrm{max}$ separately, each as a function of the
  dynamical frequency. The dramatic kink in
  Figure~\ref{fig:fpeakmod_vs_wmaxmod} is due to a sharp change in the
  behavior of $f_\mathrm{peak}$ once $2\pi
  f_\mathrm{peak}>\sqrt{G\bar{\rho}_c}$. An approximate nuclear
  saturation density of
  $\rho_\mathrm{nuc}=2.7\times10^{14}\,\mathrm{g\,cm}^{-3}$ is plotted
  as well for reference. The top panel contains the 1487 simulations
  with $T/|W|<0.17$, while the bottom panel contains all 1704
  collapsing simulations to show the decrease in $\Omega_\mathrm{max}$
  at extreme rotation rates.}
\label{fig:fdyn_vs_stuff}
\end{figure}

\begin{figure}
\center \includegraphics[width=\linewidth]{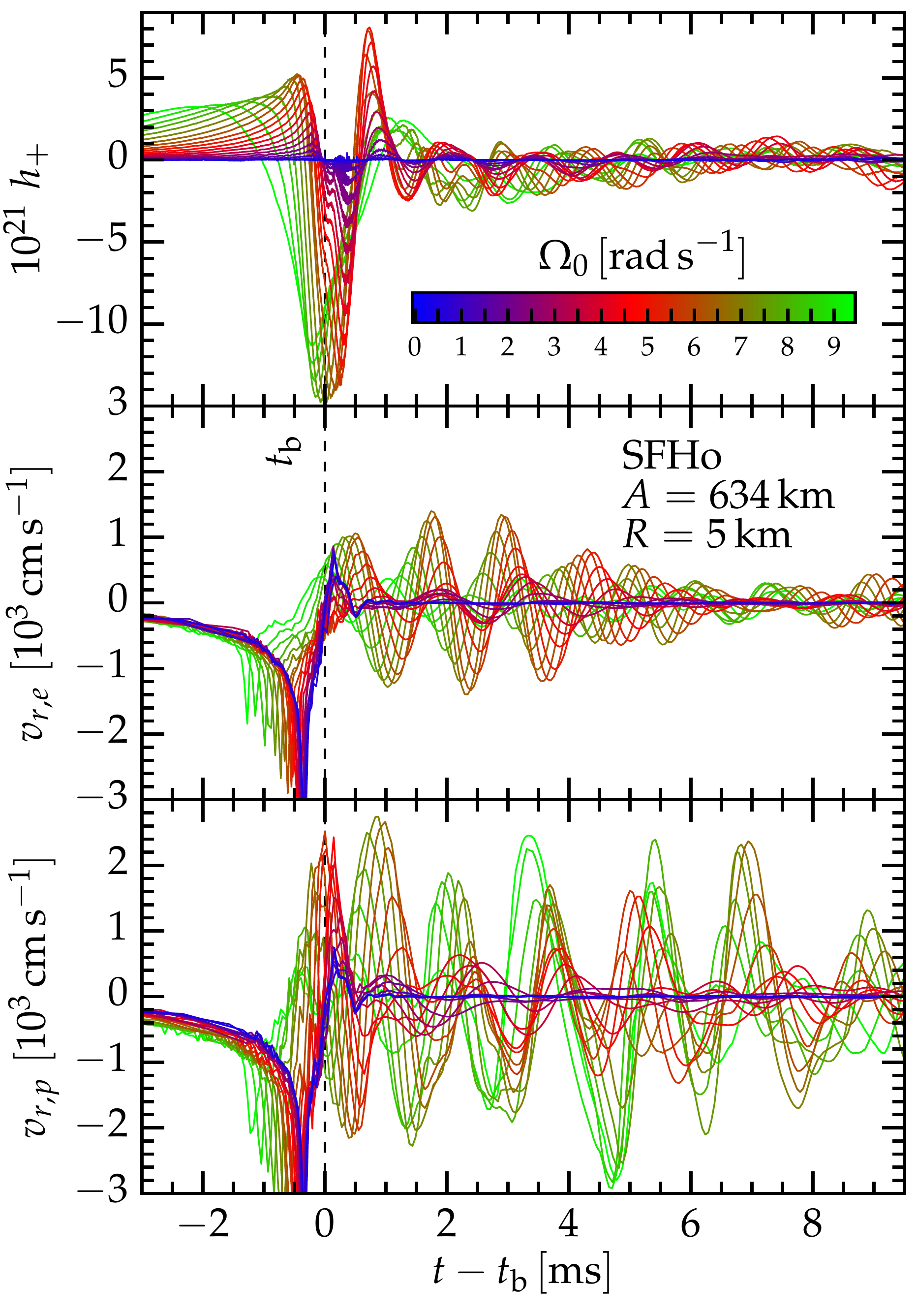}
\caption{\textbf{Rotation Changes Oscillation Mode Character}.
  In the top panel, we plot the GW signals for 20 cores collapsed with
  $A3 = 634\,\mathrm{km}$ and the SFHo EOS, color-coded according to
  their initial central rotation rate. The center and bottom panel
  show the radial velocity at $5\,\mathrm{km}$ from the origin along
  the equatorial and polar axis, respectively. We indicate core bounce
  with a vertical dashed line.  In the
  Rapid Rotation regime (starting at around the
  transition from red to green color), postbounce GW frequency and
  velocity oscillation frequency increase visibly. In the same regime,
  the oscillation mode structure changes. The polar velocities
  continue to increase, while the oscillations are damped along the
  equator.}
\label{fig:velcompare}
\end{figure}

We can gain more insight into the relationship of $f_\mathrm{peak}$
and $\Omega_\mathrm{max}$ by considering
Figure~\ref{fig:fdyn_vs_stuff}, in which we plot both
$f_\mathrm{peak}$ (top panel) and $\Omega_\mathrm{max}$ (bottom panel)
against the dynamical frequency $\sqrt{G \bar{\rho}_c}$. Since
rotation decreases $\bar{\rho}_c$, rotation rate \emph{increases from
  right to left} in the figure.

First, consider $f_\mathrm{peak}$ in the top panel of
Figure~\ref{fig:fdyn_vs_stuff}. At high $\bar{\rho}_c$ (Slow Rotation
regime), all $f_\mathrm{peak}$ cluster with EOS below the line $2\pi
f_\mathrm{peak} = \sqrt{G\bar{\rho}_c}$ with small differences between
rotation rates, just as we saw in Figures~\ref{fig:fpeak} and
\ref{fig:fpeakmod_vs_wmaxmod}.  However, as the rotation rate
  increases (and $\bar{\rho}_c$ decreases), $f_\mathrm{peak}$ rapidly
increases and exhibits the spreading with differential rotation
already observed in Figure~\ref{fig:fpeak}. Notably, this occurs in
the region where the peak PNS oscillation frequency exceeds its
dynamical frequency, $2\pi f_\mathrm{peak} > \sqrt{G\bar{\rho}_c}$.

Now turn to the $\Omega_\mathrm{max}$ -- $\sqrt{G\bar{\rho}_c}$
relationship plotted in the bottom panel of
Figure~\ref{fig:fdyn_vs_stuff}. At the lowest rotation rates, this
plot simply captures how $\bar{\rho}_\mathrm{c}$ varies between
EOS. For slowly rotating cores, $\Omega_\mathrm{max}$ is substantially
smaller than the dynamical frequency $\sqrt{G\bar{\rho}_c}$, and
$\Omega_\mathrm{max}$ points cluster in a line for each EOS. As
$\Omega_\mathrm{max}$ surpasses $\sqrt{G\bar{\rho}_c}$, this smoothly
transitions to the Rapid Rotation regime, in which $\bar{\rho}_c$ is
significantly driven down with increasing rotation rate. At the
highest rotation rates (Extreme Rotation regime),
$\Omega_\mathrm{max}$ exceeds $\sqrt{G\bar{\rho}_\mathrm{c}}$ by a few
times and centrifugal effects dominate in the final phase of core
collapse, preventing further collapse and spin-up. Faster initial
rotation (lower $\bar{\rho}_\mathrm{c}$) results in lower
$\Omega_\mathrm{max}$ in this regime, consistent with previous
work~\cite{abdikamalov:14}.

The bottom panel of Figure~\ref{fig:fdyn_vs_stuff} also allows us to
understand the effect of precollapse differential rotation. Stronger
differential rotation naturally reduces centrifugal support. Thus it
allows a collapsing core to reach higher $\Omega_\mathrm{max}$ before
centrifugal forces prevent further spin-up. This causes the spreading
branches for the different A values in our models.

Armed with the above observations on differential rotation and the
$2\pi f_\mathrm{peak}$ -- $\sqrt{G\bar{\rho}_c}$ and
$\Omega_\mathrm{max}$ -- $\sqrt{G\bar{\rho}_c}$ relationships, we now
return to Figure~\ref{fig:fpeakmod_vs_wmaxmod}. It depicts a sharp
transition in the behavior of $f_\mathrm{peak}$ at
$\Omega_\mathrm{max} = \sqrt{G\bar{\rho}_c}$. A sharp transition is
present in the $2\pi f_\mathrm{peak}$ -- $\sqrt{G\bar{\rho}_c}$
relationship, but not in the $\Omega_\mathrm{max}$ --
$\sqrt{G\bar{\rho}_c}$ relationship shown in
Figure~\ref{fig:fdyn_vs_stuff}. The variable connected to PNS
structure, $\bar{\rho}_c$, instead varies smoothly and slowly with
rotation through the $\Omega_\mathrm{max} = \sqrt{G\bar{\rho}_c}$
line. This is a strong indication that the sharp upturn of
$f_\mathrm{peak}$ at $\Omega_\mathrm{max} = \sqrt{G\bar{\rho}_c}$ in
Figure~\ref{fig:fpeakmod_vs_wmaxmod} is due to \emph{a change in the
  dominant PNS oscillation mode rather than due to an abrupt change in
  PNS structure}. The observation that centrifugal effects do not
become dominant until $\Omega_\mathrm{max}$ is several times
$\sqrt{G\bar{\rho}_c}$ corroborates this interpretation.

In Figure~\ref{fig:velcompare}, we plot the GW signals along with the
equatorial and polar radial velocities $5\,\mathrm{km}$ from the
origin for all 20 simulations using the SFHo EOS with a differential
rotation parameter $A3=634\,\mathrm{km}$. The postbounce GW frequency
clearly follows the frequency of the fluid oscillations.  Both
frequencies begin to significantly increase at around
$\Omega_0\approx5\,\mathrm{rad\,s}^{-1}$ (corresponding to
$T/|W|\approx0.06$, red-colored graphs). The polar and equatorial
velocity oscillation amplitudes initially increase with rotation rate
(colors going from blue to red), but when rotation becomes rapid
(colors going from red to green) the equatorial velocities decrease
and polar velocities continue to grow. This demonstrates that
\emph{the multi-dimensional PNS mode structure is altered at rapid
  rotation and no longer follows a simple $\ell = 2, m= 0$
  description}. This is also apparent from comparing the left and
right panels of Figure~\ref{fig:colormap}.

While the above results show that the increase in $f_\mathrm{peak}$ is
most likely a consequence of changes in the mode structure with
rotation, it is not obvious what detailed process is driving the
changes. While future work will be needed to answer this conclusively,
we can use the work of Dimmelmeier~\emph{et al.}~\cite{dimmelmeier:06}
as the basis of educated speculation. They study oscillations of
rotating equilibrium polytropes and show that the $\ell = 2, m=0$
f-mode frequency has a weak dependence on both rotation rate and
differential rotation. This is consistent with our findings for models
in the Slow Rotation regime ($T/|W| \lesssim 0.06$). They also
identify several inertial modes whose restoring force is the Coriolis
force (e.g., \cite{stergioulas:03}). The inertial mode frequency
increases rapidly with rotation and is sensitive to differential
rotation, which is what we see for our PNS oscillations in the Rapid
Rotation regime ($T/|W| \gtrsim 0.06$). Our PNS cores are also
significantly less dense than the equilibrium models
of~\cite{dimmelmeier:06}, which allows the $\ell=2$ modes in our
simulations to have lower oscillation frequencies that intersect with
the frequencies of the inertial modes in~\cite{dimmelmeier:06}. It
could thus be that in our PNS cores inertial and $\ell = 2$ f-mode
eigenfunctions overlap and couple nonlinearly, leading to an
excitation of predominantly inertial oscillations as rotation becomes
more rapid. The increase of the inertial mode frequency with rotation
would explain the trends we see in $f_\mathrm{peak}$ in
Figure~\ref{fig:fpeak}.

Coriolis forces should become dynamically important for oscillations
when the oscillation frequency is locally smaller than the Coriolis
frequency, given by $2\pi f_\mathrm{core} = 2 \Omega \sin \theta$
(e.g.,~\cite{saio:13}), where $\theta$ is the latitude from the
equator and, for simplicity, $\Omega$ is a uniform rotation
rate. Thus, we expect Coriolis effects to become locally relevant when
$\Omega \gtrsim 2 \pi f_\mathrm{peak} / (2 \sin \theta) \approx
\sqrt{G\bar{\rho}_c} / (2 \sin \theta)$. The kink in
Figure~\ref{fig:fpeakmod_vs_wmaxmod} is at $\Omega_\mathrm{max} =
\sqrt{G\bar{\rho}_c}$, and hence the behavior of
the PNS oscillations changes precisely when we
expect Coriolis effects to begin to matter. This is supports the
notion that the PNS oscillations may be transitioning to inertial
nature at high rotation rates.

\emph{Conclusions:} The effects of the EOS on the postbounce GW
frequency can be parameterized almost entirely in terms of the
dynamical frequency $\sqrt{G\rho_c}$ of the core after bounce. In the
Slow Rotation regime ($T/|W| \lesssim 0.06$), the postbounce frequency
depends little on rotation rate. In the Rapid Rotation regime ($0.06
\lesssim T/|W| \lesssim 0.17$), inertial effects modify the nature of
the oscillations, causing the frequency to increase with rotation
rate. We find that the maximum rotation rate outside of
$5\,\mathrm{km}$ is the most useful parameterization of rotation for
the purpose of understanding the oscillation frequencies. In the
Extreme Rotation regime ($T/|W| \gtrsim 0.17$), the postbounce GW
frequency decreases with rotation because centrifugal support keeps
the core very extended.

\subsection{GW Correlations with Parameters and EOS}
\label{sec:correlations}

We are interested in how characteristics of the GWs vary with
rotation, properties of the EOS, and the resulting conditions during
core collapse and after bounce. Rather than plot every variable
against each other variable, we employ a simple linear correlation
analysis. We calculate a linear correlation coefficient $\mathcal{C}$
between two quantities $U$ and $V$ that quantifies the strength of the
linear relationship between two variables:
\begin{equation}
  \mathcal{C}_{U,V} =
  \frac{\sum(\frac{U-\overline{U}}{s_U})(\frac{V-\overline{V}}{s_V})}{(N-1)}\,\,.
\end{equation}
The summation is over all $N$ simulations included in the
analysis. The sample standard deviation of a quantity $U$ is
\begin{equation}
  s_U = \sqrt{\frac{1}{N-1}\sum(U-\overline{U})^2}\,\,,
\end{equation}
 where $\overline{U} = \sum U/N$ is the average value of $U$ over all
 $N$ simulations. The correlation coefficient is always bound between
 $-1$ (strong negative correlation) and $1$ (strong positive
 correlation). This only accounts for linear correlations, so even if
 two variables are tightly coupled, nonlinear relationships will
 reduce the magnitude of the correlation coefficient and a more
 involved analysis would be necessary for
   characterizing nonlinear relationships (see, e.g.,
 \cite{engels:14}).

\begin{figure*}
\center \includegraphics[width =
  \textwidth]{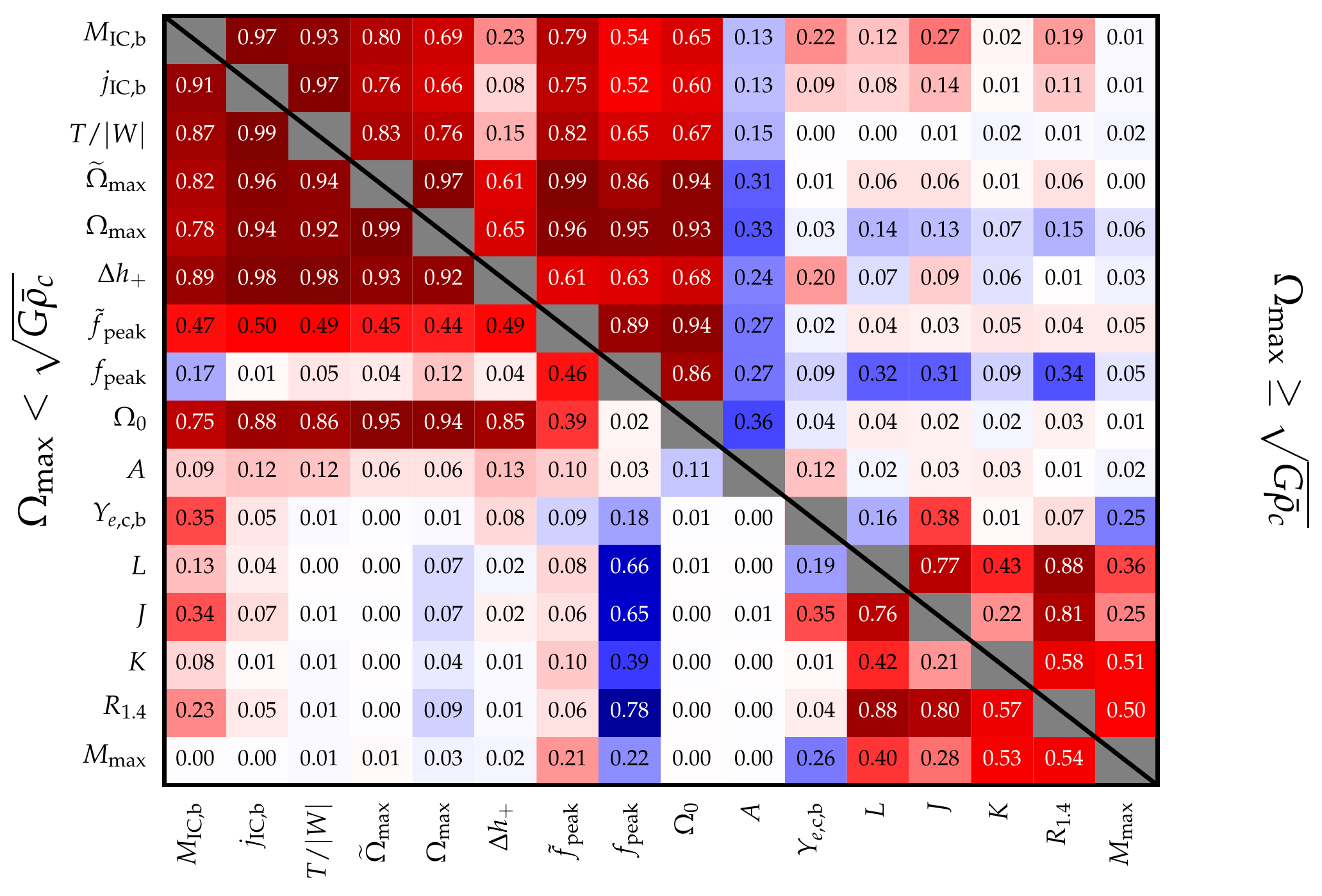}
\caption{\textbf{Correlation Coefficients.} We calculate linear
  correlation coefficients between several parameters and observables
  in our collapsing simulations. The cell color represents the number
  within the cell, with positive correlations being red and negative
  correlations blue. \textit{Bottom Left}: Correlation coefficients
  for 874 simulations with $\Omega_\mathrm{max}<\sqrt{G\bar{\rho}_c}$
  (i.e.\ slowly rotating). \textit{Top Right:} Correlation
  coefficients for 613 simulations with
  $\Omega_\mathrm{max}>\sqrt{G\bar{\rho}_c}$ (i.e.\ rapidly rotating)
  and $T/|W|<0.17$. $M_\mathrm{IC,b}$ is the mass of the inner core,
  defined by the region in sonic contact with the center, at core
  bounce. $j_\mathrm{IC}$ is the angular momentum of the inner core at
  bounce. $T/|W|$ is the inner core's ratio of rotational kinetic to
  gravitational potential energy at core bounce. $\Omega_\mathrm{max}$
  is the maximum rotation rate obtained at any time in the simulation
  outside of $R=5\,\mathrm{km}$ and
  $\widetilde{\Omega}_\mathrm{max}=\Omega_\mathrm{max}/\sqrt{G\bar{\rho}_c}$. $f_\mathrm{peak}$
  is the peak frequency of GWs from postbounce PNS oscillations, and
  $\widetilde{f}_\mathrm{peak}=f_\mathrm{peak}/\sqrt{G\bar{\rho}_c}$. $\Omega_0$
  is the precollapse maximum rotation rate and $A$ is the precollapse
  differential rotation parameter. $Y_{e,c}$ is the central electron
  fraction at core bounce. The incompressibility $K$, symmetry energy
  $J$, density derivative of the symmetry energy $L$, radius of a
  $1.4\,M_\odot$ star $R_{1.4}$, and $M_\mathrm{max}$ are properties
  of the EOS described in Section~\ref{sec:eos}.}
\label{fig:correlations}
\end{figure*}

We display the correlation coefficients of several relevant quantities
in Figure~\ref{fig:correlations}. $L$, $J$, $K$, $R_{1.4}$, and
$M_\mathrm{max}$ are all innate properties of a given EOS
(Section~\ref{sec:eos}). $A$ and $\Omega_0$ are the input parameters
that determine the rotation profile as defined in
Equation~\ref{eq:rotation}. The rest of the quantities are outputs
from the simulations. Quantities defined at the time of core bounce
are the inner core mass $M_\mathrm{IC,b}$, the central electron
fraction $Y_{e,\mathrm{c,b}}$, the inner core angular momentum
$j_\mathrm{IC,b}$, and the ratio of the inner core rotational energy
to gravitational energy $T/|W|$. Rotation is also parameterized by the
maximum rotation rate $\Omega_\mathrm{max}$ and by
$\widetilde{\Omega}_\mathrm{max}=\Omega_\mathrm{max}/\sqrt{G\bar{\rho}_c}$
(see Section~\ref{sec:postbounce} for definitions). GW characteristics
are quantified in the amplitude of the bounce signal $\Delta h_+$, the
peak frequency of the postbounce signal $f_\mathrm{peak}$, and its
variant normalized by the dynamical frequency
$\widetilde{f}_\mathrm{peak}=f_\mathrm{peak}/\sqrt{G\bar{\rho}_c}$. The
bottom left half of the plot shows the values of the correlation
coefficients for 874 simulations in the Slow Rotation regime
($\Omega_\mathrm{max}<\sqrt{G\bar{\rho}_c}$, $T/|W| \lesssim 0.06$)
and the top right half shows correlations for 613 simulations in the
Rapid Rotation regime ($\Omega_\mathrm{max}\geq \sqrt{G\bar{\rho}_c}$,
$0.06 \lesssim T/|W| \leq 0.17$).

There is a region in the bottom right corner of
Figure~\ref{fig:correlations} that shows the correlations between EOS
parameters $L$, $J$, $K$, $R_{1.4}$, and $M_\mathrm{max}$. Since we
chose to use existing EOS rather than create a uniform parameter
space, there are correlations between the input values of $L$, $J$, and
$K$ that impose some selection bias on the other correlations. In our
set of 18 EOS, there is a strong correlation between $R_{1.4}$ and
both $L$ and $J$. The maximum neutron star mass correlates most
strongly with $K$ and $L$. These findings are not new and just reflect
current knowledge of how the nuclear EOS affects neutron star
structure (e.g.,~\cite{lattimer:01,lattimer:12,oertel:17}). The small
amount of asymmetry in this corner is the effect of selection bias, as
some EOS contribute more data points to one or the other rotation
regime.

Next, we note that the central $Y_e$ at bounce ($Y_\mathrm{e,c,b}$)
exhibits correlations with EOS characteristics $J$, $L$, and
$M_\mathrm{max}$. This encodes the EOS dependence in the high-density
part of the $Y_e(\rho)$ trajectories shown in
Figure~\ref{fig:yeofrho}.  The mass of a nonrotating inner core at
bounce is sensitive to $Y_\mathrm{e,c,b}^2$ (though we note that it is
also sensitive to $Y_e$ at lower densities and to EOS properties). Our
linear analysis in Figure~\ref{fig:correlations} picks this up as a
clear correlation between $Y_\mathrm{e,c,b}$ and
$M_\mathrm{IC,b}$. This correlation is stronger in the slow to
moderately rapidly rotating models (bottom left half of the figure)
and weaker in the rapidly rotating models (top right half of the
figure) since in these models rotation strongly increases
$M_\mathrm{IC,b}$. This can also be seen in the strong correlations of
$M_\mathrm{IC,b}$ with all of the rotation variables.

As discussed in Section~\ref{sec:bounce} and pointed out in previous
work (e.g., \cite{abdikamalov:14}), the GW signal from bounce,
quantified by $\Delta h_+$, is very sensitive to mass
$M_\mathrm{IC,b}$ and $T/|W|$ of inner core at bounce. Our correlation
analysis confirms this and shows that the $\Delta h_+$ is also
correlated equally strongly with $j_\mathrm{IC,b}$ and
$\Omega_\mathrm{max}$ as with $T/|W|$.  As expected from
Figure~\ref{fig:amplitude}, correlation with the differential rotation
parameter $A$ is weak in the slow to moderately rapid rotation regime,
but there is a substantial anti-correlation with the value of $A$ in
the rapidly rotating regime (the smaller $A$, the more differentially
spinning a core is at the onset
of collapse).

Figure~\ref{fig:correlations} also shows that the most interesting
correlations of any observable from an EOS perspective are exhibited
by the peak postbounce GW frequency $f_\mathrm{peak}$. In the slow to
moderately rapidly rotating regime ($\Omega_\mathrm{max} \lesssim
\sqrt{G\bar{\rho}_c}$), $f_\mathrm{peak}$ has its strongest
correlations with EOS characteristics $K$, $J$, $L$, $R_{1.4}$
through their influence on the PNS central density
and is essentially independent of the rotation rate
(cf.~Figures~\ref{fig:fpeak} and \ref{fig:fpeakmod_vs_wmaxmod}).  For
rapidly rotating models ($\Omega_\mathrm{max} \gtrsim
\sqrt{G\bar{\rho}_c}$) there is instead a clear correlation of
$f_\mathrm{peak}$ with all rotation quantities. Note that the
correlations with EOS quantities are all but removed for the
normalized peak frequency $\widetilde{f}_\mathrm{peak} =
f_\mathrm{peak} / \sqrt{G\bar{\rho}_c}$. This supports our claim in
Section~\ref{sec:postbounce} that the influence of the EOS on the peak
frequency is parameterized essentially by the postbounce dynamical
frequency $\sqrt{G\bar{\rho}_c}$.

\emph{Conclusions:} Linear correlation coefficients show the
interdependence of rotation parameters, EOS parameters, and simulation
results. We use these to support our claims that the EOS dependence is
parameterized by the dynamical frequency and that rotation is
dynamically important for oscillations in the Rapid Rotation regime.

\subsection{Prospects of Detection and Constraining the EOS}
\label{sec:snr}

The signal to noise ratio (SNR) is a measure of the strength of a
signal observed by a detector with a given level of noise. We
calculate SNRs using the Advanced LIGO noise curve at design
sensitivity in the high-power zero-detuning
configuration~\cite{aligo,LIGO-sens-2010}. We assume optimistic
conditions where the rotation axis is perpendicular to the line of
sight and the LIGO interferometer arms are optimally oriented and
$10\,\mathrm{kpc}$ from the core collapse event. Following
\cite{flanagan:98a,abdikamalov:14}, we define the matched-filtering SNR $\rho$ of an
observed GW signal $h(t)$ as

\begin{equation}
  \rho = \frac{\langle d,x\rangle}{\langle x,x\rangle^{1/2}}\,\,,
\label{eq:snr}
\end{equation}
where $d$ is observed data and $x$ is a template waveform. When we
calculate an SNR for our simulated
signals, we take $d=x$ to mimic the GWs from the source matching a
template exactly, and this simplifies to $\rho^2=\langle
x,x\rangle$. The inner product integrals are calculated using
\begin{equation}
  \langle a,b\rangle = \int_0^\infty \frac{4\widetilde{h}_a^* \widetilde{h}_b}{S_n}df\,\,,
\end{equation}
where $S_n$ is the one-sided noise spectral density. We follow the
  LIGO convention~\cite{ligoconventions:04} for Fourier transforms,
  namely
\begin{equation}
  \widetilde{h}(f) = \int_{-\infty}^{\infty}h(t) e^{-2\pi ift}dt\,\,.
\end{equation}

Furthermore, we estimate the difference between two waveforms as seen
by Advanced LIGO with the mismatch $\mathcal{M}$ described and
implemented in Reisswig \& Pollney~\cite{reisswig:11}:
\begin{equation}
  \mathcal{M} = 1 - \max_{t_\mathrm{A}}\left[\frac{\langle x_1,x_2\rangle}{\sqrt{\langle x_1,x_1\rangle\langle x_2,x_2\rangle}}\right]\,\,,
  \label{eq:mismatch}
\end{equation}
where the latter term is the match between the two waveforms and is
maximized over the relative arrival times of the two waveforms
$t_\mathrm{A}$. Note that due to the axisymmetric nature of our
simulations, our waveforms only have the $+$ polarization, making a
maximization over complex phase unnecessary.

The simulated waveforms span a finite time and is sampled at
nonuniform intervals. To mimic real LIGO data, we resample the GW
time series data at the LIGO sampling frequency of
$16384\,\mathrm{Hz}$ before performing the discrete Fourier
transform.

\begin{figure}
\includegraphics[width=\linewidth]{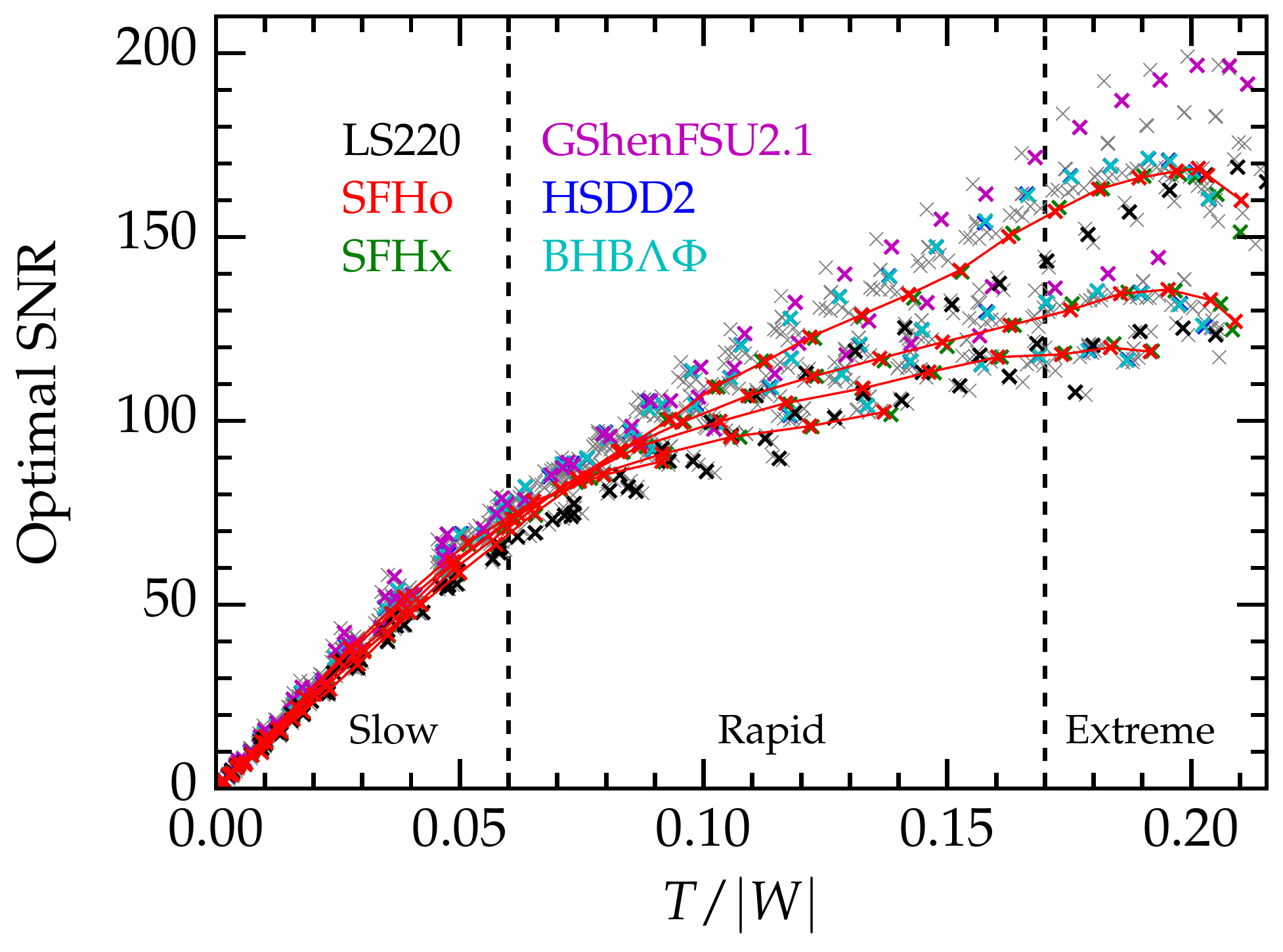}
\caption{\textbf{Signal to Noise Ratios.} The SNR for all 1704
  collapsing simulations that result in collapse and core bounce,
  assuming the rotation axis is perpendicular to the line of sight,
  the aLIGO interferometer is optimally oriented and at design
  sensitivity in the high power zero detuning configuration, and the
  source is $10\,\mathrm{kpc}$ away. A SNR of $\gtrsim10$ is
  considered detectable. The colors correspond to the EOS in
  Figure~\ref{fig:constraints}. A line is drawn through all the SFHo
  simulations to guide the eye. Each of the five branches corresponds
  to a different value of the differential rotation parameter $A$,
  where $A=300\,\mathrm{km}$ is the longest branch and
  $A=10000\,\mathrm{km}$ is the shortest.}
\label{fig:snr}
\end{figure}

In Figure~\ref{fig:snr}, we show the SNR for our 1704 collapsing cores
assuming a distance of $10\,\mathrm{kpc}$ to Earth.  Faster rotation
(higher $T/|W|$ of the inner core at bounce) leads to stronger
quadrupolar deformations, in turn causing stronger signals that are
more easily observed, but only up to a point. If rotation is too fast,
centrifugal support keeps the core more extended with lower average
densities, resulting in a less violent quadrupole oscillation and
weaker GWs. This happens at lower rotation rates for the rotation
profiles that are more uniformly rotating (e.g., the
$A5=10000\,\mathrm{km}$ series), since the large amount of angular
momentum and rotational kinetic energy created by even a small
rotation rate can be enough to provide significant centrifugal
support. The more strongly differentially rotating cases (e.g., the
$A1 = 300\,\mathrm{km}$ series) require much faster rotation before
centrifugal support becomes important at bounce. This also means that
they can reach greater inner core deformations and generate stronger
GWs.

All of the EOS result in similar SNRs for a given rotation profile.
We observe a larger spread with EOS in estimated SNR for the rapid,
strongly differentially rotating cases. The bounce part is the
strongest part of the GW signal and dominates the SNR. Hence, the
EOS-dependent differences in the bounce signal pointed out in
Section~\ref{sec:bounce} are most relevant for understanding the EOS
systematics seen in Figure~\ref{fig:snr}. For example, the LS220 EOS
yields the smallest inner core masses at bounce and correspondingly
the smallest $\Delta h_+$. This translates to the systematically lower
SNRs for this EOS.

\begin{figure}
\includegraphics[width=\linewidth]{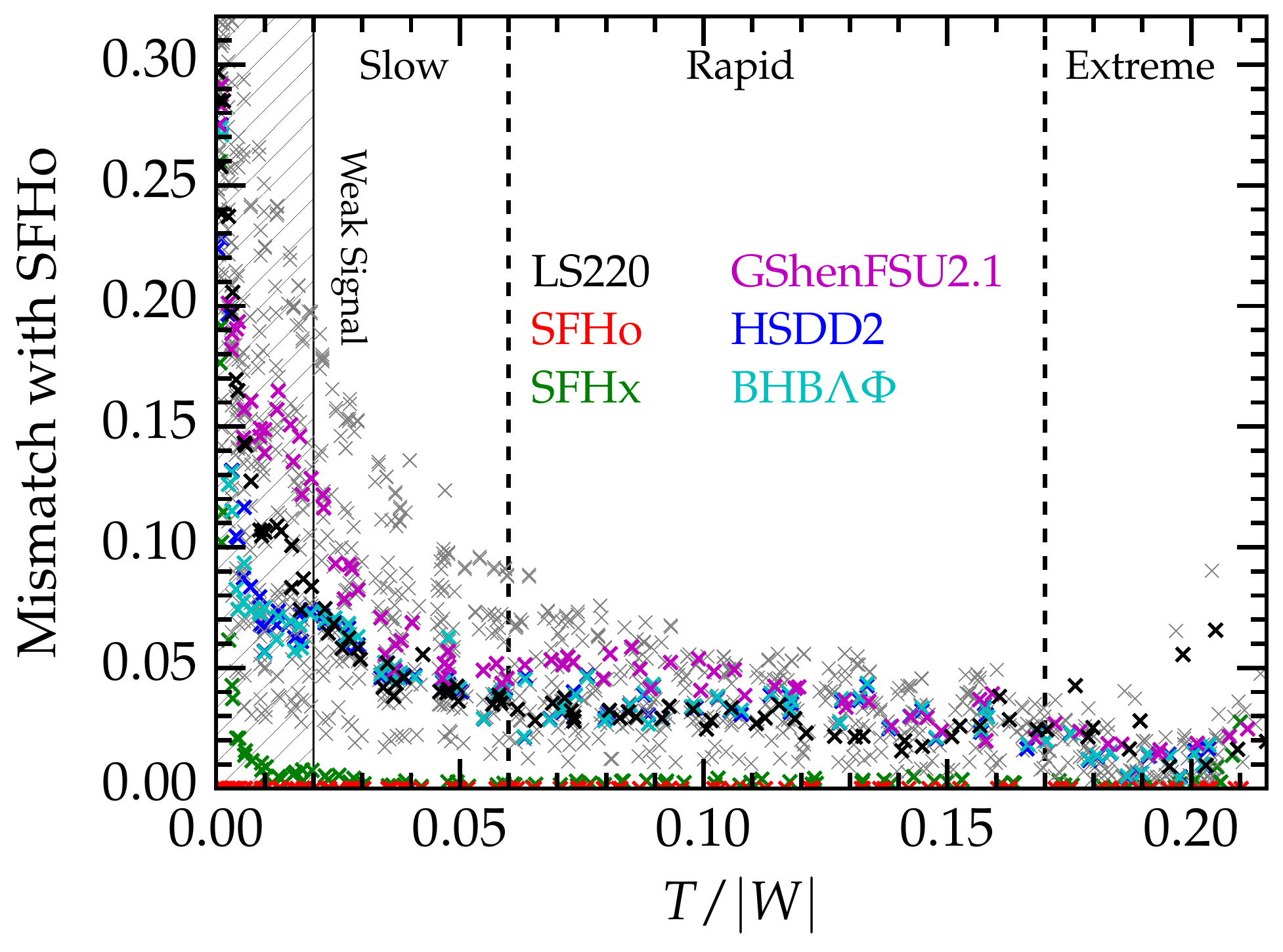}
\caption{\textbf{GW Differences due to the EOS.} The GW mismatch (see
  Equation~\ref{eq:mismatch}) integrated between SFHo and each of the
  other EOS for the same rotation parameters ($A$,$\Omega_0$) for all
  1704 collapsing simulations. Note that $T/|W|$ at bounce will be
  slightly different between simulations with the same initial
  rotation parameters due to EOS effects. Only data through
  $6\,\mathrm{ms}$ after the end of the bounce signal are used to
  avoid contributions from prompt convection. Differences between EOS
  decrease with faster rotation as the bounce signal becomes stronger
  and rotational effects become more important. The HShen and HShenH
  EOS (not identified by color and shown in gray) have the
  consistently largest mismatches with SFHo in the Slow and Rapid
  Rotation regimes. Mismatch calculations at $T/|W|\lesssim0.02$ are
  unreliable due to a very weak GW signal. In the extreme rotation
  regime, some EOS develop larger mismatches with SFHo. This occurs
  because simulations with these EOS transition to a centrifugal
  bounce at subnuclear density at lower rotation rates than SFHo. The
  resulting qualitative and quantitative change in the waveforms leads
  to larger mismatches.}
\label{fig:mismatch}
\end{figure}

We can get a rough estimate for how different the waveforms are with
the simple scalar mismatch (Equation~\ref{eq:mismatch}), which we
calculate with respect to the simulations using the SFHo EOS and the
same value of $A$ and $\Omega_0$. Simulations using different EOS but
the same initial rotation profile will result in slightly different
values of $T/|W|$ at bounce, so this measures the difference between
waveforms from the same initial conditions rather than from the same
bounce conditions. In the context of a matched-filter search, the
mismatch roughly represents the amount of SNR lost due to differences
between the template and the signal. However, note that searches for
core collapse signals in GW detector data have thus far relied on
waveform-agnostic methods that search for excess power above the
background noise (e.g., \cite{ligo:16c}).

Figure~\ref{fig:mismatch} shows the results of the mismatch
calculations. The large mismatches at $T/|W|\lesssim0.02$ are simply
due to the small amplitudes of the GWs causing large relative
errors. The mismatch results for such slowly spinning models have no
predictive power and we do not analyze them further. At higher
rotation rates, \emph{the dynamics are increasingly determined by
  rotation and decreasingly determined by the details of the EOS}, and
the mismatch generally decreases with increasing rotation rate.

An exception to this rule occurs in the Extreme Rotation regime
($T/|W| \gtrsim 0.17$) where waveforms show increasing mismatches with
SFHo simulation results (most notably, LS220 and LS180). In this
regime, the bounce dynamics changes due to centrifugal support and
bounce occurs below nuclear saturation density for some EOS. Moreover,
when centrifugal effects become dominant, bounce is also slowed down,
widening the GW signal from bounce and reducing its amplitude. The
initial rotation rate around which this occurs differs between EOS and
the resulting qualitative and quantitative changes in the waveforms
drive the increasing mismatches.

In Figure~\ref{fig:mismatch}, the HShen EOS (included in the gray
crosses) consistently shows the highest mismatch with SFHo. These two
EOS use different low-density and high-density treatments (see
Table~\ref{tab:eos} and Section~\ref{sec:eos}). It 
is insightful to compare mismatches between EOS using the same
(or similar) physics in either their high-density or low-density
treatments of nuclear matter in order to isolate the origin of
  large mismatch values. In the following, we again use the example
of the $A3 = 634\,\mathrm{km}, \Omega_0=5.0\,\mathrm{rad\,s}^{-1}$
rotation profile and compute mismatches between pairs of EOS. HShen
and HSTM1 both use the RMF TM1 parameterization for high-density
uniform matter, but deal with nonuniform lower-density matter in
different ways (see Section~\ref{sec:eos}). Their mismatch is
$\mathcal{M}=0.85\%$.  GShenNL3 and HSNL3 use the RMF NL3
parameterization for uniform matter and also differ in their
nonuniform matter treatment. They have a mismatch of
$\mathcal{M}=5.1\%$. This is comparable to the HShen-SFHo mismatch of
$\mathcal{M}=7.3\%$. We find a mismatch of $\mathcal{M}=3.2\%$ for the
GShenFSU2.1--HSFSG pair. Both use the RMF FSUGold parameterization for
uniform nuclear matter and again differ in the nonuniform parts.

The above results suggest that \emph{the treatment of low-density
  nonuniform nuclear matter is at least in some cases an important
  differentiator between EOS} in the GW signal of rotating core
collapse. While perhaps somewhat unexpected, this finding may, in
fact, not be too surprising: Previous work (e.g.,
\cite{dimmelmeier:08,abdikamalov:14}) already showed that the GW
signal of rotating core collapse is sensitive to the inner core mass
at bounce (and, of course, its $T/|W|$, angular momentum, or its
maximum angular velocity; cf.\ Section \ref{sec:correlations}). The
inner core mass at bounce is sensitive to the low-density EOS through
the pressure and speed of sound in the inner core material in the
final phase of collapse and through chemical potentials and
composition, which determine electron capture rates and thus the $Y_e$
in the final phase of collapse and at bounce.

We can also compare EOS with the same treatment of nonuniform
lower-density matter, but different high-density treatments. We again
pick the $A3 = 634\,\mathrm{km}, \Omega_0=5.0\,\mathrm{rad\,s}^{-1}$
($T/|W| \sim 0.075$) model sequence as an example for quantitative
differences. GShenFSU2.1 and GShenFSU1.7 ($\mathcal{M}=0.0031\%$)
differ only at super-nuclear densities, where GShenFSU2.1 is extra
stiff in order to support a $2\,M_\odot$ neutron star. HShenH adds
hyperons to HShen ($\mathcal{M}=0.0027\%$), BHB$\Lambda$ adds hyperons
to HSDD2 ($\mathcal{M}=0.0082\%$), and BHB$\Lambda\Phi$ includes an
extra hyperonic interaction over BHB$\Lambda$
($\mathcal{M}=0.014\%$). All of the Hempel-based EOS (HS, SFH, BHB)
use identical treatments of low-density nonuniform matter, but
parameterize the EOS of uniform nuclear matter differently. For our
example rotation profile, the mismatch with SFHo varies from 0.12\%
(for SFHx) to 7.6\% (for GShenNL3). \emph{The results are comparable
  with the mismatch induced by differences in the low-density regime.}

\emph{Conclusions:} We expect a maximum SNR of around $200$ from a
source at a distance of $10\,\mathrm{kpc}$, though this depends both
on the amount of differential rotation and the EOS. Using use a simple
scalar mismatch to calculate the differences between waveforms
generated using different EOS, we find that both the treatment of
nonuniform and uniform nuclear matter significantly affect the
waveforms, though differences at densities more than about twice
nuclear are of little importance.

\subsection{Effects of Variations in Electron Capture Rates}
\label{sec:ecapture}

Electron capture in the collapse phase is a crucial ingredient in CCSN
simulations and influences the inner core mass at bounce
($M_\mathrm{IC,b}$) by setting the electron fraction in the final
phase of collapse (e.g., \cite{hix:03,burrows:83}). As pointed out in
the literature (e.g.,
\cite{moenchmeyer:91,dimmelmeier:08,ott:12a,abdikamalov:14}), and in
this study (cf.\ Section \ref{sec:bounce}), $M_\mathrm{IC}$ at bounce
and $\rho_c$ after bounce has a decisive
influence on the rotating core collapse GW
signal.

In order to study how variations in electron capture rates affect our
GW predictions, we carry out three additional sets of simulations
using the SFHo EOS, $A3 = 634\,\mathrm{km}$, and all 20 corresponding
values of $\Omega_0$ listed in Table~\ref{tab:rotation}.

\begin{figure}
\center
\includegraphics[width=\linewidth]{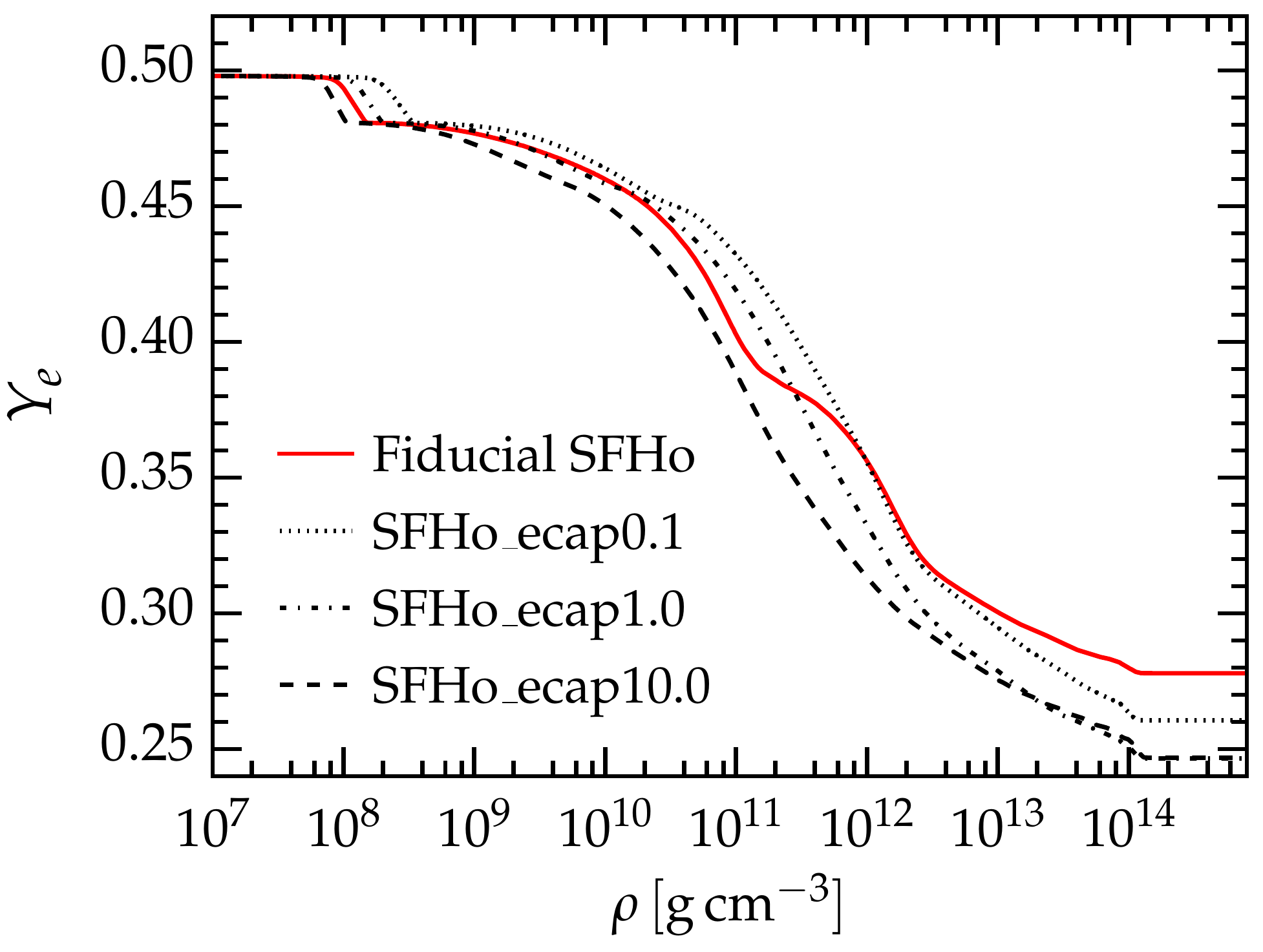}
\caption{\textbf{$\mathbf{Y_e(\rho)}$ Profiles From Variations in
    Electron Capture Treatment.}  We plot our fiducial $Y_e(\rho)$
  profile for the SFHo EOS along with $Y_e(\rho)$ profiles obtained
  with the approach of Sullivan~\emph{et al.}~\cite{sullivan:16} for
  the SFHo EOS using detailed tabulated nuclear electron capture rates
  (SFHo\_ecap1.0) and also rates multiplied by 0.1 (SFHo\_ecap0.1) and
  10 (SFHo\_ecap10.0) as a proxy for systematic uncertainties in
  the actual rates. Note that these $Y_e(\rho)$ profiles differ
  substantially from our fiducial profile, leading to different inner
  core masses and GW signals.}
\label{fig:yeofrho_ecapture}
\end{figure}

In one set of simulations, SFHo\_ecap1.0, we employ a $Y_e(\rho)$
parameterization obtained from \texttt{GR1D} simulations using the
approach of Sullivan~\emph{et al.}~\cite{sullivan:16} that
incorporates detailed tabulated electron capture rates for individual
nuclei. This is an improvement over the prescriptions of
\cite{bruenn:85,langanke:03} that operates on an average
$(\bar{A},\bar{Z})$ nucleus. Sullivan~\emph{et al.}~\cite{sullivan:16}
found that randomly varying rates for individual nuclei has little
effect, but systematically scaling rates by all nuclei with a global
constant can have a large effect on the resulting deleptonization
during collapse. In order to capture a factor $100$ in uncertainty,
the other two additional sets of simulations use $Y_e(\rho)$
parameterizations, obtained by scaling the detailed electron capture
rates by $0.1$ (SFHo\_ecap0.1) and $10$ (SFHo\_ecap10.0).

In Figure~\ref{fig:yeofrho_ecapture}, we plot the three new
$Y_e(\rho)$ profiles together with our fiducial SFHo $Y_e(\rho)$
profile. All of the new $Y_e(\rho)$ profiles predict substantially
lower $Y_e$ at high densities than our fiducial profiles for the SFHo
EOS. However, the SFHo\_ecap0.1 profile, and to a lesser extent the
SFHo\_ecap1.0 profile, have higher $Y_e$ at intermediate densities of
$10^{11}-10^{12}\,\mathrm{g\,cm}^{-3}$ than the fiducial profile.
This is relevant for our analysis here, since in the final phase of
collapse, a large part of the inner core passes this density range
less than a dynamical time from core bounce. Thus, the higher $Y_e$ in
this density range can have an influence on the inner core mass at
bounce.

In the nonrotating case, the fiducial SFHo inner core mass at bounce
is $M_\mathrm{IC,b} = 0.582\,M_\odot$ and we find $0.562\,M_\odot$,
$0.506\,M_\odot$, and $0.482\,M_\odot$, for SFHo\_0.1x\_ecap,
SFHo\_1x\_ecap, and SFHo\_10x\_ecap, respectively.  Note that
SFHo\_1x\_ecap and SFHo\_10x\_ecap give the same $Y_e(\rho)$ at $\rho
\gtrsim 10^{13}\,\,\mathrm{g\,cm}^{-3}$, but SFHo\_1x\_ecap predicts
higher $Y_e$ at $\rho \sim 10^{11}-10^{12}\,\mathrm{g\,cm}^{-3}$
(cf.~Figure~\ref{fig:yeofrho_ecapture}) and thus has a larger inner
core mass at bounce.

\begin{figure}
\center
\includegraphics[width=\linewidth]{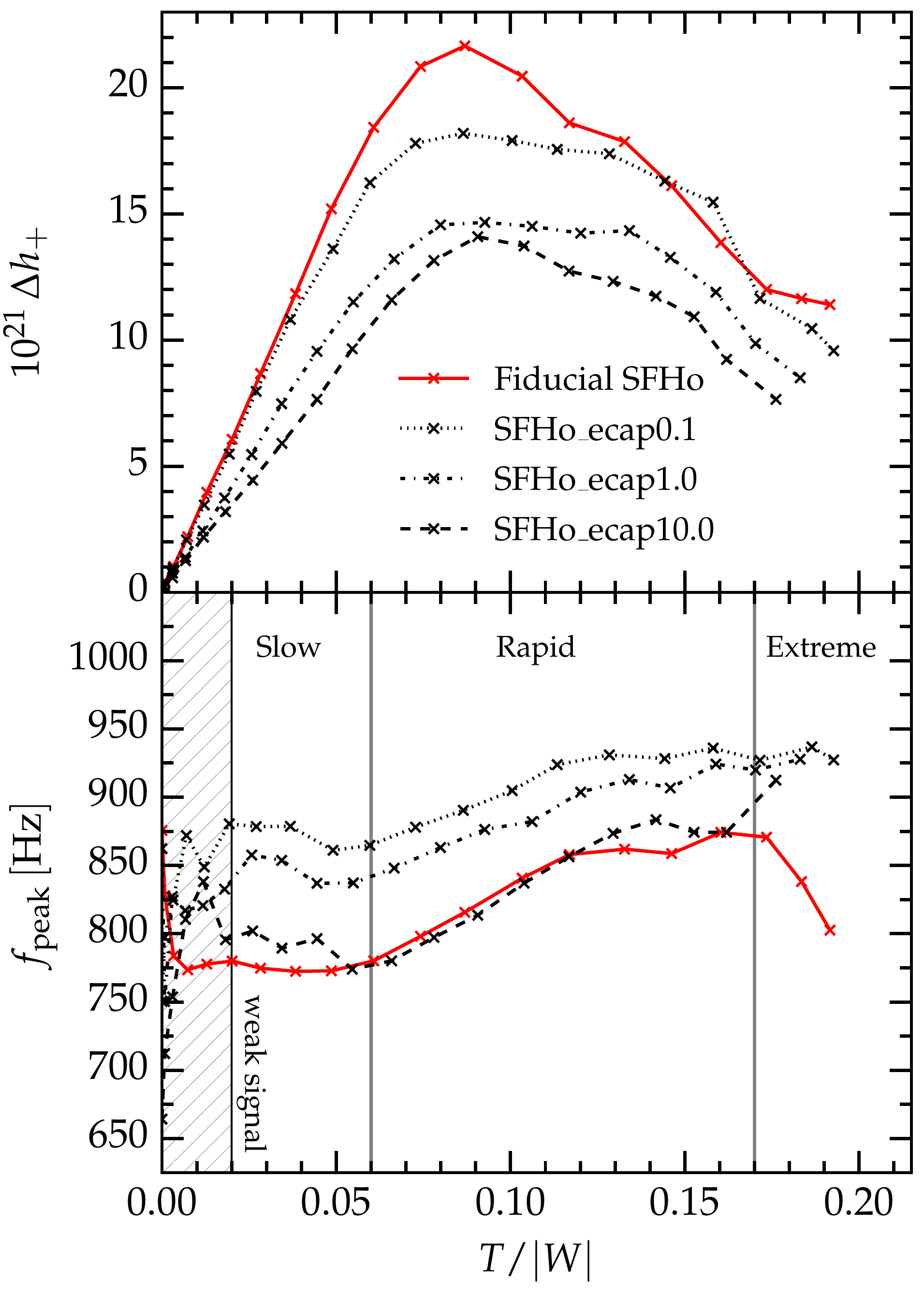}
\caption{\textbf{Changes in GW Observables with Variations in Electron
    Capture Rates.} We show results for $\Delta h_+$ (at
  $10\,\mathrm{kpc}$, top panel) and $f_\mathrm{peak}$ for SFHo EOS
  simulations with $A3 = 634\,\mathrm{km}$ with our fiducial
  $Y_e(\rho)$ profile and with the new $Y_e(\rho)$ profiles from
  simulations with detailed tabulated nuclear electron capture rates
  (cf.~Figure~\ref{fig:yeofrho_ecapture}). Differences in electron
  capture treatment and uncertainties in capture rates lead to
  differences in the key GW observables that are as large as those
  induced by switching EOS.}
\label{fig:ecap}
\end{figure}

In Figure~\ref{fig:ecap}, we present the key GW observables $\Delta
h_+$ and $f_\mathrm{peak}$ resulting from our rotating core collapse
simulations with the new $Y_e(\rho)$ profiles. We also plot our
fiducial SFHo results for comparison.  The top panel shows $\Delta
h_+$ and we note that the differences between the fiducial SFHo
simulations and the runs with the SFHo\_ecap1.0 base profile are
substantial and larger than differences between many of the EOS
discussed in Section~\ref{sec:bounce}
(cf.~Figure~\ref{fig:amplitude}). The differences with SFHo\_ecap10.0
$\Delta h_+$ are even larger. The SFHo\_ecap0.1 simulations produce
$\Delta h_+$ that are very close to the fiducial SFHo results in the
Slow Rotation regime. This is a consequence of the fact that the inner
core masses of the fiducial SFHo and SFHo\_ecap0.1 simulations are
very similar in this regime (cf.~Section
\ref{sec:bounce}). SFHo\_ecap1.0 and SFHo\_ecap10.0 produce smaller
$\Delta h_+$, because their inner cores are less massive at bounce.

The bottom panel of Figure~\ref{fig:ecap} shows $f_\mathrm{peak}$, the
peak frequencies of the GWs from postbounce PNS oscillations.  Again,
there are large differences in $f_\mathrm{peak}$ between the fiducial
SFHo simulations and those using $Y_e(\rho)$ obtained from detailed
nuclear electron capture rates. \emph{These differences are as large
  as the differences between many of the EOS} shown in
Figure~\ref{fig:fpeak}. In the Slow Rotation regime and into the Rapid
Rotation regime, the SFHo\_ecap1.0 base simulations have
$f_\mathrm{peak}$ that are systematically $50-75\,\mathrm{Hz}$ higher
than the fiducial SFHo simulations. For the SFHo\_ecap0.1 the
difference is $\sim$$100\,\mathrm{Hz}$ and in the SFHo\_ecap10.0 case,
the difference is surprisingly only $\lesssim$$25\,\mathrm{Hz}$.

For the SFHo\_ecap0.1 runs, we find a higher time-averaged postbounce
central density $\bar{\rho}_c$ than in the fiducial case. Hence, the
higher $f_\mathrm{peak}$ we observe fits our expectations from
Section~\ref{sec:postbounce}. Explaining $f_\mathrm{peak}$ differences
for SFHo\_ecap1.0 and SFHo\_ecap10.0 is more challenging: We find that
SFHo\_ecap1.0 runs have $\bar{\rho}_c$ that are similar or slightly
lower than those of the fiducial SFHo simulations, yet SFHo\_ecap1.0
$f_\mathrm{peak}$ are systematically higher. Similarly, SFHo\_ecap10.0
$\bar{\rho}_c$ are systematically lower than the fiducial
$\bar{\rho}_c$, yet the predicted $f_\mathrm{peak}$ are about the
same. These findings suggest that not only $\bar{\rho}_c$, but also
other factors, e.g., possibly the details for the $Y_e$ distribution in
the inner core or the immediate postbounce accretion rate play a role
in setting $f_\mathrm{peak}$.

As a quantitative example, we choose the previously considered $\Omega
= 5.0\,\mathrm{rad\,s}^{-1}$ case and compare our fiducial results
with those of the detailed electron capture runs. For the fiducial
SFHo run, we find $\Delta h_+ = 20.8\times10^{-21}$ (at
$10\,\mathrm{kpc}$) and $f_\mathrm{peak} = 798 \,\mathrm{Hz}$, with
$M_\mathrm{IC,b}=0.708\,M_\odot$ and
$\bar{\rho}_c=3.45\times10^{14}\,\mathrm{g\,cm}^{-3}$. The
corresponding detailed electron capture runs yield $\Delta h_+ =
\{17.8, 13.2, 11.6\}\times10^{-21}$, $f_\mathrm{peak} = \{878, 848,
780\} \,\mathrm{Hz}$, $M_\mathrm{IC,b}=\{0.707, 0.611,
0.561\}\,M_\odot$, and $\bar{\rho}_c=\{3.58, 3.43,
3.28\}\times10^{14}\,\mathrm{g\,cm}^{-3}$ for SFHo\_ecap\{0.1, 1.0,
10.0\}, respectively. The differences between these fiducial and
detailed electron capture runs are comparable to the differences
between the fiducial SFHo EOS and the fiducial LS220 EOS simulations
discussed in Sections~\ref{sec:bounce} and \ref{sec:postbounce}.

When considering the GW mismatch for the $\Omega_0 =
5.0\,\mathrm{rad\,s}^{-1}$ case between fiducial SFHo, and
SFHo\_ecap0.1, SFHo\_ecap1.0, and SFHo\_ecap10.0, we find we find
6.2\%, 6.2\%, and 4.9\%, respectively. These values are larger than
the mismatch values due to EOS differences shown in
Figure~\ref{fig:mismatch}.

\emph{Conclusions:} The results of this exercise clearly show that the
GW signal is very sensitive to the treatment of electron capture
during collapse. Differences in this treatment and in capture rates
can blur differences between EOS. Though a systematic uncertainty in
electron capture rates by a factor as large as 10 in either direction
is unlikely, the differences caused by variations in $Y_e(\rho)$
described in this section are major issues if one seeks to extract EOS
information from an observed rotating core collapse GW signal.

\section{Conclusions}
\label{sec:conclusions}

We carried out more than 1800 two-dimensional rapidly rotating
general-relativistic hydrodynamic core collapse simulations to
investigate the effects the nuclear EOS has on GW signals from rapidly
rotating stellar core collapse, using 18 microphysical EOS and 98
different rotation profiles.

We distinguish three rotation regimes based on the ratio of rotational
kinetic to gravitational energy $T/|W|$ of the inner core at bounce:
Slow Rotation ($T/|W|<0.06$), Rapid Rotation ($0.06<T/|W|<0.17$), and
Extreme Rotation ($T/|W|>0.17$). We find that in the Slow Rotation
regime, the behavior of the GW bounce signal is nearly independent of
the EOS and is straightforwardly explained by an order of magnitude
perturbative analysis. The amplitude of the bounce signal varies
linearly with the rotation rate, parameterized by $T/|W|$ of the inner
core at bounce, in agreement with previous work
(e.g.,~\cite{abdikamalov:14,dimmelmeier:08}). The differences between
bounce signals from different EOS are due largely to corresponding
differences in the mass of the inner core at bounce.  The GWs from
postbounce oscillations of the protoneutron star are almost
independent of the rotation rate in the Slow Rotation regime. The
effects of the EOS on the GW frequency can be parameterized almost
entirely in terms of the dynamical frequency $\sqrt{G\rho_c}$ of the
core after bounce.

In the Rapid Rotation regime, the maximum rotation rate at bounce
exceeds the dynamical frequency (above $T/|W|\approx0.06$), and
inertial (i.e.\ Coriolis and centrifugal forces) effects become
significant and fundamentally change the character of the
oscillations. The bounce amplitudes depart from their linear
relationship with $T/|W|$ and depend on both the EOS and the degree of
precollapse differential rotation. The variations due to the EOS are
significantly smaller than those due to differing rotation
profiles. Inertial effects confine oscillations to the poles and
increase the oscillation frequency approximately linearly with the
maximum rotation rate. Even in this regime, the dynamical time of the
postbounce core parameterizes the effects of the EOS on top of the
effects of rotation.

In the Extreme Rotation regime ($T/|W|\gtrsim0.17$) the stellar cores
undergo a centrifugally-supported bounce. Increasing the rotation rate
in this regime leads to \textit{smaller} rotational kinetic energy at
bounce as centrifugal support keeps the collapsed cores more
extended. The bounce GW signal correspondingly weakens, and the
postbounce GW frequency appears to decrease, though weaker
protoneutron star oscillations make positively identifying the peak
frequency less reliable.

Our results show that EOS differences in the collapse phase are as
important as the high-density parameterization in determining
characteristics of the GWs. Different treatments of low-density matter
produce differences in the bounce signal, postbounce oscillation
frequency, and overall signal (as measured by the GW mismatch) that
are comparable to those produced by differences in high-density
parameterizations or differences in the treatment of the transition
from nonuniform to uniform nuclear matter. Densities do not exceed
around twice nuclear density in the bounce and brief postbounce phases
of core collapse that we study. Hence, the GW signal from these phases
does not probe exotic physics or conditions in very massive neutron
stars.

We demonstrate that using detailed electron capture rates for
individual nuclei as opposed to the fiducial single nucleus approach
to electron capture results in differences in the bounce and
postbounce GWs comparable to those caused by using a different
EOS. The GW characteristics are also sensitive to systematic
uncertainties in the electron capture rates, producing similarly large
variations when scaling the capture rates by a factor of 10 in either
direction. We also demonstrate that a density-parameterization of the
electron fraction $Y_e(\rho)$ during the collapse phase lacks the
precision required for detailed interpretation of observed GW
signals. Variations in the way the parameterization is implemented
produce changes in the GWs comparable to those produced by different
EOS. This leads us to the conclusion that for quantitatively reliable
GW predictions full multi-dimensional neutrino radiation-hydrodynamic
simulations that include realistic weak interactions will be needed.

\begin{figure}
\center \includegraphics[width=\linewidth]{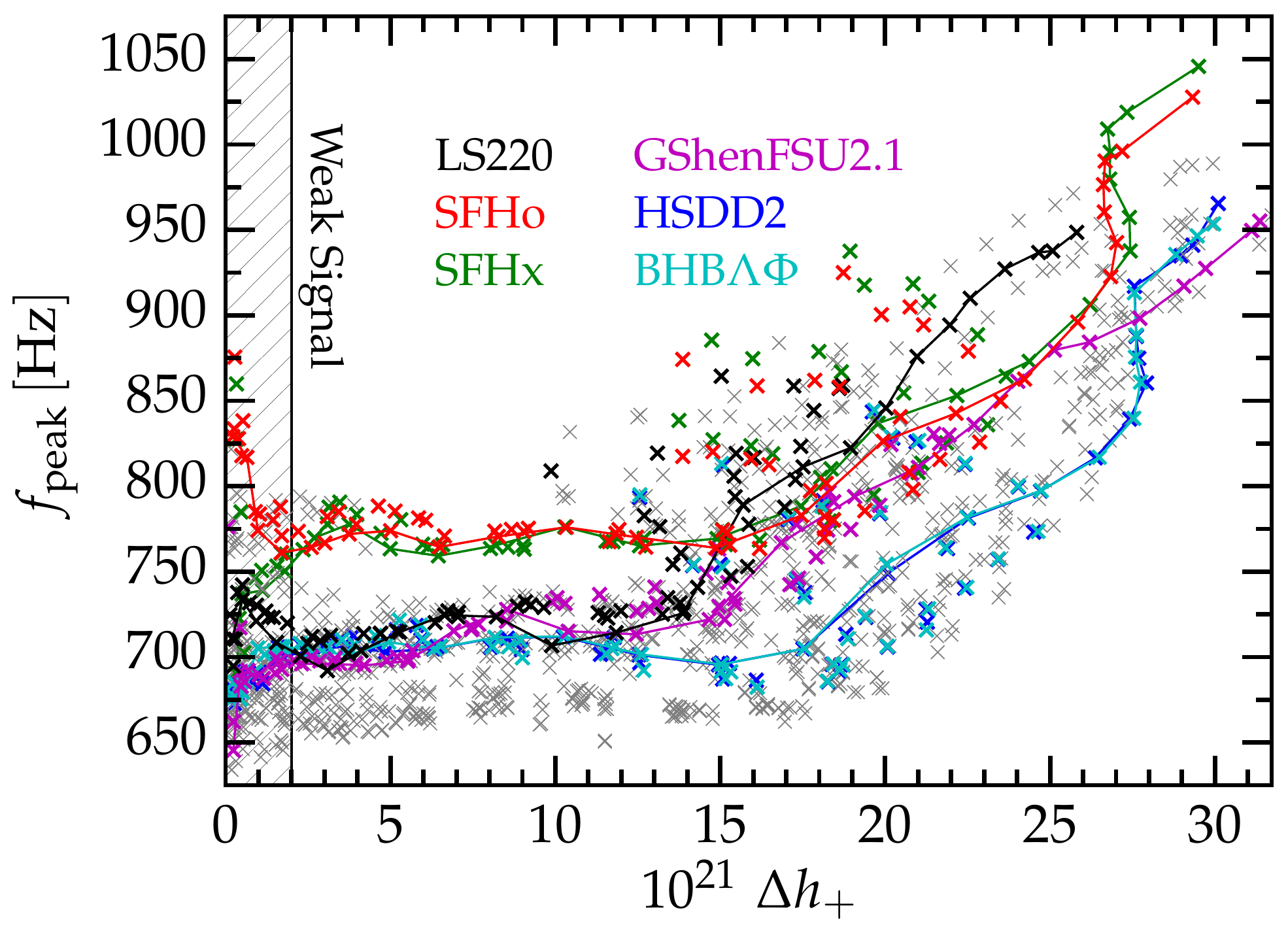}
\caption{\textbf{Discerning the EOS.} We plot the GW peak frequency
  against the bounce signal amplitude for each of our 1704 collapsing
  cores. Data from the $A1=300\,\mathrm{km}$ simulations are connected
  with lines to guide the eye. We predict a region of parameter space
  where we can reasonably expect rapidly rotating core collapse GW
  bounce and early post-bounce signals to lie given uncertainties in
  the nuclear EOS. For signals with $\Delta h_+ \lesssim
  15\times10^{-21}$ (at $10\,\mathrm{kpc}$), we may be able to
  distinguish the EOS from GW signals if the distance and orientation
  can be accurately determined. Peak frequencies at the slowest
  rotation rates (corresponding to $\Delta h_+ \lesssim
  2\times10^{-21}$ in the figure) are unreliable due to extremely weak
  GW signals.}
\label{fig:fpeak_vs_amp}
\end{figure}
In Figure~\ref{fig:fpeak_vs_amp}, we plot the GW bounce signal
amplitude against the frequency of GWs from postbounce oscillations to
show that different EOS occupy different, though partially overlapping
regions in this observable space. This effectively maps uncertainties
in the nuclear EOS to uncertainties in predicted GW signals from
rapidly-rotating core collapse. Signals observed from the bounce and
early postbounce phases of rotating core collapse outside of this
region would be of great interest, since they would indicate
unanticipated EOS physics and/or collapse dynamics. It may be possible
to use the bounce amplitude to determine how quickly the star is
rotating at bounce. The peak frequency could then constrain the EOS if
there is enough core rotation to produce a reliable postbounce
oscillation peak and little enough for the collapse to be in the Slow
Rotation regime.

However, we must note that there are large uncertainties in the
measured distances and orientations of nearby core-collapse
supernovae, and also in the errors introduced by approximations made
in the simulations. GW strain decreases inversely with distance, so
the bounce amplitude is known only as well as the distance. Since the
observed GW strain varies roughly with $h\sim\sin^2(\theta)$, where
$\theta$ is the angle between the rotation axis and the line of sight,
an accurate determination of the source orientation is required to be
able to map the GW strain to a rotation rate. Inferring the peak
frequency does not require distance or orientation measurements, but
is subject to other observational uncertainties, e.g., the GW detector
phase accuracy.  Parameter estimation and model selection studies with
more sophisticated data analysis tools, like those used by
\cite{abdikamalov:14,roever:09,logue:12}, are required to evaluate the
feasibility of extracting EOS properties given real detector
characteristics and noise.

It should also be noted that GWs from rotating core collapse will only
be detectable from sources out to the Magellanic Clouds. Furthermore,
even those cores that are in our Slow Rotation regime are still very
rapidly spinning from a stellar evolution point of view and produce
protoneutron stars with spin periods of $\lesssim
5\,\mathrm{ms}$. Massive stars with rapidly spinning cores are
expected to be exceedingly rare. These caveats and the above
limitations, combined with the relatively small differences in the GW
characteristics and protoneutron star oscillations induced by EOS
variations, mean that we are unlikely to be able to use a GW signal
from rotating core collapse to discern the EOS with current GW
detectors and simulation methods.

The present study has elucidated the various ways in which the nuclear
EOS can impact the rotating core collapse GW signal. While we are
confident that our qualitative findings are robust, our GW signal
predictions are not quantitatively reliable. The most important
limitation to be removed by future work is the lack of 2D neutrino
radiation-hydrodynamics in the collapse phase. Our results on
differences caused by differing treatments of various regimes of the
same underlying EOS parameterization also suggest that more work in
nuclear theory may be needed. In particular, there is an important
need for consistent EOS frameworks with which only differences in EOS
physics, but not differences in methods, cause differences in the GW
signal. In addition, though previous studies have shown that different
progenitors result in only slightly different inner core masses
\cite{janka:12b} and GW signal characteristics (assuming the same
resulting inner core mass and angular momentum) \cite{ott:12a}, a
quantitative understanding of progenitor-induced uncertainties will
require a much more exhaustive study of progenitor dependence of GW
signals from rotating CCSNe.

While axisymmetry is a good approximation for collapse, bounce, and
the early postbounce phase ($\lesssim10\,\mathrm{ms}$ after bounce),
rotating core collapse is host to rich three dimensional (3D)
postbounce dynamics that can drive GW emission, including rotational
instabilities and the nonaxisymmetric standing accretion shock
instability. 3D simulations of rotating core collapse and postbounce
GW emission have been carried out
(e.g.,~\cite{ott:07prl,scheidegger:10,kuroda:14}), but the EOS
dependence of the GWs generated by 3D dynamics has yet to be
explored. GWs from prompt and neutrino-driven convection and from the
standing accretion shock instability in both rotating and nonrotating
core collapse~\cite{mueller:13gw,yakunin:15,andresen:16} have some EOS
dependence as well~\cite{marek:09b,kuroda:16b}, but the EOS parameter
space has thus far been only sparsely sampled. Future studies of GWs
emitted by these dynamics may yet provide alternate means of
discerning the nuclear EOS.

\section*{acknowledgements}

We would like to thank Jim Fuller, Hannah Klion, Peter Goldreich,
Hiroki Nagakura, Pablo Cerd\'a-Dur\'an, Hajime Sotani, Luke Roberts,
Andr\'e da Silva Schneider, Chuck Horowitz, Jim Lattimer, Sarah Gossan, and Bill
Engels for many insightful discussions. The authors wish to thank
Remco Zegers for supporting the development of the (EC) weak rate
library, which was instrumental in the completion of this work. SR was
supported by the DOE CSGF, which is provided under grant number
DE-FG02-97ER25308, and the NSF Blue Waters Graduate Fellowship. This
research is part of the Blue Waters sustained-petascale computing
project, which is supported by the National Science Foundation (awards
OCI-0725070 and ACI-1238993) and the State of Illinois. Blue Waters is
a joint effort of the University of Illinois at Urbana-Champaign and
its National Center for Supercomputing Applications. These simulations
were performed on the Stampede cluster of the NSF XSEDE network under
allocation TG-PHY100033 and benefited from access to Blue Waters
under allocation NSF PRAC ACI-1440083. This research is supported by the
NSF under award numbers CAREER PHY-1151197, AST-1212170, and
PHY-1404569, by the International Research Unit of Advanced Future
Studies, Kyoto university, and by the Sherman Fairchild Foundation. EA
acknowledges support from NU ORAU and Social Policy grants. CS
acknowledges support from the National Science Foundation under grant
No.\ PHY-1430152 (JINA Center for the Evolution of the Elements) and
No.\ PHY-1102511 and from the Department of Energy National Nuclear
Security Administration under award number DE-NA0000979. EO
acknowledges support for this work by NASA through Hubble Fellowship
grant \#51344.001-A awarded by the Space Telescope Science Institute,
which is operated by the Association of Universities for Research in
Astronomy, Inc., for NASA, under contract NAS 5-26555.



\appendix

\section{$Y_e(\rho)$ Fits}
\label{app:yeofrho}
In the simulations presented in the main body of the paper we use and
interpolate $Y_e(\rho)$ profiles directly from a 1D simulation
snapshot. A commonly used alternative is to fit a function to this
profile and evaluate the function rather than interpolating data in a
profile. For convenience and for use in the numerics study in
Appendix~\ref{app:numerics}, we also generate functional fits for
these profiles. Following~\cite{liebendoerfer:05fakenu} with a tweak
at high densities, we fit our 1D $Y_e(\rho)$ profiles using the
fitting function

\begin{equation}
\begin{split}
Y_e &= \begin{cases}
  \begin{aligned}
    0.5&(Y_{e,2}+Y_{e,1}) \\
    +x/2&(Y_{e,2}-Y_{e,1}) \\
    +Y_{e,c}&[1-|x|\\
    +4|x|&(|x|-0.5)(|x|-1)]
  \end{aligned}  & \rho\leq\rho_2 \\[3ex]
  Y_{e,2} + m(\log_{10}\rho - \log_{10}\rho_2) & \rho > \rho_2\,\,, \\
  \end{cases}\\[3ex]
x &= \begin{aligned} \max\left(-1,\min\left(1,\frac{2
    \log_{10}\rho-\log_{10}\rho_2-\log_{10}\rho_1}{\log_{10}\rho_2-\log_{10}\rho_1}\right)\right)\,\,,\\
\end{aligned}\\[3ex]
m &= \frac{Y_{e,H}-Y_{e,2}}{\log_{10}\rho_H - \log_{10}\rho_2}\,\,.
\end{split}
\end{equation}
The parameters $\rho_H=10^{15}\,\mathrm{g\ cm}^{-3}$ and $Y_{e,1}=0.5$
are fixed. The parameters $\{\rho_1,\rho_2,Y_{e,2},Y_{e,c},Y_{e,H}\}$
are fit using the Mathematica {\tt MyFit} function, subject to the
constraints
\begin{equation}
\begin{aligned}
  10^7 \leq &\frac{\rho_1}{\mathrm{g\,cm}^{-3}} \leq 10^{8.5}\,\,, \\
  10^{12} \leq &\frac{\rho_2}{\mathrm{g\,cm}^{-3}} \leq 10^{14}\,\,, \\
  0.2 \leq &Y_{e,2} \leq 0.4\,\,, \\
  0.02 \leq &Y_{e,c} \leq 0.055\,\,, \\
  &\frac{dY_e}{d\rho}<0\,\,.\\
\end{aligned}
\label{eq:deleptonization_fit}
\end{equation}

\begin{figure}
\includegraphics[width=\linewidth]{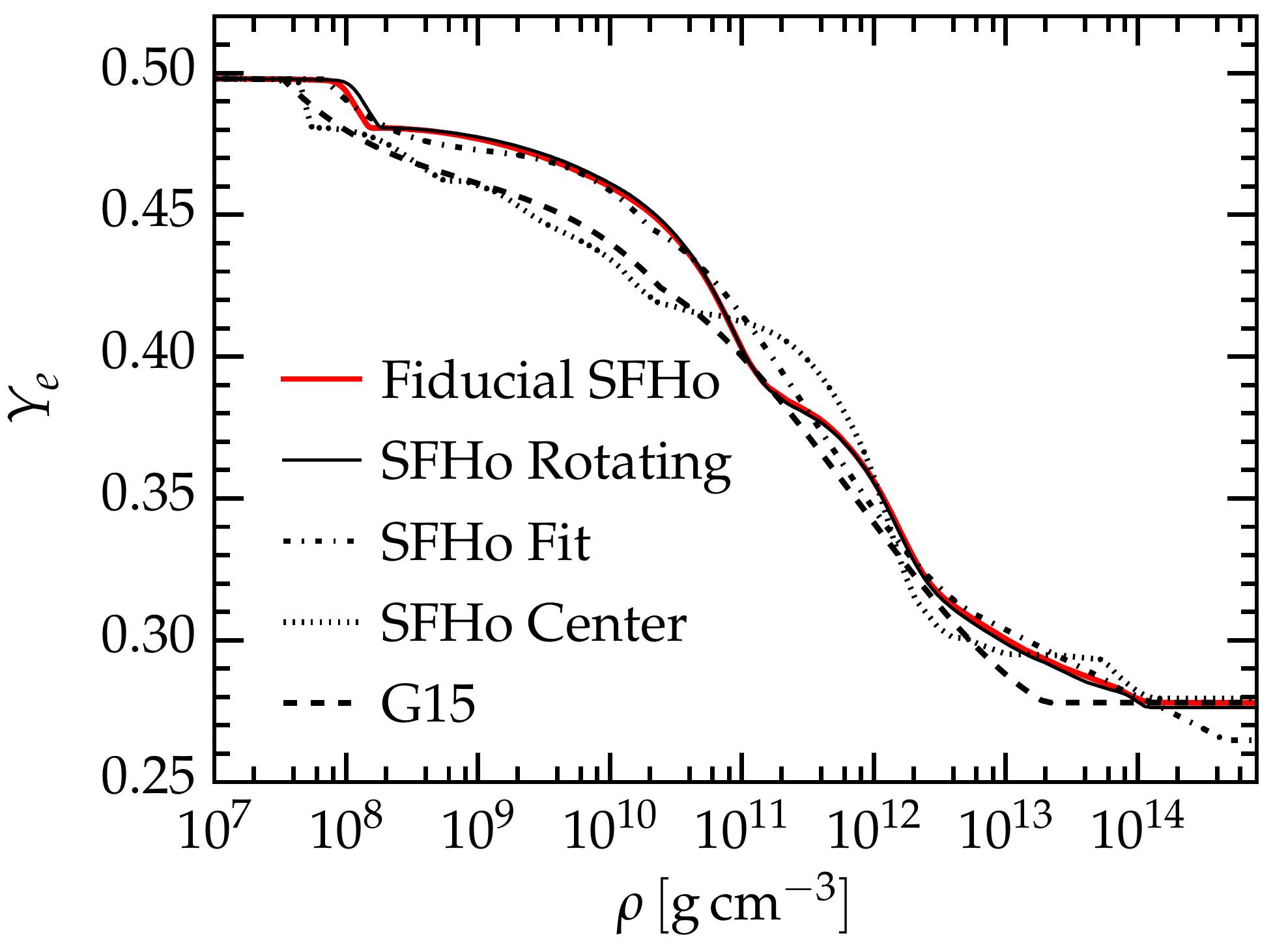}
\caption{\textbf{Test $\mathbf{Y_e(\rho)}$ Profiles.} We plot the
  different possibilities for deleptonization functions one might
  input into the 2D GRHD simulations. The solid red line is the
  $Y_e(\rho)$ directly taken from the radial profile at the moment
  when the central $Y_e$ is lowest. The solid black line is also
  directly taken from the radial data of a {\tt GR1D} simulation using
  ``shellular'' rotation with $A=634\,\mathrm{km}$,
  $\Omega_0=5.0\,\mathrm{rad\,s}^{-1}$. The dot-dashed line is a fit
  to the nonrotating $Y_e(\rho)$ using the same parameters
  as~\cite{liebendoerfer:05fakenu} in addition to a high-density
  slope. The dashed line is the G15 fit
  from~\cite{liebendoerfer:05fakenu}. The dotted line is a record of
  the central $Y_{e,c}(\rho_c)$ throughout nonrotating collapse,
  appended to the $Y_e(\rho)$ profile at $t=0$.}
\label{fig:yeofrho_test}
\end{figure}
\begin{table}
\caption{\textbf{Fitted $\mathbf{Y_e(\rho)}$ Profiles.}  We provide
  results for the fitting parameters in
  Equation~\ref{eq:deleptonization_fit} for each EOS. We provide these
  fits for convenience, but do not use them in our 2D simulations
  presented in the main body of the paper and instead interpolate from
  the numerical \texttt{GR1D} results.}
\begin{tabular}{llllll}
\hline\hline
EOS               & $\log_{10}\rho_1$ & $\log_{10}\rho_2$ & $Y_{e,2}$ & $Y_{e,c}$   & $Y_{e,H}$\\ \hline
SFHo              & 7.795           & 12.816           & 0.308  & 0.0412   & 0.257 \\
SFHx              & 7.767           & 12.633           & 0.323  & 0.0380   & 0.275 \\
SFHo\_ecap0.1     & 8.210           & 13.053           & 0.291  & 0.0493   & 0.237 \\
SFHo\_ecap1.0     & 8.022           & 12.882           & 0.281  & 0.0528   & 0.224 \\
SFHo\_ecap10.0    & 7.743           & 12.405           & 0.294  & 0.0473   & 0.226 \\
LS180             & 7.738           & 13.034           & 0.290  & 0.0307   & 0.243 \\
LS220             & 7.737           & 12.996           & 0.292  & 0.0298   & 0.245 \\
LS375             & 7.755           & 12.901           & 0.295  & 0.0279   & 0.251 \\
HShen             & 7.754           & 13.124           & 0.303  & 0.0398   & 0.267 \\
HShenH            & 7.751           & 13.124           & 0.303  & 0.0397   & 0.267 \\
GShenFSU1.7       & 7.939           & 12.935           & 0.305  & 0.0403   & 0.257 \\
GShenFSU2.1       & 7.939           & 12.935           & 0.305  & 0.0403   & 0.257 \\
GShenNL3          & 7.917           & 13.104           & 0.299  & 0.0412   & 0.247 \\
HSDD2             & 7.797           & 12.813           & 0.308  & 0.0411   & 0.259 \\
HSNL3             & 7.798           & 12.808           & 0.308  & 0.0409   & 0.253 \\
HSIUF             & 7.792           & 12.777           & 0.311  & 0.0403   & 0.257 \\
HSTMA             & 7.793           & 12.787           & 0.310  & 0.0408   & 0.252 \\
HSTM1             & 7.799           & 12.812           & 0.308  & 0.0411   & 0.253 \\
HSFSG             & 7.792           & 12.784           & 0.311  & 0.0404   & 0.256 \\
BHB$\Lambda$      & 7.794           & 12.815           & 0.308  & 0.0412   & 0.259 \\
BHB$\Lambda\Phi$  & 7.794           & 12.814           & 0.308  & 0.0412   & 0.259 \\
Liebend\"orfer G15& 7.477           & 13.301           & 0.278  & 0.0350   & 0.278 \\
\hline\hline
\end{tabular}
\label{tab:yeofrho}
\end{table}
The resulting fit parameters are listed in Table~\ref{tab:yeofrho} for
each EOS. In Figure~\ref{fig:yeofrho_test}, we plot the $Y_e(\rho)$
profiles for the SFHo EOS that we use in the SFHo 2D simulations,
along with our fit. We also plot the G15 fit
from~\cite{liebendoerfer:05fakenu}, and the $Y_e(\rho)$ profile
obtained by tracking the density and electron fraction of the center
during collapse in the {\tt GR1D} simulation and appending this to the
$Y_e(\rho)$ at $t=0$ profile for low densities. We describe the
results of test simulations using each of these profiles in
Appendix~\ref{app:numerics}.

\section{Numerics Study}
\label{app:numerics}
We attempt to quantify the errors resulting from the various numerical
and physical approximations in our approach by performing a
sensitivity study with various parameters in all simulation phases.
We employ the SFHo EOS for these tests and adopt
$A3=634\,\mathrm{km}$, $\Omega=5.0\,\mathrm{rad\,s}^{-1}$ as the
fiducial rotation setup in rotating test simulations. Key quantitative
results from the fiducial 1D and 2D simulations used for comparison
are listed in bold at the top of Tables \ref{tab:GR1D_tests} and
\ref{tab:coconut_tests}.

\subsection{1D Tests}
\label{app:GR1Dtests}
\begin{table}
\caption{\textbf{\texttt{GR1D} Test Results.} Key diagnostic quantities from 1D
  simulation tests are listed, along with corresponding quantities
  from select 2D simulations for comparison. $t_\mathrm{b}$ is the
  time from simulation start to core bounce. $M_\mathrm{IC,b}$,
  $\rho_{c,b}$, $T_{c,b}$, and $Y_{e,c,b}$ are the mass of the inner
  core, the central density, the central temperature, and the central
  electron fraction, respectively, at core bounce. Note that we
  average $\rho_{c,b}$ in the interval $[t_\mathrm{b}, t_\mathrm{b} +
    0.2\,\mathrm{ms}]$ to filter out spurious oscillations that are
  purely numerical in this single-point quantity at the origin.
  Bolded rows are fiducial simulations, and the two {\tt CoCoNuT} rows
  are the same quantities from two of the 2D simulations. In the {\tt
    NuLib} block, we vary only the input physics and resolution for
  the neutrino interaction table used in the 1D simulations. In the
  {\tt GR1D} block, we vary only {\tt GR1D} simulation resolution and
  rotation. In the $Y_e(\rho)$ block, we experiment with using
  different prescriptions for the deleptonization profile, including
  the G15 fit from~\cite{liebendoerfer:05fakenu} (see
  Figure~\ref{fig:yeofrho_test}).}
\begin{tabular}{lccccc}
\hline \hline
Test & $t_\mathrm{b}$ & $M_\mathrm{IC,b}$ & $\rho_\mathrm{c,b}$ & $T_\mathrm{c,b}$ & $Y_\mathrm{e,c,b}$ \\
     & $(\mathrm{ms})$     & $(M_\odot)$         & $(\mathrm{g\,cm}^{-3})$ & $(\mathrm{MeV})$ & \\
\hline
\textbf{{\tt GR1D} Nonrot.}    & \textbf{180} & \textbf{0.583} & \textbf{4.31} & \textbf{14.9} & \textbf{0.288} \\
\textbf{{\tt CoCoNuT} Nonrot.} &\textbf{174}  & \textbf{0.582} & \textbf{4.38} & \textbf{14.8} & \textbf{0.278} \\
\textbf{{\tt CoCoNuT} Fiducial}    &\textbf{200}  & \textbf{0.708} & \textbf{4.16} & \textbf{12.8} & \textbf{0.278} \\
\hline
{\tt GR1D} $n_r=1500$              & 180          & 0.583          & 4.26          & 14.9          & 0.288          \\
{\tt GR1D} Rotating                & 202          & 0.674          & 3.95          & 13.9          & 0.286          \\
\hline
{\tt GR1D} $Y_e(\rho)$ Direct                 & 210          & 0.583          & 4.37          & 14.1          & 0.278          \\
{\tt GR1D} $Y_e(\rho)$ Fit                    & 211          & 0.592          & 4.43          & 14.2          & 0.265          \\
{\tt GR1D} $Y_e(\rho)$ Center                 & 174          & 0.610          & 4.26          & 17.3          & 0.279          \\
{\tt GR1D} $Y_e(\rho)$ G15                    & 189          & 0.547          & 4.22          & 12.5          & 0.279          \\
\hline
{\tt NuLib} $n_E=36$               & 180          & 0.582          & 4.25          & 15.0          & 0.288          \\
{\tt NuLib} $n_\rho=123$           & 180          & 0.583          & 4.27          & 14.7          & 0.288          \\
{\tt NuLib} $n_T=150$              & 180          & 0.582          & 4.25          & 14.9          & 0.288          \\
{\tt NuLib} $n_{Y_e}=150$          & 180          & 0.583          & 4.28          & 14.8          & 0.288          \\
\hline \hline
\end{tabular}
\label{tab:GR1D_tests}
\end{table}

As described in Section~\ref{sec:methods}, we use {\tt GR1D}
simulations to generate $Y_e(\rho)$ profiles for the 2D simulations,
and so these profiles encode the effects of the EOS during the
collapse phase of the 2D simulations. Here we check the various levels
of physical and numerical approximations made in calculating the
profiles used in the main text. We also check whether using one of
these profiles produces results consistent with full transport. In
Table~\ref{tab:GR1D_tests}, we list the time to bounce $t_b$, the mass
of the inner core at bounce $M_\mathrm{IC,b}$, and the central
density, temperature, and electron fraction at bounce.

Table~\ref{tab:GR1D_tests} shows that the nonrotating 1D \texttt{GR1D}
radiation-hydrodynamic simulation and the 2D \texttt{CoCoNuT}
hydrodynamic simulation agree well in key collapse results and in
particular in $M_\mathrm{IC,b}$. This confirms that the $Y_e(\rho)$
parameterization captures deleptonization and its effect on the
collapsing core well, as previously shown by
\cite{liebendoerfer:05fakenu}.  The difference in the central $Y_e$ at
bounce ($0.288$ in the \texttt{GR1D} run vs.\ $0.278$ in the
\texttt{CoCoNuT} simulation) is due to our use of $Y_e(\rho)$ from the
\texttt{GR1D} simulation at the time of minimum central $Y_e$, which
occurs just before bounce. Due to shifts in the local beta
equilibrium, the central $Y_e$ in the radiations-hydrodynamic
simulation increases again after its global minimum.

An important open question is to what extent rotation affects the
validity of the $Y_e(\rho)$ for deleptonization during collapse. While
we cannot currently carry out detailed multi-D radiation-hydrodynamic
simulations to answer this conclusively, we can include rotation
approximately in \texttt{GR1D} 1D radiation-hydrodynamic simulations,
using the ``shellular rotation'' approximation
(cf.~\cite{oconnor:10,thompson:05}). We employ the fiducial rotation
profile specified by $A3=634\,\mathrm{km}$ and
$\Omega_0=5\,\mathrm{rad\,s}^{-1}$ as in the 2D case, though the
radial coordinate relevant for the rotational setup is the spherical
radius in \texttt{GR1D}.

The ``\texttt{GR1D} Rotating'' row in Table~\ref{tab:GR1D_tests} shows
that the effects of rotation on the collapse dynamics are
qualitatively similar between 1D ``shellular rotation'' and 2D
rotation: $t_\mathrm{b}$ and $M_\mathrm{IC,b}$ increase and
$\rho_\mathrm{c,b}$ decreases. However, in 1D, the quantitative changes
are smaller than in 2D, which is consistent with the findings of
\cite{ott:06spin}, who more extensively compared 1D ``shellular
rotation'' with 2D rotation.

Figure~\ref{fig:yeofrho_test} compares the $Y_e(\rho)$ profile
obtained from the rotating \texttt{GR1D} simulation with the fiducial
$Y_e(\rho)$ profile and other possible profiles. As expected
(cf.~Section~\ref{sec:methods}), rotation in the ``shellular''
approximation leads to only minor differences in $Y_e(\rho)$ between
the nonrotating case and the fiducial rotational
setup.

In the first row of the {\tt GR1D} block of
Table~\ref{tab:GR1D_tests}, we list results from a \texttt{GR1D}
simulation with 1.5 times the standard resolution. The differences
with the standard resolution run are very small, giving us confidence
that ours \texttt{GR1D} simulation results are numerically
converged.

The $Y_e(\rho)$ profiles extracted from the 1D radiation-hydrodynamic
simulations should give a good approximation to collapse-phase
deleptonization and its impact on collapse and bounce
dynamics~\cite{liebendoerfer:05fakenu}. We test this assertion by
re-running the \texttt{GR1D} 1D simulations with various choices for
the $Y_e(\rho)$ profiles (see Figure~\ref{fig:yeofrho_test}) rather
than using neutrino transport.  The results are listed in the third
block of Table~\ref{tab:GR1D_tests}.

We find that our fiducial $Y_e(\rho)$ profile
(cf.~Section~\ref{sec:yeofrho}, row ``\texttt{GR1D} $Y_e(\rho)$
Direct'' in Table~\ref{tab:GR1D_tests}) leads to inner core masses,
bounce densities, and thermodynamics that approximate the
radiation hydrodynamics results very well. Using a fit to the fiducial
$Y_e(\rho)$ (``\texttt{GR1D} $Y_e(\rho)$ Fit'') or generating the
$Y_e(\rho)$ profile from the central value of $Y_e$ (``\texttt{GR1D}
$Y_e(\rho)$ Center'') leads to larger differences in all quantities
(e.g., $\gtrsim$$5\%$ in $M_\mathrm{IC,b}$). These quantitative
differences are of the same order as those due to differences in EOS
and electron capture treatment (cf.~Section~\ref{sec:ecapture} and the
``\texttt{GR1D} $Y_e(\rho)$ G15'' row). For instance, different EOS
lead to inner core masses at bounce in the range of
$0.549-0.618\,M_\odot$. Hence, the $Y_e(\rho)$ parameterization can
lead to a systematic error that muddles the interpretation of results
from simulations using different EOS. For quantitatively reliable
predictions, full 2D radiation-hydrodynamic
simulations will be necessary.

The entries in the {\tt NuLib} block of Table~\ref{tab:GR1D_tests}
give results for test simulations with different resolutions of our
neutrino interaction table. These are to be compared with the fiducial
neutrino interaction table that has resolution $n_E=24$ (number of
energy groups), $n_\rho=82$, $n_T=100$, $n_{Y_e}=100$. All tables span
the range
\begin{equation}
\begin{aligned}
  0     &< E/(\mathrm{MeV})       &< 287\,\,\,,\\
  10^6  &< \rho/(\mathrm{g\,cm}^{-3}) &< 10^{15}\,\,,\\
  0.05  &< T/(\mathrm{MeV})       &< 150\,\,\,,\\
  0.035 &< Y_e                    &< 0.55\,\,.\\
\end{aligned}
\end{equation}
The energy, density, and temperature points in the table are
logarithmically spaced and the electron fraction points are evenly
spaced. Increasing the table resolution has negligible impact on the
\texttt{GR1D} results.

\subsection{2D Tests}
\label{app:coconut_tests}
\begin{table}
\caption{\textbf{Waveform Test Results.} In the {\tt NuLib}, {\tt
    GR1D}, and $Y_e(\rho)$ blocks, we simply run the fiducial {\tt
    CoCoNuT} simulation using the $Y_e(\rho)$ profiles extracted from
  the {\tt GR1D} tests listed in Table~\ref{tab:GR1D_tests}. In the
  {\tt CoCoNuT} block, we only modify 2D simulation
  parameters. $M_\mathrm{IC,b}$ is the mass of the inner core at
  bounce, $\mathcal{M}_\mathrm{fid}$ is the GW mismatch with the
  fiducial simulation, $f_\mathrm{peak}$ is the peak frequency of the
  GWs from postbounce oscillations, and $\Delta h_+$ is the difference
  between the largest positive and negative GW strain values of the
  bounce signal.}
\begin{tabular}{lccccc}
\hline \hline
Test & $M_\mathrm{IC,b}$ & $\mathcal{M}_\mathrm{fid}$ & $f_\mathrm{peak}$ & $\Delta h_+$  \\
     & $(M_\odot)$         &                            & $(\mathrm{Hz})$   & $(10^{-21})$  \\
\hline
\textbf{{\tt CoCoNuT} Fiducial} & \textbf{0.718} & \textbf{0} & \textbf{793} & \textbf{20.9} \\
\hline              
{\tt NuLib} $n_E=36$        & 0.717          & 2.10(-5)       & 794          & 20.9 \\
{\tt NuLib} $n_\rho=123$    & 0.718          & 2.91(-5)       & 794          & 21.0 \\
{\tt NuLib} $n_T=150$       & 0.717          & 4.63(-5)       & 794          & 21.0 \\
{\tt NuLib} $n_{Y_e}=150$   & 0.718          & 1.48(-5)       & 794          & 20.9 \\
\hline              
{\tt GR1D} $n_r=1500$       & 0.716          & 1.23(-5)       & 794          & 21.0 \\
{\tt GR1D} Rotating         & 0.711          & 9.21(-5)       & 794          & 20.6 \\
\hline              
$Y_e(\rho)$ Fit             & 0.729          & 9.53(-3)       & 812          & 20.6 \\
$Y_e(\rho)$ Center          & 0.747          & 4.87(-2)       & 810          & 23.3 \\
$Y_e(\rho)$ G15             & 0.655          & 7.86(-2)       & 752          & 14.1 \\
\hline              
{\tt CoCoNuT} $n_r=500$     & 0.718          & 1.79(-3)       & 795          & 21.5 \\
{\tt CoCoNuT} $n_\theta=80$ & 0.718          & 1.03(-4)       & 794          & 21.1 \\
{\tt CoCoNuT} Eq. Bounce    & 0.714          & 4.40(-3)       & 789          & 21.6 \\
{\tt CoCoNuT} rk3           & 0.716          & 3.34(-3)       & 797          & 20.9 \\
\hline \hline
\end{tabular}
\label{tab:coconut_tests}
\end{table}

In Table~\ref{tab:coconut_tests}, we list the inner core mass at
bounce, the GW mismatch (see Section~\ref{sec:snr}) with the fiducial
2D simulation, the peak frequency, and the bounce signal amplitude for
several 2D tests. The results of the fiducial 2D simulation are bolded
at the top for comparison.

The {\tt NuLib} and {\tt GR1D} blocks of Table~\ref{tab:coconut_tests}
use the $Y_e(\rho)$ profile generated by the corresponding 1D test
simulation in a 2D simulation otherwise identical to the fiducial
one. These all produce negligible differences in all
quantities. Rotation is multidimensional, so the ``shellular
rotation'' approximation in \texttt{GR1D} does not take into account
multidimensional effects. The lack of impact of approximate 1.5D
rotation on the collapse deleptonization suggests that using a
$Y_e(\rho)$ profile from a nonrotating 1D simulation in
moderately-rapidly rotating 2D collapse simulation is acceptable. The
choice of $Y_e(\rho)$ parameterization, however, leads to significant
differences, as already pointed out in the previous
Appendix~\ref{app:GR1Dtests}. The GW mismatch for the ``Fit'' and
``Center'' choices with the fiducial approach is $\sim$1\% and
$\sim$5\%, respectively. The peak frequencies differ by
$\sim$2\%. Using the G15 $Y_e(\rho)$ fit of
\cite{liebendoerfer:05fakenu} leads to even larger mismatch of
$\sim$$8\%$ and a peak frequency differing by as much as
$\sim$$40\,\mathrm{Hz}$. These differences are as large or larger than
differences between many EOS discussed in \S\ref{sec:results}.  We do
not expect this to affect the universal trends we establish in the
main text, since differences in EOS already produce different
$Y_e(\rho)$ profiles yielding simulation results that consistently
follow the universal trends. However, it reaffirms that for
quantitatively reliable GW signal predictions, a detailed and
converged treatment of prebounce deleptonization with radiation
hydrodynamics is vital.

In the final block of Table~\ref{tab:coconut_tests}, we summarize
results of simulations in which we increase the resolution and order
of the time integrator in \texttt{CoCoNuT} simulations. These lead to
waveform mismatches of up to 0.4\%, significantly smaller than those
from systematic errors induced by the prebounce deleptonization
treatment. As pointed out in Section~\ref{sec:coconut}, we transition
from the $Y_e(\rho)$ deleptonization prescription to neutrino leakage
when the entropy along the polar axis exceeds
$3\,k_b\,\mathrm{baryon}^{-1}$. In rotating models, this is a fraction
of a millisecond before this occurs on the equatorial axis, which is
our definition of the time of core bounce. The row labeled
``\texttt{CoCoNuT} Eq.\ Bounce'' shows that having the trigger on the
equatorial axis results in negligible differences.

To summarize, our 1D and 2D simulation results are essentially
independent of the neutrino interaction table resolution and of the 1D
grid resolution. There is a weak dependence on the 2D grid resolution
(below 1\% mismatch in all resolution tests). However, the results are
sensitive to the treatment of prebounce deleptonization at the level
of several percent GW mismatch. Again, future GR radiation
hydrodynamic simulations with detailed nuclear electron capture rates
will be needed for reliable predictions of gravitational waveforms
from rotating core collapse.

\end{document}